\documentclass[aps,prd,twocolumn,showpacs,superscriptaddress,nofootinbib,10pt]{revtex4-2}

\usepackage{acronym}
\usepackage{amsfonts}
\usepackage{amsmath}
\usepackage{amssymb}
\usepackage{bm}
\usepackage{cases}
\usepackage{color}
\usepackage{comment}
\usepackage[dvipsnames]{xcolor}
\usepackage{enumerate}
\usepackage{epsfig}
\usepackage{etoolbox}
\usepackage{fancyref}
\usepackage{fancyhdr}
\usepackage[T1]{fontenc} 
\usepackage{graphicx}
\usepackage{hyperref}
\usepackage{indentfirst}
\usepackage{latexsym}
\usepackage{mathrsfs}
\usepackage{mciteplus}
\usepackage{multirow}
\usepackage{rotating}
\usepackage{scrextend}
\usepackage{soul}
\usepackage{subfigure}
\usepackage{verbatim}
\usepackage{amsthm}

\newcommand{\jeven}{\mbox{\rm\j}}

\begin{document}

\title[Gauge-Independent Metric Reconstruction of Perturbations of Vacuum Spherically-Symmetric Spacetimes]{Gauge-Independent Metric Reconstruction of Perturbations of Vacuum Spherically-Symmetric Spacetimes}

\author{Michele Lenzi}
\email{lenzi@ice.csic.es}
\affiliation{Institut de Ci\`encies de l'Espai (ICE, CSIC), Campus UAB, Carrer de Can Magrans s/n, 08193 Cerdanyola del Vall\`es, Spain}
\affiliation{Institut d'Estudis Espacials de Catalunya (IEEC), Edifici Nexus, Carrer del Gran Capit\`a 2-4, despatx 201, 08034 Barcelona, Spain}

\author{Carlos F. Sopuerta}
\email{carlos.f.sopuerta@csic.es}
\affiliation{Institut de Ci\`encies de l'Espai (ICE, CSIC), Campus UAB, Carrer de Can Magrans s/n, 08193 Cerdanyola del Vall\`es, Spain}
\affiliation{Institut d'Estudis Espacials de Catalunya (IEEC), Edifici Nexus, Carrer del Gran Capit\`a 2-4, despatx 201, 08034 Barcelona, Spain}

\date{\today}

\begin{abstract}
Perturbation theory of vacuum spherically-symmetric spacetimes (including the cosmological constant) has greatly contributed to the understanding of black holes, relativistic compact stars and even inhomogeneous cosmological models. 
The perturbative equations can be decoupled in terms of (gauge-invariant) master functions satisfying $1+1$ wave equations.
In this work, building on previous work on the structure of the space of master functions and equations, we study the reconstruction of the metric perturbations in terms of the master functions.
To that end, we consider the general situation in which the perturbations are driven by an arbitrary energy-momentum tensor. 
Then, we perform the metric reconstruction in a completely general perturbative gauge.  
In doing so, we investigate the role of Darboux transformations and Darboux covariance, responsible for the isospectrality between odd and even parity in the absence of matter sources and also of the physical equivalence between the descriptions based on all the possible master equations.
We also show that the metric reconstruction can be carried out in terms of any of the possible master functions and that the expressions admit an explicitly covariant form.
\end{abstract}

\maketitle

\section{Introduction}

General Relativity (GR) is a classical theory of gravity that can accommodate all the physical observations/experiments involving gravitational fields, despite a better theoretical understanding of some phenomena, like dark matter and dark energy, may be desirable. It also predicts the generation of gravitational waves (GWs), whose existence has been confirmed by recent observations made with the ground-based laser-interferometer detectors LIGO and Virgo~\cite{LIGOScientific:2016aoc,LIGOScientific:2018mvr,LIGOScientific:2021usb}, and more recently by the LIGO-Virgo-KAGRA scientific collaboration~\cite{KAGRA:2021vkt}. This new type of Astronomy, usually known as Gravitational Wave Astronomy, constitutes one of the best tools we have to test gravity in general and GR, specially in dynamical situations involving the strongest possible gravitational field configurations corresponding to objects like neutron stars and black holes (BHs) (see~\cite{LIGOScientific:2016lio} for tests of GR with the first GW event ever detected~\cite{LIGOScientific:2016vlm}). The expectation is that the precision of these tests will improve as the sensitivity of the current detectors increases with the different instrument upgrades.  However, we may need to wait for third-generation ground detectors, as Einstein Telescope~\cite{Sathyaprakash:2012jk} or Cosmic Explorer~\cite{Evans:2021gyd}, to be able to make groundbreaking discoveries in fundamental physics with GW observations. Alternatively, we can also expect revolutionary science from space-based GW detectors operating in the low-frequency band like LISA~\cite{LISA:2017pwj,Barausse:2020rsu,LISA:2022kgy,Colpi:2024xhw}, which is expected to be launched around 2035. 

To a large extent, GW science requires the use of precise theoretical models of the expected GW signals, the gravitational waveforms (see~\cite{LISAConsortiumWaveformWorkingGroup:2023arg} for the case of LISA) which, in turn, require the development of a number of exact, approximate, and numerical techniques. Of particular relevance is relativistic perturbation theory, which provides a good theoretical basis for the description of the generation and propagation of GWs as well as for the computation of associated physical quantities like GW energy and angular momentum fluxes (see~\cite{Misner:1973cw} for textbook accounts).  When applied to BHs, the so-called black hole perturbation theory (BHPT) can successfully describe important physical phenomena such as scattering processes (we can compute the black hole greybody factors) and their quasinormal oscillations (the BH \emph{ringdown} in the context of the coalescence of binary BH systems).  See~\cite{Gerlach:1979rw,Gerlach:1980tx,Sarbach:2001qq,Clarkson:2002jz} for BHPT formalisms (and~\cite{Mukohyama:2000ui,Kodama:2003jz} for D-dimensional maximally-symmetric background spacetimes) and~\cite{Nollert:1999re,Kokkotas:1999bd,Andersson:2000mf,Kokkotas:2003mh} for reviews describing perturbations of BHs and neutron stars. Applications of BHPT to gravitational wave astronomy can be found in~\cite{Sasaki:2003xr,Ferrari:2007dd,Berti:2018vdi}, and to fundamental physics in~\cite{Cardoso:2012qm,Brito:2015oca,Barack:2018yly}.

BHPT is also the main tool to describe one of the most important sources of GWs for space-based observatories like LISA, the so-called extreme-mass-ratio inspirals (EMRIs), where a stellar-mass compact object is captured by a supermassive BH and performs a long inspiral before it plunges into the big BH (see~\cite{Amaro-Seoane:2012lgq,Babak:2017tow,Berry:2019wgg,Amaro-Seoane:2022rxf,Cardenas-Avendano:2024mqp} for details on the physics of and science from EMRIs). The inspiral is driven by gravitational backreaction which, in the context of BHPT, is described by the so-called self-force~\cite{Mino:1997bx} (see~\cite{Poisson:2011nh,Barack:2009ux,Barack:2018yvs,Pound:2021qin} for accounts on the self-force programme). The computation of the self-force requires regularization (it becomes singular at the stellar-mass compact object location) of the metric perturbations as it is made out of their gradients and evaluated at the particle location. It also requires, in principle, the knowledge of all the components of the metric perturbations. This means that we need to have a good understanding of BHPT.

The first steps in BHPT for non-spinning BHs were given in the 1950s by Regge and Wheeler~\cite{Regge:1957td}. Using the spherical symmetry of the BH background, they decomposed the perturbations in scalar, vector, and tensor harmonics, and showed that the harmonics decouple, and also that modes with different parity decouple. Then, they managed to study the odd-parity perturbations of a Schwarzschild BH~\cite{Schwarzschild:1916uq} and were able to decouple the perturbative Einstein equations for each harmonic mode in terms of a single complex function, the \emph{master} function, which satisfies a wave type equation with a potential that represents the response of the BH to the perturbations. It turns out that the master functions carries the true degrees of freedom of the gravitational field and hence, we can extract from it all the physical information encoded in the perturbations, in particular the gravitational radiation emitted and its energy-momentum content.  As a consequence, it has been shown that master functions are gauge invariant. The even-parity sector of the metric perturbations of Schwarzschild was not put in the same status until the 1970s, when Zerilli~\cite{Zerilli:1970se,Zerilli:1970la} managed to decouple the perturbative equations and find master functions and master equations with a different potential (see also~\cite{Moncrief:1974vm} and also~\cite{Vishveshwara:1970cc,Vishveshwara:1970zz}).   

In the case of spinning BHs, given that the Kerr metric was found in the 1960s~\cite{Kerr:1963ud}, a similar framework as in the non-spinning case was developed in the 1970s from the work of Teukolsky~\cite{Teukolsky:1972le,Teukolsky:1973ha}. The main difference is that while in the Schwarzschild case the master functions are functions of the metric perturbations and its first-order derivatives, in the Kerr case, they are given in terms of gauge-invariant components of the perturbative Weyl curvature tensor, which are expressed in a compact way if one uses the Newman-Penrose formalism~\cite{Newman:1961qr} with a basis of vectors adapted to the main null principal directions of the Kerr Weyl tensor (it has a special algebraic structure corresponding to the so-called type D in the Petrov classification~\cite{Stephani:2003tm}). 

Although master functions carry the physical information of the perturbations, there are situations where we may need to know all the metric perturbations, independently on whether they are gauge invariant or not.  This is the case of the self-force, which tells us how an object moves around a BH under its own self-gravity. The self-force is a gauge-dependent quantity, but it is necessary to derive the perturbed trajectory, from where we can compute gauge-invariant gravitational wave polarizations (at the next perturbative order).  Therefore, we need in BHPT a procedure to recover the metric perturbations from the master functions. 

The goal of this paper is to generalize existing metric reconstruction procedures for perturbations of vacuum spherically-symmetric backgrounds~\cite{Spiers:2023mor,Thompson:2018lgb,Hopper:2010uv, Brizuela:2009qd,Martel:2003jj}, including a cosmological constant (throughout the paper we refer to these spacetimes as $\Lambda$-vacuum spacetimes). In our analysis, we allow for energy-momentum distributions appearing at the perturbative level (i.e., not affecting the background).  We will provide metric reconstruction formulae that are independent of the gauge (they are derived in a completely general gauge).  We will also consider the case where there is a general energy-momentum tensor at the perturbative level (at first-order in the perturbations), which is the situation we have when computing the perturbations induced by a point-like object moving in the background geometry and whose rest-mass can be considered small so that the modified gravitational field can be treated in a perturbative way.  Finally, we also consider the metric reconstruction procedure in the framework of the symmetries of the perturbative equations, studied by the authors in previous works. In particular, we consider here the metric reconstruction procedure in the context of what we called \emph{Darboux covariance}, a symmetry in the space of possible master equations that tells us that the physics they describe is equivalent.

The starting point of this work is a previous study~\cite{Lenzi:2021wpc} where we analyzed the phase space of master functions and equations in $\Lambda$-vacuum background spacetimes. To that end we assumed: (i) The master functions are linear combinations of the metric perturbations and its first-order derivatives. (ii) The coefficients of these linear combinations are  time-independent, i.e. they only depend on the radial area coordinate $r$. (iii) The master functions satisfy a wave equations of the form:
\begin{equation}
\left(\square^{}_2 - \Omega^{}_\ell(r) \right)\Psi^{}_{\ell m} = 0 \,,    \label{covariant-master-equation-vacuum}
\end{equation}
where $\square_2$ is the d'Alambertian operator constructed from the metric on the two-dimensional Lorentzian manifold $M^2$ orthogonal to the 2-spheres of symmetry (i.e. the orbits of the rotation group of Killing symmetries of the background spacetime), and $\Omega(r)$ is essentially the potential which, in the study of~\cite{Lenzi:2021wpc} is initially arbitrary and its form is only determined by the perturbative Einstein equations. It is important to remark that the operator in~\eqref{covariant-master-equation-vacuum}, acting on the master function $\Psi_{\ell m}$, is constructed only from the background geometry.

Assumptions (i) and (ii) are shared by the known master functions, in particular those introduced by Regge-Wheeler and by Zerilli. The first one is a simplification that limits the space of master functions. The second one is a consequence of the structure of the master functions: Since they are linear in the perturbations, the coefficients of the linear combination have to be functional of the background metric.  The outcome of the study of~\cite{Lenzi:2021wpc} can be summarized as follows: There are two branches of master equations: (i) The \textit{standard} branch, which contains master equations with the known potentials, i.e. the Regge-Wheeler potential~\cite{Regge:1957td} for odd-parity perturbations and the Zerilli potential~\cite{Zerilli:1970la} for even-parity perturbations. (ii) The \textit{Darboux} branch, where there is an infinite set of master equations with new potentials. 

Some interesting consequences to remark from the study are: (i) All the resulting master functions  turn out to be gauge-invariant despite it was not imposed. This is consistent with the fact that master functions encode the true degrees of freedom of the gravitational perturbations.
(ii) All the master functions admit a fully covariant form with respect to the two-dimensional Lorentzian metric. 
(iii) In the same way that the Regge-Wheeler and Zerilli potentials coincide for the case of a maximally-symmetric background, the equations for the potentials in the second branch also coincide. 

On the other hand, in~\cite{Lenzi:2021njy} it was shown that, for a given harmonic, all the master functions are physically equivalently, independently of the parity. The main reason behind this it that all the master equations/functions can be related by means of Darboux transformations (see also~\cite{Glampedakis:2017rar}). A direct consequence is that Darboux transformations preserves the spectrum, that is, they are isospectral transformations. More specifically, Darboux \emph{covariance} implies that all these master equations lead to the same reflection and transmission coefficients, which is generalization of the result of Refs.~\cite{1980RSPSA.369..425C,Chandrasekhar:1992bo}. In~\cite{Lenzi:2021njy} it was also shown that there is another type of isospectral symmetries of the master equations, actually an infinite number of them, associated with deformations along the flow of the Korteweg-de Vries (KdV) equation~\cite{doi:10.1137/1018076}. The associated conservation laws produce an infinite set of conserved quantities~\cite{Miura:1968JMP.....9.1204M, Zakharov:1971faa, Lax:1968fm}, the KdV integrals. These integrals were also shown to be invariant under Darboux transformations. In~\cite{Lenzi:2022wjv}, it was shown that one can determine the reflection and transmission coefficients for BH scattering only in terms of the KdV integrals via a moment problem. In~\cite{Lenzi:2023inn}, practical computations are described for the case of the Schwarzschild potentials and also for a P\"oschl-Teller potential. 

In this paper, we consider the problem of reconstructing all the metric perturbations in terms of the solutions of the master equation in the case of spherically-symmetric $\Lambda$-vacuum backgrounds spacetimes, without fixing the perturbative gauge, i.e. in a gauge-independent way\footnote{Note the difference between gauge independence and gauge invariance. The first one referes to perturbative computations in a generic gauge, while the second is a property of a perturbative object which does not change its functional form in terms of metric perturbations under changes in the perturbative gauge.}. This reconstruction process is of great importance for the computation of the self-force~\cite{Poisson:2011nh,Barack:2009ux,Barack:2018yvs}, and it can be very helpful for any other application requiring the knowledge of the metric perturbations beyond the gauge-invariant sector.  This includes Hamiltonian formulations of the perturbations (see~\cite{Brizuela:2008sk} for a treatment of the odd-parity sector), in particular towards quantization in a semiclassical context (see, e.g.~\cite{Bernar:2016zgq,Bernar:2017kug,Bernar:2018nww}). In our metric reconstruction we have included a generic energy-momentum tensor sourcing the perturbations, generalizing in this way several results in our previous papers~\cite{Lenzi:2021wpc,Lenzi:2021njy}. We also show a systematic way of carrying the reconstruction, which may be useful for generalizing this to other situations where we may have different (spherically-symmetric) background (for instance, for the treatment of compact relativistic stars), or even a different theory of gravity.  Finally, we put the metric reconstruction process in the context of the hidden symmetry of the space of master functions and equations studied in~\cite{Lenzi:2021njy}. Actually, we explain how to realize the metric reconstruction in terms of master functions different from the ones known in the broad literature on the subject.

\noindent\textit{Plan of the paper}. In Sec.~\ref{relativistic-perturbation-theory} we introduce the main elements of BHPT that use in this work: The perturbative Einstein equations; the background spacetimes; the harmonic decomposition; and gauge invariant quantities.  In Sec.~\ref{master-functions} we describe the space of master functions and equations introduced in~\cite{Lenzi:2021wpc}, and whose structure was studied in~\cite{Lenzi:2021njy}, extending the results to the case in which we have an arbitrary energy-momentum tensor at the first perturbative order. We also extend the study of Darboux covariance clarifying some open questions in~\cite{Lenzi:2021njy}. In Sec.~\ref{metric-reconstruction-procedure}, we describe a systematic procedure to carry out the metric reconstruction and provide general results taking in account the full space of master functions introduced previously. In Sec.~\ref{discussion} we discuss these results, compare them with other formalisms in General Relativity and consider possible applications and extensions to other situations.  We have the following appendices: In Appendix~\ref{background-spacetime-metric} we briefly discuss the family of background spacetime geometries. In Appendix~\ref{sphericalharmonics} we list some properties of scalar, vector, and tensor spherical harmonics that are relevant for this paper.  In Appendix~\ref{harmonic-components-perturbative-equations}, we list Einstein perturbative equations, including a generic energy-momentum tensor contribution at first-order, both in terms of the original metric perturbations and in terms of the gauge invariant metric perturbations. Finally, in Appendix~\ref{energy-momentum-harmonic-components-equations} we list the harmonic components of the energy-momentum conservation equations.

\noindent\textit{Conventions}. Through this paper we use units in which $c=8\pi G=1$, where $c$ is the speed of light and $G$ the Newtonian gravitational constant.  We use a definition of the Einstein tensor that includes the cosmological constant term [see Eq.~\eqref{efes-background}]. Regarding indices: Greek letters are used for spacetime indices; capital Latin letters are used for indices in the 2-sphere ($S^2$); small-case Latin letters are used for indices in the Lorentzian two-dimensional manifold $M^2$.  The metric on $S^2$ is written as $\Omega_{AB}$ while in $M^2$ it is $g_{ab}$.
Covariant differentiation in the four-dimensional background spacetime is either denoted by a semicolon or by the nabla operator $\widehat{\nabla}$. Instead, in 
$S^2$ is denoted using a vertical bar ($\Omega_{AB|C} = 0$) and in $M^2$ by a colon ($g_{ab:c} = 0$). The antisymmetric covariant unit tensor associated with the volume form (Levi-Civita tensor) in $S^2$ is denoted by $\epsilon_{AB}$, and the corresponding one on the Lorentzian manifold $M^{2}$ is denoted by $\varepsilon_{ab}\,$.

\section{Black Hole Perturbation Theory in the non-rotating case}
\label{relativistic-perturbation-theory} 

BHPT is a particular case of relativistic perturbation theory (see~\cite{Stewart:1974uz,Wald:1984cw}), in which a physically realistic system can be described as a deviation from an idealized situation. We assume the existence of a one-parameter family of spacetimes, $(\mathcal{M}_\lambda, g_\lambda)$, with $\lambda=0$ corresponding to the \textit{background} spacetime describing the idealized situation, and construct the physical spacetime by Taylor expanding around $\lambda=0$. In this way, the parameter $\lambda$ formally controls the magnitude of the perturbations (see e.g.~\cite{Stewart:1974uz,Bruni:1996im,Bruni:2002sma,Sopuerta:2003rg}). Since in our case $\lambda$ is a dummy parameter we ignore it through the paper.

In our case, the background spacetime is taken to be a $\Lambda$-vacuum spherically symmetric spacetime. Then, the background spacetime metric, $\widehat{g}^{}_{\mu\nu}$, satisfies the following version of Einstein's field equations~\footnote{We use a hat to denote quantities associated with the background spacetime, like $\widehat{Q}$.}:
\begin{eqnarray}
\widehat{G}^{}_{\mu\nu} = \widehat{R}^{}_{\mu\nu} -\frac{1}{2}\widehat{g}^{}_{\mu\nu} \widehat{R} + \Lambda\,\widehat{g}^{}_{\mu\nu} = 0\,,
\label{efes-background}
\end{eqnarray}
where $\widehat{R}^{}_{\mu\nu}$ and $\widehat{G}^{}_{\mu\nu}$ denote the Ricci and Einstein tensors of the background metric respectively, $\widehat{R} = \widehat{g}^{\mu\nu} \widehat{R}^{}_{\mu\nu}$ is the background scalar curvature, and $\Lambda$ is the cosmological constant. Since the background is $\Lambda$-vacuum we have from Eq.~\eqref{efes-background}:
\begin{equation}
\widehat{R}^{}_{\mu\nu} = \Lambda \widehat{g}^{}_{\mu\nu}\,, \quad
\widehat{R} = 4 \Lambda \,.    
\label{ricci-and-scalar-curvature-background}
\end{equation}
After a correspondence between the physical (perturbed) and background spacetimes is established (see Sec.~\ref{Ss: gauge-invariance}), we can write the physical spacetime metric $g^{}_{\mu\nu}$ as:
\begin{equation}
g^{}_{\mu\nu} = \widehat{g}^{}_{\mu\nu} + h^{}_{\mu\nu} \,,
\label{fundamental-perturbative-equation}
\end{equation}
where $h_{\mu\nu}$ ($|h_{\mu\nu}|\ll|\widehat{g}_{\mu\nu}|$) are the metric perturbations.

Let us now fix the notation we adopt. Given any quantity $Q$ in the perturbed spacetime, a $\delta$ in front of it denotes its perturbative part, i.e. $\delta Q = Q - \widehat{Q}$. For the example, in the case of the metric we have: $h_{\mu\nu} = \delta g_{\mu\nu} = g_{\mu\nu} - \widehat{g}_{\mu\nu}$.  One can then find that the perturbation of the Christoffel symbols at first order can be expressed in terms of the background covariant derivatives (denoted here by a semicolon) of the metric perturbations as follows:
\begin{eqnarray}
\delta\Gamma_{\mu\nu}^\rho= \frac{1}{2}\widehat{g}^{\rho\sigma}_{}\left( h^{}_{\mu\sigma;\nu} + h^{}_{\nu\sigma;\mu} - h^{}_{\mu\nu;\sigma} \right)\,.
\label{perturbed-Christoffel}
\end{eqnarray}
While the Christoffel symbols of the background are not tensors with respect to  changes of the coordinates of the background spacetime, their perturbations are since they are the subtraction of two Christoffel symbols.  
Then, we can write the perturbations of the Riemann tensor in terms of the perturbed Christoffel symbols:
\begin{equation}
\delta R^{\mu}{}^{}_{\nu\rho\sigma} = \delta\Gamma^{\mu}_{\nu\sigma;\rho} -  \delta\Gamma^{\mu}_{\nu\rho;\sigma} = 
2\delta\Gamma^{\mu}_{\nu[\sigma;\rho]} \,,
\label{perturbed-Riemann}
\end{equation}
and the perturbations of the Ricci tensor follows directly:
\begin{eqnarray}
\delta R^{}_{\mu\nu} = \delta\Gamma^{\rho}_{\mu\nu;\rho} - \delta\Gamma^{\rho}_{\rho\mu;\nu}\,.
\label{perturbed-Ricci0}
\end{eqnarray}
The Einstein tensor can be decomposed as: $G_{\mu\nu} = \widehat{G}^{}_{\mu\nu} + \delta G_{\mu\nu} $, and we can write $\delta G_{\mu\nu}$ in terms of the previous perturbative objects as follows [see Eq.~\eqref{efes-background}]:
\begin{equation}
\delta G^{}_{\mu\nu} = \delta R^{}_{\mu\nu} - \frac{1}{2}\widehat{g}^{}_{\mu\nu} \delta R - \Lambda h^{}_{\mu\nu} \,.
\end{equation}
Then, the perturbation of the Einstein tensor can finally be written as
\begin{eqnarray}
\delta G^{}_{\mu\nu} & = & -\frac{1}{2}\widehat{\square}\,\bar{h}^{}_{\mu\nu} -\widehat{R}^{\rho}_{}{}^{}_\mu{}^{\sigma}{}^{}_\nu \bar{h}^{}_{\rho\sigma}  + \widehat{\nabla}^{}_{(\mu}\mathcal{L}^{}_{\nu)} \nonumber \\ 
& - & \frac{1}{2}\widehat{g}^{}_{\mu\nu}\left(\widehat{\nabla}^{}_{\rho}\mathcal{L}^{\rho}\right) - \Lambda \bar{h}^{}_{\mu\nu} \,,
\label{perturbative-einstein-tensor}
\end{eqnarray}
where $\widehat{R}^{\rho}_{}{}^{}_\mu{}^{\sigma}{}^{}_\nu$ is the background Riemann tensor, and we have introduced the trace-reversed metric perturbations:
\begin{equation}
\bar{h}^{}_{\mu\nu} = h^{}_{\mu\nu} - \frac{1}{2}\widehat{g}^{}_{\mu\nu}  h\,,
\end{equation}
with $h$ being the trace of $h_{\mu\nu}$ with respect to the background metric: $h = \widehat{g}^{\mu\nu}h^{}_{\mu\nu} = -\widehat{g}^{\mu\nu}\bar{h}^{}_{\mu\nu}$. 
The operator $\widehat{\square}$ is the d'Alambertian associated with the background metric, thus:
\begin{equation}
\widehat{\square}\, \bar{h}^{}_{\mu\nu} = \bar{h}^{}_{\mu\nu;\rho}{}^{;\rho}\,.
\end{equation}
Finally, in Eq.~\eqref{perturbative-einstein-tensor} we have introduced the quantity:
\begin{equation}
\mathcal{L}^{}_{\mu} = \widehat{g}^{\rho\sigma}\widehat{\nabla}^{}_\rho \bar{h}^{}_{\sigma\mu} \,,
\label{lorenz-gauge-quantity}
\end{equation}
to isolate terms which vanish in the Lorentz gauge, i.e. $\mathcal{L}^{}_{\mu} = 0$. Nevertheless, in this paper we work in a completely general gauge.

In the following, are going to consider the situation in which the perturbations are sourced by matter fields whose nature we do not specify and hence, it is arbitrary. In other words, we consider a completely general energy-momentum tensor $T_{\mu\nu}$, which only affects the metric perturbations but not the background. In this way, the perturbative field equations become:
\begin{equation}
\delta G^{}_{\mu\nu} =  
T^{}_{\mu\nu} \,,
\label{efes-full}
\end{equation}
with $\delta G_{\mu\nu}$ given by Eq.~\eqref{perturbative-einstein-tensor}. Until here, we have only imposed the background to be a $\Lambda$-vacuum solution [Eq.~\eqref{efes-background}]. In what follows, we summarize the ingredients involved in the treatment of perturbations of spherically-symmetric background spacetimes.

\subsection{Structure of the Background Spacetime}
\label{vacuum-spherically-spacetime}

Our background spacetimes are spherically-symmetric solutions of the Einstein vacuum equations with a cosmological constant [see Eq.~\eqref{efes-background} and Appendix~\ref{background-spacetime-metric}].  They have a special geometric structure as they are the \emph{warped} product of two manifolds: $M^2\times_r S^2$, where $S^2$ is the $2$-sphere, $r$ is the radial area coordinate defining the warp factor, and  $M^2$ is a two-dimensional Lorentzian manifold. Therefore, the background metric is the semidirect product of a Lorentzian metric on $M^2$, $g_{ab}$, and the unit curvature metric on $S^2$, $\Omega_{AB}$:
\begin{eqnarray}
ds^2=\widehat{g}^{}_{\mu\nu}dx^{\mu}dx^{\nu} = g^{}_{ab} dx^a dx^b + r^2\Omega^{}_{AB} d\Theta^A d\Theta^B  \,,
\label{warped-metric-background}
\end{eqnarray}
where
\begin{eqnarray}
g^{}_{ab} dx^a dx^b & = & -f(r)\,dt^2+\frac{dr^2}{f(r)} \,, 
\label{b-metric-1} \\
\Omega^{}_{AB} d\Theta^A d\Theta^B & = & d\theta^2 +\sin^2\theta d\varphi^2 \,,
\label{b-metric-2}
\end{eqnarray}
where $(x^{a})=(t,r)$ and $(\Theta^{A})=(\theta,\varphi)$ are coordinates on $M^2$ (Schwarzschild coordinates) and $S^2$ respectively, and $f(r)$ is a function parameterising time translations (related to the redshift).  Using the background Einstein's equations~\eqref{efes-background} one derive ordinary differential equations (ODEs) for $f(r)$:
\begin{eqnarray}
r f' + f + \Lambda r^2 - 1 = 0 \,, 
\label{efesbackground-1}\\
r \left(f'' + 2 \Lambda \right) + 2 f' = 0 \,.
\label{efesbackground-2}
\end{eqnarray}
From these equations we can write the cosmological constant in terms of $f(r)$ and its derivatives in two different forms:
\begin{eqnarray}
\Lambda & = & \frac{1}{r^2}\left[1-\left(rf\right)'\right] = -\frac{1}{2r^2}\left(r^2 f' \right)' \,.
\label{exp-cosmological-constant} 
\end{eqnarray}
Now, combining this expression for $\Lambda$ with Eq.~\eqref{efesbackground-1} we can write 
\begin{eqnarray}
&& \left[r\left(1-f-\frac{\Lambda}{3}r^2\right)\right]'= 0 \quad \Rightarrow \nonumber \\
&& \Rightarrow \quad r\left(1 - f -\frac{\Lambda}{3}r^2 \right) = 2M = \text{const.} \,, 
\label{exp-mass}
\end{eqnarray}
where $M$ is the spacetime mass.  Actually, from this equation we obtain the following well-known expression for $f(r)$:
\begin{equation}
f(r) = 1 -\frac{2M}{r} -\frac{\Lambda}{3}r^2 \,.
\label{f-background-function}
\end{equation}
See Appendix~\ref{background-spacetime-metric} for the different well-known metrics included. 

On the Lorentzian two-dimensional spacetime $M^2$ we can introduce a basis of vectors that is going to be useful in order to express some of the results we are presenting in this work:
\begin{eqnarray}
r^{}_a = r^{}_{:a} \,, \qquad
t^a = - \varepsilon^{ab} r^{}_b \,. 
\label{definition-ra-ta}
\end{eqnarray}
The first one, $r_a$, is the gradient of the area radial coordinate, an invariant of the rotation group of symmetries. The second one is a timelike Killing vector, which makes the spacetime static (it is irrotational and hence, it generates orthogonal hypersurfaces). 
Moreover, they are orthogonal:
\begin{equation}
t^a r^{}_a = 0 \,,    
\end{equation}
and their norms are:
\begin{eqnarray}
r^a r^{}_a = f \,, \qquad  t^a t^{}_a = -f \,.
\end{eqnarray}

On the other hand, it is convenient for future purposes to introduce the d'Alambertian operator associated with the metric of $M^2$, i.e. $g_{ab}$:
\begin{equation}
\square^{}_2 \Psi \equiv g^{ab} \Psi^{}_{:ab} = \frac{1}{\sqrt{-g}}\partial^{}_a\left(\sqrt{-g}g^{ab}\partial^{}_b\Psi\right)\,.
\label{dalambertian-M2}
\end{equation}
In Schwarzschild coordinates we have the following expression for $\square_2$:
\begin{equation}
\square^{}_2\phi  =  -\frac{1}{f}\frac{\partial^{2}\phi}{\partial t^{2}} +\frac{\partial}{\partial r}\left(f\frac{\partial\phi}{\partial r}\right)\,.
\label{expression-box-M2}
\end{equation}
%

\subsection{Expansion of the Perturbations in Spherical Harmonics} \label{multipolar-expansion}

The warped geometry of the background leads to the separation of solutions of the wave equations whose operator is built with the background metric. In particular, we can separate the dependence on the coordinates of $M^2$ from the angular dependence on $S^2$ for the perturbative Einstein equations~\eqref{efes-full}.  In practice, this is done by expanding the metric perturbations using the eigenfunctions of the Laplacian,
the scalar spherical harmonics, and the vector and tensor objects we derive from them. In turn, we can distinguish between even- and odd-parity harmonics depending on how they transform under parity transformation given by: $(\theta,\phi)$ $\rightarrow$ $(\pi-\theta, \phi+\pi)$.  If a given harmonic object $\mathcal{O}^{\ell m}$ transforms as: $\mathcal{O}^{\ell m} \rightarrow (-1)^{\ell}\mathcal{O}^{\ell m}$ it is said to be of the even-parity type; while if it transforms as $\mathcal{O}^{\ell m} \rightarrow (-1)^{\ell+1}\mathcal{O}^{\ell m}$ it is said to be of the odd-parity type.  

With this in mind, the scalar, vector and tensor spherical harmonics are:  \textit{Scalar spherical harmonics}: $Y^{\ell m}$. As we have just mentioned,  they are eigenfunctions of the Laplace operator on $S^2$ (see Appendix~\ref{sphericalharmonics}):
\begin{eqnarray}
\Omega^{AB}Y^{\ell m}_{|AB} = -\ell( \ell +1)Y^{\ell m}\,.
\label{laplacian-of-Ylm}
\end{eqnarray}
\textit{Vector Spherical Harmonics}: They are defined for $\ell \geqslant 1$, are given by:
\begin{eqnarray}
Y^{\ell m}_{A} \equiv Y^{\ell m}_{|A} \qquad~ & & \quad \text{Even (polar) parity}\,, \\[2mm]
X^{\ell m}_{A} \equiv -\epsilon^{}_A{}^B Y^{\ell m}_{B} & & \quad \text{Odd (axial) parity}\,.
\label{vectorharmonics}
\end{eqnarray}
\textit{Tensor Spherical Harmonics}: The basis of symmetric $2$nd-rank tensor spherical harmonics, which are defined for $\ell \geqslant 2$, are given by 
\begin{eqnarray}
T_{AB}^{\ell m} \equiv Y^{\ell m}\,\Omega_{AB}  \qquad\qquad\qquad\quad & & \quad \text{Even parity}\,, \\[2mm]
Y^{\ell m}_{AB} \equiv Y_{|AB}^{\ell m} + \frac{\ell(\ell+1)}{2}Y^{\ell m}\,\Omega_{AB} & & \quad \text{Even parity}\,, \\[2mm]
X_{AB}^{\ell m} \equiv X^{\ell m}_{(A|B)}  \qquad\qquad\qquad\qquad & & \quad \text{Odd parity}\,, 
\label{tensorharmonics}
\end{eqnarray}
with the following traces
\begin{equation}
\Omega^{AB}T^{\ell m}_{AB} = 2 Y^{\ell m}\,,\quad
\Omega^{AB}Y^{\ell m}_{AB} = \Omega^{AB} X^{\ell m}_{AB} = 0\,. 
\label{tensorharmonics-trace-properties}
\end{equation}
Some orthogonality and differential properties of these spherical harmonics, necessary to manipulate and separate the perturbed Einstein equations, can be found in Appendix~\ref{sphericalharmonics}.

The expansion of the metric perturbations in scalar, vector, and tensor spherical harmonics leads to the decoupling of the equations, not only for each harmonic $(\ell,m)$, but also for modes with different parity~\cite{Gerlach:1979rw, Gerlach:1980tx}. Then, we write
\begin{eqnarray}
h^{}_{\mu\nu} = \sum_{\ell ,m} h^{\ell m, \mathrm{odd}}_{\mu\nu} + h^{\ell m, \mathrm{even}}_{\mu\nu} \,,
\label{eq:2.7}
\end{eqnarray}
with
\begin{equation}
h^{\ell m, \mathrm{odd}}_{\mu\nu}  = \begin{pmatrix} 0 & h_a^{\ell m}X^{\ell m}_A \\[2mm]
  					   		                 \ast & h_2^{\ell m}X^{\ell m}_{AB}  
							    \end{pmatrix}\,, 
\label{metric-perturbation-odd}
\end{equation}
and
\begin{equation}					
h^{\ell m, \mathrm{even}}_{\mu\nu}  = \begin{pmatrix} h_{ab}^{\ell m}Y^{\ell m} & \jeven^{\ell m}_aY^{\ell m}_A \\[2mm]
\ast    & r^2\left( K^{\ell m}T^{\ell m}_{AB} + G^{\ell m}Y^{\ell m}_{AB}\right)\end{pmatrix}\,,
\label{metric-perturbation-even}
\end{equation}
where the asterisk tells us that the corresponding component is obtain form the fact that the metric perturbations is a symmetric tensorial object. Moreover, $K^{\ell m}$, $G^{\ell m}$ and $h_2^{\ell m}$ denote the scalar perturbations; $h_a^{\ell m}$ and $j^{\ell m}_a$ the vector perturbations; and $h_{ab}^{\ell m}$ the tensorial ones. All these quantities depend only on the coordinates of $M^2$, i.e. on $\{x^a\}$.  

Since we are considering the case in which the perturbations are sourced by a distribution of energy and momentum, we also need to decompose the energy-momentum tensor in spherical harmonics. We can follow closely the decomposition of the metric perturbations as both are symmetric tensor.  First, we separate even- and odd-parity harmonics:  
\begin{eqnarray}
T^{}_{\mu\nu} = \sum_{\ell ,m} T^{\ell m, \mathrm{odd}}_{\mu\nu} + T^{\ell m, \mathrm{even}}_{\mu\nu} \,,
\label{Tmunu-split}
\end{eqnarray}
The odd-parity sector has the following structure
\begin{equation}
T^{\ell m, \mathrm{odd}}_{\mu\nu}  = \begin{pmatrix} 0 & \mathcal{S}_a^{\ell m}X^{\ell m}_A \\[2mm]
\ast & \mathcal{S}^{\ell m}X^{\ell m}_{AB}  							    \end{pmatrix}\,, 
\label{Tmunu-odd}
\end{equation}
where the odd-parity harmonic modes are defined as follows:
\begin{eqnarray}
\mathcal{S}^a_{\ell m} & = & \frac{r^2}{\ell(\ell +1)}\int_{S^2}d\Omega\, T^{a A}\bar{X}_A^{\ell m}\,,
\label{Tmunu-salm}
\\
\mathcal{S}^{}_{\ell m} &=& 2r^4\frac{(\ell-2)!}{(\ell+2)!}\int_{S^2}d\Omega\, T^{AB}\bar{X}_{AB}^{\ell m}\,,
\label{Tmunu-slm}
\end{eqnarray}
where $d\Omega$ is the two-sphere \emph{volume} element.
The even-parity sector has a different structure given by:
\begin{equation}					
T^{\ell m, \mathrm{even}}_{\mu\nu}  = \begin{pmatrix} \mathcal{Q}_{ab}^{\ell m}Y^{\ell m} & \mathcal{P}^{\ell m}_aY^{\ell m}_A  \\[2mm]
\ast    & r^2\left( \mathcal{T}^{\ell m}T^{\ell m}_{AB} + \mathcal{P}^{\ell m}Y^{\ell m}_{AB}\right)\end{pmatrix}\,,
\label{Tmunu-even}
\end{equation}
and the harmonic components read:
\begin{eqnarray}
\mathcal{Q}^{ab}_{\ell m} &=& \int_{S^2}d\Omega\, T^{ab}\bar{Y}^{\ell m} \,,
\label{Tmunu-qablm}
\\
\mathcal{P}_{\ell m}^a &=& \frac{r^2}{\ell(\ell +1)}\int_{S^2}d\Omega\, T^{a A}\bar{Y}_A^{\ell m}\,,
\label{Tmunu-palm}
\\
\mathcal{P}_{\ell m} &=& 2r^2\frac{(\ell-2)!}{(\ell+2)!}\int_{S^2}d\Omega\, T^{AB}\bar{Y}_{AB}^{\ell m}\,,
\label{Tmunu-plm}
\\
\mathcal{T}_{\ell m} &=& \frac{r^2}{2} \int_{S^2}d\Omega\, T^{AB}\bar{T}_{AB}^{\ell m} \,.
\label{Tmunu-tlm}
\end{eqnarray}
In summary, $(\mathcal{P}^{\ell m}(x^a), \mathcal{T}^{\ell m}(x^a), \mathcal{S}^{\ell m}(x^a))$ are the scalar (polar) harmonics; $(\mathcal{P}^{\ell m}_a(x^b), \mathcal{S}^{\ell m}_a(x^b))$ are the vector harmonics; and  $\mathcal{Q}_{ab}^{\ell m}(x^c)$ are the tensor ones. 

We can now insert the harmonic decomposition of the metric perturbations and of the energy-momentum tensor into the perturbative Einstein equations~\eqref{efes-full} to obtain the decoupled equations for each harmonic and parity (see Appendix~\ref{harmonic-components-perturbative-equations}). In addition, it is important to take into account the energy-momentum tensor conservation equations, which are a consequence of the perturbative second Bianchi identities:
\begin{equation}
\widehat{\nabla}^\mu \delta G^{}_{\mu\nu} = 0 
\quad \Rightarrow \quad \widehat{\nabla}^{\mu} T^{}_{\mu\nu} = 0\,.
\label{perturbative-second-bianchi-identities}
\end{equation}
That is, the local energy-momentum conservation law is with respect to the background Levi-Civita connection.  Introducing here the harmonic decomposition of $h_{\mu\nu}$, $T_{\mu\nu}$, and derived quantities, we obtain the decoupled equations for each harmonic and parity (see Appendix~\ref{energy-momentum-harmonic-components-equations}).

\subsection{Gauge Invariance}
\label{Ss: gauge-invariance}

A key relation in relativistic perturbation theory~\cite{Stewart:1974uz,Bruni:1996im} is represented by Eq.~\eqref{fundamental-perturbative-equation}, where the perturbed metric is written as the sum of a background metric and the perturbations. 
As we already mentioned in Sec.~\ref{relativistic-perturbation-theory}, this relation requires the prescription of a correspondence (a diffeomorphism) between the background and the physical spacetimes. 
Given one of the possible (infinite) correspondences, we can pull back the physical metric and related tensorial structure into the background tensorial structure. 
Therefore, the relation established by Eq.~\eqref{fundamental-perturbative-equation} depends on the choice of correspondence. 
Fixing the correspondence between background and physical spacetimes is what we refer to as fixing the perturbative gauge.
To understand this better, consider a point $p$ in the physical spacetime and two different diffeomorphisms to the background (two different gauges), say $\phi_1$ and $\phi_2$, such that $q_1=\phi_1(p)$ and $q_2=\phi_2(p)$ are two points in the background spacetime (in principle they are different unless $\phi_1=\phi_2$). Since they are diffeomorphisms, we can write $q_2=\phi_2 \circ \phi^{-1}_1 (q_1)$. The composed mapping, $\Phi = \phi_2 \circ \phi^{-1}_1$ is what we call a gauge transformation, from gauge $\phi_1$ to gauge $\phi_2$, and acts by changing the reference point in the background to which we associate the perturbation. In perturbation theory we assume that the gauge tranformation induces a motion on the background point such that, when written in a given coordinate system, the coordinate version of the gauge transformation can be Taylor-expanded in $\lambda$ (see the beginning of Sec.~\ref{relativistic-perturbation-theory}), so that it looks, at first-order in $\lambda$, as follows (see, e.g.~\cite{Bruni:1996im}):
\begin{equation}
x^{\mu}\quad\longrightarrow\quad  x'^{\mu} = {x}^{\mu} + \lambda\xi^{\mu}\,,
\label{gauge-transformation}
\end{equation}
where ${x}^{\mu}$ and $x'^{\mu}$ are the coordinates of two points $q_2$ and $q_1$ respectively,
and the vector field $\xi^{\mu}$ is the first-order generator of the gauge transformation (see~\cite{Bruni:2002sma,Sopuerta:2003rg} for the analogous equation in multi-parameter perturbation theory). In what follows, we absorb the perturbative parameter $\lambda$ into the generator $\xi^\mu$, and assume it is small in the same way as we do with the metric perturbations, i.e. $|\xi^{\mu}|\ll|\widehat{g}_{\mu\nu}|$.

The gauge transformation in Eq.~\eqref{gauge-transformation} generates the following transformation of the metric perturbations:
\begin{equation}
h^{}_{\mu\nu}\quad\longrightarrow\quad {h'}^{}_{\mu\nu} = h^{}_{\mu\nu}  -2\,\xi^{}_{(\mu ;\nu)}\,.
\label{gauge-transformation-metric}
\end{equation}
Given that we have split the metric perturbations and energy-momentum tensor in harmonics,
we should also split the gauge transformation generator $\xi^\mu$. 
The even-parity sector harmonic $(\ell,m)$ of the gauge generator has the form:
\begin{equation}
\xi^{\ell m, \mathrm{even}}_{\mu}  = \left( \alpha^{\ell m}_{a}(x^{b})Y^{\ell m}\,,  r^{2}\beta^{\ell m}(x^{a})Y^{\ell m}_{A}\right) \,,
\label{gauge-vector-xi-1}
\end{equation}
and the corresponding odd-parity harmonic mode
\begin{equation}
\xi^{\ell m, \mathrm{odd}}_{\mu} =  \left( 0\,, r^{2}\gamma^{\ell m}(x^{a})X^{\ell m}_{A} \right) \,.
\label{gauge-vector-xi-2}
\end{equation}
Note that there are three gauge functions for even-parity perturbations and one for the odd-parity ones.  

Introducing the multipolar decomposition of the metric perturbations and the gauge vector into Eq.~\eqref{gauge-transformation-metric}, we obtain the transformation rules for even-parity metric perturbations under general perturbative gauge transformations:
\begin{eqnarray}
h'^{\ell m}_{ab} & =  & h^{\ell m}_{ab} - 2\,\alpha^{\ell m}_{(a:b)} \,, 
\label{gauge-transformation-even-1} \\[2mm]
\jeven'^{\ell m}_a & = & \jeven^{\ell m}_a - \left(\alpha^{\ell m}_{a} + r^{2}\beta^{\ell m}_{:a} \right)\,, 
\label{gauge-transformation-even-2} \\[2mm]
K'^{\ell m} & = & K^{\ell m} + \ell(\ell+1)\beta^{\ell m} -2\frac{r^{:a}}{r}\alpha^{\ell m}_a \,, 
\label{gauge-transformation-even-3} \\[2mm]
G'^{\ell m} & = & G^{\ell m} - 2\beta^{\ell m}\,.
\label{gauge-transformation-even-4}
\end{eqnarray}
and the ones for odd-parity metric perturbations:
\begin{eqnarray}
h'^{\ell m}_{a} & =  & h^{\ell m}_{a} - r^{2}\gamma^{\ell m}_{:a} \,, 
\label{gauge-transformation-odd-1} \\[2mm]
h'^{\ell m}_{2} & =  & h^{\ell m}_{2} - 2\, r^{2}\gamma^{\ell m} \,.
\label{gauge-transformation-odd-2}
\end{eqnarray}
It turns out that there are combinations of the metric perturbations and its derivatives that are invariant under gauge transformations. In the case of even-parity metric perturbations there are four independent gauge-invariant quantities, which can be written as:
\begin{eqnarray}
\tilde{h}^{}_{ab} & = & h^{}_{ab} - \kappa^{}_{a:b} - \kappa^{}_{b:a}\,, 
\label{expression-hathab} \\
\tilde{K}         & = & K + \frac{\ell(\ell+1)}{2} G - 2\frac{r^{a}}{r}\kappa^{}_{a} \,,
\label{expression-hatK}
\end{eqnarray}
where
\begin{eqnarray}
\kappa^{}_{a} & = & \jeven^{}_{a} - \frac{r^{2}}{2}G^{}_{:a}\,, \quad
r^{}_a = r^{}_{:a}~\Rightarrow~r^a=g^{ab}r^{}_b\,.
\end{eqnarray}
In the case of odd-parity metric perturbations there are two independent gauge-invariant quantities:
\begin{equation}
\tilde{h}^{}_{a} = h^{}_{a} -\frac{1}{2}h^{}_{2:a} + \frac{r^{}_{a}}{r} h^{}_{2} \,,
\label{expression-hta}
\end{equation}

We can also analyze the changes in the components of the energy-momentum tensor  $T^{}_{\mu\nu}$ under the gauge transformation of Eq.~\eqref{gauge-transformation}.  In analogy with what happens with the components of the metric perturbations [see Eq.~\eqref{gauge-transformation-metric}], we can show that the components of $T_{\mu\nu}$ transform as follows:
\begin{equation}
T^{}_{\mu\nu}\quad\longrightarrow\quad {T'}^{}_{\mu\nu} = T^{}_{\mu\nu}  - \pounds^{}_\xi \widehat{T}^{}_{\mu\nu}\,,
\label{gauge-transformation-Tmunu}
\end{equation}
where $\pounds_\xi$ denotes Lie differentiation (see, e.g.~\cite{Misner:1973cw,Wald:1974np}) with respect to the vector field generating the gauge transformation, $\xi^\mu$, and $\widehat{T}_{\mu\nu}$ is the background energy-momentum tensor, which in our case is identically zero. And hence it follows that the components of the energy-momentum tensor are gauge-invariant. This is exactly what the well-known Stewart-Walker lemma~\cite{Stewart:1974uz} tell us, and we could have just invoked it. But this discussion gives more relevance to the fact that we introduce the energy-momentum distribution already at first perturbative order.

\section{The Space of Master Functions and Equations}\label{master-functions}

The basic idea leading to master equations and master functions is to find linear functionals of the metric perturbations (for each harmonic and parity mode with $\ell\geq 2$), the master functions, containing \textit{only} the gauge-invariant metric perturbations themselves and their derivatives
\begin{equation}
\Psi^{}_{\ell m} = \Psi^{}_{\ell m}[r,\vec{H}^{\ell m},\partial^{}_a\vec{H}^{\ell m},\partial^{2}_{ab}\vec{H}^{\ell m}\,,\cdots]\,, \label{concept-master-function}    
\end{equation}
where
\begin{equation}
\vec{H}^{\ell m} = (\tilde{h}^{\ell m}_{ab},\tilde{K}^{\ell m}; \tilde{h}^{\ell m}_a)\,,
\label{master-function-dependence}
\end{equation}
so that the (vacuum) perturbative Einstein equations reduce to  wave equations for the master functions of the form: 
\begin{equation}
\left(\square^{}_2 - \frac{1}{f}V^{}_\ell \right)\Psi^{}_{\ell m} = 0  \,,  
\label{covariant-master-equation-vacuum-bis}  
\end{equation}
where $V_\ell(r)$ is the $\ell$-dependent potential, which will be in general different for even- and odd-parity modes.  From a more physical point of view, the master functions capture the true degrees of freedom of the gravitational field at the perturbative level, and hence they must be gauge invariant as reflected in the previous equations. 

In~\cite{Lenzi:2021wpc}, the full space of master functions and equations, for $\Lambda$-vacuum spherically-symmetric spacetimes, was studied in a systematic way. All the possible master equations were found by assuming that the master functions are linear combinations, with coefficients depending only on $r$, of the metric perturbations and their first-order derivatives. Two branches of possible pairs of potentials/master functions, $\{(V,\Psi)\}$ were identified:

\noindent (i) \textit{The standard branch}.  We call it standard because it contains a single potential for each parity, and these potentials are the known ones:
The Regge-Wheeler potential, $V_{\mathrm{RW}}$, for odd-parity perturbations~\cite{Regge:1957td} and the Zerilli potential, $V_{\mathrm{Z}}$, for even-parity perturbations~\cite{Zerilli:1970la}. 

\noindent (ii) \textit{The Darboux branch}.  This is a completely new branch for both parities, where we find an infinite set of potentials and master functions, $\{(V,\Psi)\}$. In particular, the set of allowed potentials is determined by a non-linear second-order ODE. 

There are some relevant comments about this construction: First, in Eq.~\eqref{concept-master-function} there is no explicit dependence in time. As we mentioned before, this is a consequence of the existence of a timelike Killing vector in the background, $K^{\mu} = (t^a,0)$. Second, in the study of~\cite{Lenzi:2021wpc} we did not impose gauge invariance [in the ansatz for the master function we included all the metric perturbations, in contrast to Eq.~\eqref{concept-master-function}, where only gauge invariant ones are considered], it came as a logical result of the analysis. We can again invoke the Stewart-Walker lemma and apply it to the perturbed Einstein equations, which are gauge invariant as the background Einstein tensor vanishes identically. Since the master equations are linear combinations of the perturbed Eintein equations (and their derivatives), they must also be gauge invariant. And since they only contain the master function, it has to be gauge invariant too.

The structure of the space of master functions and equations was analyzed in Ref.~\cite{Lenzi:2021njy} (see also~\cite{Lenzi:2022wjv}), where we found that all the master equations are connected by \emph{Darboux transformations}, i.e. a particular set of derivative transformations preserving the spectrum of the system (see Sec.~\ref{darboux-covariance}). Therefore, there are infinite possible descriptions of the perturbations of $\Lambda$-vacuum spherically-symmetric spacetimes in terms of master equations, all of them with the same spectrum.  
This hidden symmetry of the master equations has been called \textit{Darboux covariance}~\cite{Lenzi:2021njy}, since it actually represent a covariance in the description of gravitational perturbations.

In this section, we generalize the work of Ref.~\cite{Lenzi:2021wpc} to include generic matter sources that enter at first-order, i.e. without affecting the background spacetime, which is still a $\Lambda$-vacuum solution.  The generalization goes along the lines of the derivations of~\cite{Lenzi:2021wpc}, in the sense that the same combinations of the perturbed Einstein equations lead to master equations with the same operators acting on the master functions but with a source term that is build as linear combinations of the energy-momentum tensor $T_{\mu\nu}$ and its derivatives. That is, we obtain master equations of the form:
\begin{equation}
\left(\square^{}_2 - \frac{1}{f}V^{}_\ell \right)\Psi^{}_{\ell m} = \mathcal{F}^{}_{\ell m}  \,,    
\label{covariant-master-equation}
\end{equation}
where the source term $\mathcal{F}_{\ell m}$ has the following structure:
\begin{equation}
\mathcal{F}^{}_{\ell m} =  \mathcal{F}^{}_{\ell m}[r,g^{}_{ab},\vec{\mathcal{J}}^{}_{\ell m},\partial^{}_a\vec{\mathcal{J}}^{}_{\ell m}] \,, 
\end{equation}
where
\begin{equation}
\vec{\mathcal{J}}^{}_{\ell m} = (\mathcal{Q}^{\ell m}_{ab},\mathcal{P}^{\ell m}_a,\mathcal{T}^{}_{\ell m},\mathcal{P}^{}_{\ell m};\mathcal{S}^{\ell m}_{a},\mathcal{S}^{}_{\ell m}) \,,    
\end{equation}
are the energy-momentum tensor harmonic components, which are gauge invariant as we have argued in the previous section. 

These master equations, using Schwarzschild coordinates, $(x^a) = (t,r)$, and transforming to the radial tortoise coordinate
\begin{equation}
\frac{dx}{dr} = \frac{1}{f(r)} \,,    
\label{tortoise-coordinate}
\end{equation} 
can be written as a wave equation in one spatial dimension with a potential and a source term:
\begin{equation}
\left( -\partial^2_t + L^{}_V \right) \Psi^{}_{\ell m} = f \,\mathcal{F}^{}_{\ell m} \,,  
\label{master-wave-equation}
\end{equation}
where the operator $L_V$ is a Schr\"odinger-type operator
\begin{equation}
L^{}_V = \partial^2_x - V^{}_\ell \,.    
\label{schroedinger-operator}
\end{equation}
In the rest of the paper, for the sake of simplicity, we omit the harmonic numbers in the notation.

\subsection{Master Functions and Equations in the Standard Branch}\label{master-functions-standard branch}

The standard branch is characterized by master equations determined by the known potentials (Regge-Wheeler and Zerilli). The most general master function is a linear combination of two master functions:
\begin{equation}
{}^{}_{S}\Psi^{\mathrm{even}}_{\mathrm{odd}} = \mathcal{C}^{}_1 \Psi^{\mathrm{ZM}}_{\mathrm{CPM}} + \mathcal{C}^{}_2\Psi^{\mathrm{NE}}_{\mathrm{RW}} \,, 
\label{master-function-standard-branch}
\end{equation}
where $\mathcal{C}^{}_1$ and $\mathcal{C}^{}_2$ are two arbitrary constants. We present them in a explicitly gauge-invariant and covariant (with respect to coordinates changes in $M^2$) form.  
In the odd-parity case, the two master functions can be taken to be the well-known Regge-Wheeler~\cite{Regge:1957td} master function
\begin{equation}
\Psi^{}_{\mathrm{RW}} = \frac{r^{a}}{r}\tilde{h}^{}_{a}\,.
\label{regge-wheeler-master-function}
\end{equation}
and the Cunningham-Price-Moncrief master function~\cite{Cunningham:1978cp,Cunningham:1979px,Cunningham:1980cp} (see also~\cite{Jhingan:2002kb,Martel:2005ir}): 
\begin{equation}
\Psi^{}_{\mathrm{CPM}} = \frac{2 r}{(\ell-1)(\ell+2)}\varepsilon^{ab} \left( \tilde{h}^{}_{b:a} -\frac{2}{r}r^{}_{a}\tilde{h}^{}_{b} \right)\,.
\label{cunningham-price-moncrief-master-function}
\end{equation}
In the absence of matter sources ($T_{\mu\nu}=0\Rightarrow \mathcal{F}=0$), the Regge-Wheeler master function is exactly the time derivative of the Cunningham-Price-Moncrief master function (see e.g.~\cite{Martel:2005ir}). With matter sources it is modified in the following way
\begin{equation}
t^a \Psi^{}_{{\mathrm{CPM}}\,:a} = 2\,\Psi^{}_{\mathrm{RW}}
-\frac{4r}{(\ell+2)(\ell-1)}r^a \mathcal{S}^{}_a \,. 
\label{CPMdot-RW}
\end{equation}
In the standard branch of the odd-parity case the potential is the Regge-Wheeler potential~\cite{Regge:1957td}:
\begin{eqnarray}
V^{}_{\mathrm{RW}}(r)= f\left\{\Lambda + \frac{\ell(\ell+1) + 3(f-1)}{r^2}\right\} \,.
\label{potential-axial}
\end{eqnarray}
The source term of the master equation~\eqref{covariant-master-equation}] depends on the choice of master function within the family given in Eq.~\eqref{master-function-standard-branch}. For the Regge-Wheeler master function it is:
\begin{eqnarray}
\label{sourceRW}
\mathcal{F}^{\mathrm{RW}} = \frac{1}{r} \left[ r^a\left( \mathcal{S}^{}_{:a} - 2 \mathcal{S}^{}_a \right) + \frac{1- 3f -  \Lambda r^2}{r} \mathcal{S} \right] \,,
\end{eqnarray}
while for the Cunningham-Price-Moncrief master function it is:
\begin{eqnarray}
\label{sourceCPM}
\mathcal{F}^{\mathrm{CPM}} = \frac{4 r}{(\ell+2)(\ell-1)} \varepsilon^{ab}\mathcal{S}^{}_{a:b} \,.
\end{eqnarray}

In the even-parity case, the two master functions can be taken to be: (i) The Zerilli-Moncrief master function $\Psi_{\mathrm{ZM}}$~\cite{Zerilli:1970se,Zerilli:1970la,Moncrief:1974vm},
introduced first by Zerilli~\cite{Zerilli:1970la} and later by Moncrief~\cite{Moncrief:1974vm} (see also~\cite{Lousto:1996sx,Martel:2005ir}): 
\begin{equation}
\Psi^{}_{\mathrm{ZM}} = \frac{2 r}{\ell(\ell+1)}\left\{ \tilde{K} + \frac{2}{\lambda}\left(r^{a}r^{b}\tilde{h}^{}_{ab} - r r^{a}\tilde{K}^{}_{:a}\right) \right\}\,,
\label{zerilli-moncrief-master-function}
\end{equation}
where 
\begin{eqnarray}
\lambda(r) & = & r f'-2 \left(f-1\right)+(\ell+2)(\ell-1) \nonumber \\
           & = & (\ell+2)(\ell-1) - \Lambda r^2   -3\left( f-1\right)  \,.
\label{Lambda-defi} 
\end{eqnarray}
(ii) The following less known master function (using the name given in~\cite{Lenzi:2021wpc}): 
\begin{equation}
\Psi^{}_{\mathrm{NE}}(t,r) = \frac{1}{\lambda(r)}t^a\left(r \tilde{K}^{}_{:a} - \tilde{h}^{}_{ab}r^b \right) \,. 
\label{new-even-parity-master-function}
\end{equation}
This master function was previously unknown to us. But it turns out that a non-covariant form in the Regge-Wheeler gauge was used in Ref.~\cite{Gleiser:1998rw}, where the authors found it more convenient for second-order perturbative calculations in the context of the close-limit approximation~\cite{Price:1994pm,Gleiser:1996yc,Khanna:1999mh,Pullin:1999rg}. Moreover, more recently it has also appeared as the time derive of the Zerilli-Moncrief master function~\cite{Hopper:2017iyq}. In~\cite{Lenzi:2021wpc} we provided the explicitly gauge-invariant and covariant form for perturbations of  $\Lambda$-vacuum spherically-symmetric spacetimes.

In the absence of matter sources, the master function $\Psi^{}_{\mathrm{NE}}$ is exactly the time derivative of the Zerilli-Moncrief master function (see~\cite{Hopper:2017iyq} and~\cite{Lenzi:2021wpc} for the covariant version). When considering matter sources, we find that it is modified with the addition of harmonic components of the energy-momentum tensor in the following way:
\begin{equation}
t^a \Psi^{}_{\mathrm{ZM}:a} = 2\,\Psi^{}_{\mathrm{NE}} + \frac{4 r^2}{\ell(\ell+1)\lambda(r)} \mathcal{Q}^{}_{ab}t^a r^b \,. 
\label{ZMdot-NE}
\end{equation}
where $\lambda(r)$ was introduced in Eq.~\eqref{Lambda-defi}. In the standard branch of the even-parity case the potential is the Zerilli potential~\cite{Zerilli:1970se}
\begin{widetext}
\begin{equation}
V^{}_{\mathrm{Z}}(r) = f(r)\frac{ \lambda^3(r) - 2 \Lambda r^2\left[\lambda(r) - (\ell+2)(\ell-1)\right]^2 +2\left(\ell+2\right)^2\left(\ell-1\right)^2 \left(\ell^2+\ell+1\right)}{3 r^2 \lambda^2(r)}\,.
\label{potential-even}
\end{equation}
\end{widetext}

As in the odd-parity sector, the source term of the master equation~\eqref{covariant-master-equation} depends on the choice of master function. For the Zerilli-Moncrief master function we find the following source term:
\begin{widetext}
\begin{eqnarray}
\label{sourceZM}
\mathcal{F}^{\mathrm{ZM}} & = &- 2 r \mathcal{P} + \frac{4}{\lambda}r^a\mathcal{P}^{}_a + \frac{16 r}{\ell(\ell+1)\lambda}\left[1-\frac{(\ell+2)(\ell-1)}{2\lambda}\right] Q^{}_{ab}t^a t^b + \frac{2r}{\ell(\ell+1)}\left[1+\frac{2(r^2\Lambda-1+2f)}{\lambda}\right]\mathcal{Q}^{a}{}^{}_a \nonumber \\[2mm] 
& + & \frac{4r^2}{\ell(\ell+1)\lambda}t^c\varepsilon^{ab}\mathcal{Q}^{}_{cb:a} \,,
\end{eqnarray}
\end{widetext}
which is equivalent to the covariant expression in~\cite{Martel:2005ir} after considering the energy-momentum conservation equations (see Appendix~\ref{energy-momentum-harmonic-components-equations}). For the other even-parity master function, $\Psi_{\mathrm{NE}}$ we find the following expression for the source term
\begin{widetext}
\begin{eqnarray}
\label{sourceNE}
\mathcal{F}^{\mathrm{NE}} & = & \left[\frac{3\ell(\ell+1)}{r\lambda}-  \frac{2}{r}  + \frac{8f}{r\lambda}\left(1\frac{(\ell+2)(\ell-1)}{2\lambda}-1 \right)\right] t^a \mathcal{P}^{}_a + \frac{6}{\lambda^2}\left(f-1+\frac{\Lambda r^2}{3} \right)\mathcal{Q}^{}_{ab}t^a r^b \nonumber \\[2mm]
& - & r t^a \mathcal{P}^{}_{:a} + \frac{2f}{\lambda}\varepsilon^{ab}\mathcal{P}^{}_{a:b} - \frac{r}{\lambda} r^c\varepsilon^{ab}\mathcal{Q}^{}_{cb:a} \,.
\end{eqnarray}
\end{widetext}
This completes the description of master equations in the standard branch.

\subsection{Master Functions and Equations in the Darboux Branch} \label{master-functions-darboux-branch}

The Darboux branch contains a family of potentials that are different from the  ones of the standard branch, i.e. the Regge-Wheeler and Zerilli potentials. This family of potentials is characterized by the fact that they have to satisfy the following non-linear ODE: 
\begin{equation}
\left(\frac{\delta{V}^{}_{,x}}{\delta{V}} \right)^{}_{,x}
+ 2 \left(\frac{V^{\mathrm{RW/Z}}_{,x}}{\delta{V}} \right)^{}_{,x} - \delta{V} = 0 \,,
\label{xdarboux-odd-even}
\end{equation}
which is valid for both parities: For the even-parity case one has to select the Zerilli potential, $V_\mathrm{Z}$, in the second term, while for the odd-parity case we have to select the Regge-Wheeler potential, $V_{\mathrm{RW}}$. Moreover, we have introduced the following definition:
\begin{eqnarray}
\delta V & = &  V^{\mathrm{odd/even}} -V^{\mathrm{RW/Z}}\,,
\label{delta-potential}
\end{eqnarray}  
The master functions in the Darboux branch are also different from the ones in the standard branch and they depend on the particular potential, solution of Eq.~\eqref{xdarboux-odd-even}, that we consider. 
In order to set the ground for extending the analysis of~\cite{Lenzi:2021njy} to the presence of perturbative sources (see Sec.~\ref{darboux-covariance} for this), it is convenient to revisit the results of Ref.~\cite{Lenzi:2021wpc} for the Darboux branch, clarifying an issue with the derivation made there. 
The Darboux master functions presented in~\cite{Lenzi:2021wpc} were obtained from a systematic procedure and they appear as a linear combination of master functions with two arbitrary constants. However, one of these constants (the one that multiples a master function from the standard branch) should have been set to zero by one of the conditions of the systematic analysis that was not used [See, in Ref.~\cite{Lenzi:2021wpc}, equation (86) for the odd-parity case and equation (145) for the even-parity one].
Then, in the Darboux branch, the most general master function can be written as:
\begin{equation}
{}^{}_{D}\Psi^{\mathrm{even}}_{\mathrm{odd}} = \mathcal{C}\left( \Sigma^{\mathrm{even}}_{\mathrm{odd}}\, \Psi^{\mathrm{ZM}}_{\mathrm{CPM}} + \Phi^{\mathrm{even}}_{\mathrm{odd}}  \right)
\,,   \label{darboux-master-functions}
\end{equation}
with $\mathcal{C}$ being the only arbitrary constant. The functions $\Phi^{\mathrm{even}}_{\mathrm{odd}}$ are gauge-invariant linear combinations of the metric perturbations and their first-order derivatives, but only the combination with $\Psi^{\mathrm{ZM}}_{\mathrm{CPM}}$ in Eq.~\eqref{darboux-master-functions} is a true master function. In the odd-parity case, $\Phi^{\mathrm{odd}}$ is given by:
\begin{eqnarray}
\Phi^{}_{\mathrm{odd}}(x^a) = -\varepsilon^{ab} \tilde{h}^{}_{a:b} \,.
\label{odd-new-function}
\end{eqnarray}
while in the even-parity case $\Phi_{\mathrm{even}}$ is:
\begin{eqnarray}
\Phi^{}_{\mathrm{even}}(x^a) = \tilde{K} \,.
\label{even-new-function}
\end{eqnarray}
On the other hand, in Eq.~\eqref{darboux-master-functions} $\Sigma^{\mathrm{even}}_{\mathrm{odd}} = \Sigma^{\mathrm{even}}_{\mathrm{odd}}(x)$ is a function that contains the integral of the potential. In the odd-parity case it is given by:
\begin{eqnarray}
\Sigma^{}_{\mathrm{odd}}(x) = \frac{1}{2}\left[ \frac{1-f(r)+\Lambda\,r^2}{2\,r} + \int^{x}_{x_0} dx' V^{\mathrm{odd}}(x')\right] \,,
\end{eqnarray}
and in the even-parity case it is
\begin{eqnarray}
\nonumber
\Sigma^{}_{\mathrm{even}}(x) &=&  \frac{1}{2}\left[ \frac{(\ell+2)(\ell-1) -\lambda(r)}{2\,r} -\mathcal{C}_{\rm E} \right.
\\
&+& \left.\int^{x}_{x_0} dx' V^{\mathrm{even}}(x') \right]\,,
\end{eqnarray}
where $\mathcal{C}_{\rm E}$ is a constant proportional to the cosmological constant and the mass:
\begin{eqnarray}
\mathcal{C}_{\rm E} = \frac{4 M \Lambda}{(\ell+2)(\ell-1)} \,.
\label{CE}
\end{eqnarray}
The lower limit of integration appearing in these equations, $x_0$, depends on the particular choice of background spacetime (see Appendix~\ref{background-spacetime-metric}, where the notation is introduced).  For the BH background cases (Sch, SchdS, SchAdS), where $M\neq 0$, it is $x_0\rightarrow -\infty$. And for the maximally symmetric cases (M, dS, AdS), where $M=0$, it is $x_0=0$. The different values is a consequence of the different properties of the tortoise coordinate for the different background cases.

Finally, the general master equation in the Darboux branch is:
\begin{eqnarray}
\left( -\partial^2_t + \partial^2_x - V^{\mathrm{even/odd}}_{\ell} \right) {}^{}_{D}\Psi^{\mathrm{even}}_{\mathrm{odd}}  = f \,\mathcal{F}^{\mathrm{even}}_{\mathrm{odd}}  \,,  \label{master-equation-darboux}
\end{eqnarray}
and the source terms $\mathcal{F}^{\mathrm{even}}_{\mathrm{odd}}$, associated to the master functions in the Darboux branch, are discussed in the next section, where we write them in a way that avoids lengthy and cumbersome expressions.

\subsection{Darboux Covariance of the Metric Perturbations}
\label{darboux-covariance}

We are now going to study the hidden symmetry that joins the elements of the space of master functions and equations, and which is realized in the form of Darboux transformations~\cite{Darboux:89,1999physics...8003D,Matveev:1991ms}. In particular, we study whether Darboux covariance is still present when we have matter sources and what it requires to accommodate it, extending in this way the study of Ref.~\cite{Lenzi:2021njy}.
To that end, let us first review what exactly we mean by a Darboux transformation in the context of master equations.  Then, let us consider two master wave equations of the type given in Eq.~\eqref{master-wave-equation}, that is
\begin{equation}
\left( -\partial^2_t + L^{}_v \right) \Phi = \sigma\,,  \quad
\left( -\partial^2_t + L^{}_V \right) \Psi = S \,.  
\label{two-master-equation}
\end{equation}
The novelty here is that we are considering non-homogeneous equations with sources $\sigma$ and $S$. 
The study of Darboux transformation for the homogeneous case $S = \sigma = 0$, in the context of the space of master functions and equations, was carried out in~\cite{Lenzi:2021njy} where it was introduced as a mapping between the two pairs of master functions and potentials [see Eq.~\eqref{two-master-equation} above]:
\begin{eqnarray}
(v,\Phi)\rightarrow(V,\Psi):\quad \left\{ \begin{array}{lcl}
\Psi & = & \Phi^{}_{,x}  + g\,\Phi \,, \\[2mm]
V & = & v + 2\,g^{}_{,x}\,,
\end{array} \right.
\label{Darboux-transformation}
\end{eqnarray}  
where $g$ is the Darboux transformation generating function. In the case where there are no sources, $S=\sigma =0$, the mapping between master equations requires that the following consistency condition on $g$ is fulfilled: 
\begin{eqnarray}
 g^{}_x- g^2 + v = \mathcal{C}^{}_R \,,   
\label{riccati-g}
\end{eqnarray}
where $\mathcal{C}_R$ is an arbitrary integration constant, as the original consistency condition is the derivative of this equation. This is a Riccati equation that restricts the set of possible Darboux transformation generating functions $g$. When the source terms in Eq.~\eqref{two-master-equation} are present, we find that an additional condition, on the sources $\sigma$ and $S$, is required
\begin{eqnarray}
S  = \sigma^{}_{,x} + g \sigma\,, \label{Darboux-transformation-source-term} 
\end{eqnarray}
This new relation tells us how the source terms have to transform in order for the two master equations in Eq.~\eqref{two-master-equation} to be Darboux related. This condition can taken as part of the definition of the Darboux transformation when source terms are present.  
From these definitions and equations it is clear that the inverse Darboux transformation is generated by $-g$ (see Ref.~\cite{Matveev:1991ms} for details on the invertibility of Darboux transformations). 

The link to our infinite collection of master functions and equations is based on a mapping between the Darboux generating function and the potentials.
This comes from the Darboux transformation of the potential [Eq.~\eqref{Darboux-transformation}] and from the derivative of the Riccati equation [Eq.~\eqref{riccati-g}] for the Darboux transformation generating function. Indeed, these two equations can be manipulated to obtain the following expressions for $g_{,x}$ and $g$ in terms of the potentials:
\begin{equation}
g^{}_{,x} = \frac{V - v}{2} \,, \quad
g  =  \frac{(V+v)_{,x}}{2 (V- v)}\,.
\label{darboux-g-gx}
\end{equation}
It is straighforward to see that the consistency between these two equations for $g$ and $g_{,x}$, that is, asking that the derivative of the second equation is equal to the first one, yields 
\begin{equation}
\left(\frac{\delta V^{}_{,x}}{\delta V} \right)^{}_{,x}
+ 2 \left(\frac{v^{}_{,x}}{\delta V} \right)^{}_{,x} - \delta V = 0 
\,, 
\label{consistency-deriv}
\end{equation}
where $\delta V = V-v\,$. But, if we make the correspondence $v\longleftrightarrow V^{\mathrm{RW/Z}}$, this is precisely Eq.~\eqref{xdarboux-odd-even}. In other words, Eq.~\eqref{consistency-deriv} coincides with the condition that any potential belonging to the Darboux branch should satisfy~\cite{Lenzi:2021wpc}.
In the absence of sources, this is enough to establish that all master equations in the Darboux branch are connected with the standard branch via Darboux transformations. The corresponding Darboux generating function can be obtained by integrating the first expression in Eq.~\eqref{darboux-g-gx}, i.e.
\begin{eqnarray}
g^{\mathrm{Z}\rightarrow\mathrm{even}}_{\mathrm{RW}\rightarrow\mathrm{odd}} 
&=& \frac{1}{2}\int^{} _x dx'\, \left(V^{\mathrm{even}}_{\mathrm{odd}} - V^{\mathrm{Z}}_{\mathrm{RW}} \right) \nonumber \\
&=& \frac{1}{2}\int^{x}_{x^{}_0} dx'\, \left(V^{\mathrm{even}}_{\mathrm{odd}} - V^{\mathrm{Z}}_{\mathrm{RW}} \right) + \mathcal{C}_{\mathrm{S}\rightarrow\mathrm{D}}
\,.
\label{exp-g-odd-and-even}
\end{eqnarray}
The choice of finding $g$ in integral form, instead of the derivative one given by the second expression in Eq.~\eqref{darboux-g-gx}, is motivated by the fact that the master functions in the Darboux branch~\eqref{darboux-master-functions} depend on integral functions of the potential. At first glance, it seems like there is an arbitrary integration constant, $\mathcal{C}_{\mathrm{S}\rightarrow\mathrm{D}}$, in this definition. However, this constant is fixed by requiring that the expression for $g$ in the second relation in Eq.~\eqref{darboux-g-gx} has to be equal to the integral of the expression for $g_{,x}$ in the first one. That is, we can write
\begin{eqnarray}
\frac{(V+v)^{}_{,x}}{(V- v)} = \int^{}_x dx' \left(V-v\right) \,.
\label{consistency-integ}
\end{eqnarray}

Something similar happens between the odd- and even-parity sectors in the standard branch. Indeed, we can introduce a Darboux generating function that transforms the Regge-Wheeler potential into the Zerilli in the same way, i.e.
\begin{eqnarray}
g^{\mathrm{RW}\rightarrow\mathrm{Z}} 
&=& \frac{1}{2}\int^{x}_{x^{}_0} dx' \, \left(V^{\mathrm{Z}} - V^{\mathrm{RW}} \right)  + \mathcal{C}_{\mathrm{RW}\rightarrow\mathrm{Z}} \,.
\label{exp-g-RW-to-ZM}
\end{eqnarray}
The constant $\mathcal{C}_{\mathrm{RW}\rightarrow\mathrm{Z}} $ is fixed by Eq.~\eqref{consistency-integ} and reads:
\begin{eqnarray}
\label{CRWZ}
\mathcal{C}_{\mathrm{RW}\rightarrow\mathrm{Z}} = \alpha -  \frac{\mathcal{C}_{\rm E} }{2} \,,
\end{eqnarray}
where $\alpha$ is given by:
\begin{eqnarray} \label{alphaRWZ}
\alpha = \frac{1}{12 M}\frac{(\ell+2)!}{(\ell-2)!}  \,,
\label{definition-of-alpha}
\end{eqnarray}
and $\mathcal{C}_{\rm E}$ is given by Eq.~\eqref{CE}. We will go back to this generating function between odd- and even-parity standard branches in Sec.~\ref{Ss:special-cases}, where we analyze the consequences of Darboux transformation when considered in the frequency domain (the typical arena for this kind of transformations~\cite{Matveev:1991ms}).

Before getting deeper into the general details of the Darboux transformation between master functions, let us discuss a particular feature of it that only applies in the case of the maximmally-symmetric background spacetimes (Minkowski, de Sitter and anti-de Sitter).
By looking back at Eqs.~\eqref{exp-g-RW-to-ZM},~\eqref{CRWZ} and~\eqref{alphaRWZ}, it is clear that the Darboux generating function is not defined in the limit $M\to 0$, corresponding to the maximmally-symmetric backgrounds. This is reflected in the fact that, in this case, the even- and odd-parity potentials~\eqref{potential-axial} and~\eqref{potential-even} in the standard branch, are identical and reduce to:
\begin{equation}
V^{\rm odd}_{\ell} = V^{\rm even}_{\ell} = f\frac{\ell(\ell+1)}{r^{2}} \,.
\label{maximally-omega-potential-odd-even} 
\end{equation}
If we now look at the equation that potentials in the Darboux branch have to satisfy, Eq.~\eqref{xdarboux-odd-even}, it is a second-order non-linear ODE which is not trivial to solve analytically. However, it turns out that in the maximally-symmetric case it admits some simple solutions. Indeed, if we start from the potential of the standard branch, Eq.~\eqref{maximally-omega-potential-odd-even}, we obtain the following two solutions:
\begin{eqnarray}
V^{}_{\ell +} = f\frac{(\ell+1)(\ell+2)}{r^{2}} \,, \quad  V^{}_{\ell -} = f\frac{\ell(\ell-1)}{r^{2}} \,.
\label{potentials-ladder-ell}
\end{eqnarray}
The Darboux generating functions that produces the transformations to these potentials are:
\begin{equation}
g^{}_{+} = -\frac{\ell+1}{r}\,, \qquad g^{}_{-} = \frac{\ell}{r} \,.     
\label{generating-functions-maximally-symmetric}
\end{equation}
One can check that they satisfy the Riccati equation~\eqref{riccati-g} and the consistency condition~\eqref{consistency-integ} .
As we can see, the first one, $g_+$, correspond to shifting the harmonic number $\ell$ to $\ell+1$, while the second one, corresponds to the shift: $\ell\rightarrow\ell-1$.
This means that, in the case of maximally-symmetric background spacetimes, the Darboux transformation acts as a sort of \textit{ladder operator}, increasing or decreasing by one the value of the angular momentum number $\ell$. 
This is a very interesting result, since it means that for these backgrounds, Darboux transformation not only connects solutions of the master equations with the same harmonic numbers $(\ell,m)$ but also solutions corresponding to different number $\ell$. In the case where the master equations do not have sources, the solutions are in principle independent of $m$, so we can connect all of them.

Another point of view consists in realizing that the Darboux transformations generated by the functions in Eq.~\eqref{generating-functions-maximally-symmetric} do not change the shape of the potential. To be more specific, a Darboux transformation is said to change the potential in a \emph{shape invariant} way when it induces in the potential [see Eq.~\eqref{Darboux-transformation}] a transformation that can be written in the form (see, e.g.~\cite{Gendenshtein:1983jt,Organista:2006jo} for a detailed exposition):
\begin{equation}
V(x;f(\vec{\lambda})) = v(x;\vec{\lambda}) + R(\vec{\lambda}) \,,    
\end{equation}
where the vector $\vec{\lambda}$ contains parameters on which the potential depends. In our case, it is just $\ell$ so that $f(\vec{\lambda}) \equiv f(\ell) = \ell +1$ and $R(\vec{\lambda})  \equiv 0$ in the case in which we raise $\ell$ by one. A very interesting consequence of this property is that the spectrum can be reconstructed in an analytical way. However, in our case, this property only happens for the case of maximmally-symmetric backgrounds. 

We now get back to the general case, but still without sources ($T_{\mu\nu}=0$), and we are going to show that the master functions of the Darboux branch can be written in a form that shows their explicit connection with Darboux transformations. In doing so, we are going to point to some subtle aspects that were not addressed in~\cite{Lenzi:2021njy} and which are very relevant for extending all this formalism to the case with sources ($T_{\mu\nu}\neq 0$). First, as it was already noticed by Chandrasekhar~\cite{Chandrasekhar:1992bo}, the (Darboux) transformation between odd- and even-parity master functions is a formal transformation in the sense that it connects the master equations or, equivalently, the potentials and the solutions of the master equations. But it is not a transformation that can be formulated at level of the metric perturbations, which is obvious because there is no possibility of transforming odd-parity metric perturbations into even-parity ones.  Following this line of thought, we can see that things are different in the Darboux branch of master functions and equations, as we are going to see that the master functions come from explicit Darboux transformations of the standard branch master functions. To see this, let us first consider the even-parity sector. The first step is to realize that the gauge-invariant function $\Phi_{\mathrm{even}}$ can be written as the following transformation of the Zerilli-Moncrief master function:
\begin{equation}
\Phi^{}_{\mathrm{even}} = \Psi^{}_{\mathrm{ZM},x} + {g}_{\ast}^{\mathrm{even}} \Psi^{}_{\mathrm{ZM}} \,,
\label{darboux-like-psi-even}
\end{equation}
where
\begin{equation}
{g}_{\ast}^{\mathrm{even}} = \frac{3 M}{2 r^2} + \frac{\mathcal{C}^{}_{\rm E} }{2} - \frac{1}{2}\int^{x}_{x^{}_0} dx'\,V^{\mathrm{Z}} \,.
\end{equation}
This transformation looks like a Darboux transformation but it is actually not because the function ${g}_{\ast}^{\mathrm{even}}$ does not satisfy the consistency condition~\eqref{riccati-g} for a Darboux transformation generating function. Nevertheless, if we introduce this transformation into the even-parity version of Eq.~\eqref{darboux-master-functions} we find (for simplicity we set $\mathcal{C} = 1$):
\begin{eqnarray}
{}^{}_{D}\Psi^{}_{\mathrm{even}} =  \Psi^{}_{\mathrm{ZM},x} + g^{\mathrm{even}} \Psi^{}_{\mathrm{ZM}} \,,
\label{darboux-ZM}
\end{eqnarray}
where $g^{\mathrm{even}}$ is given by the expression in Eq.~\eqref{exp-g-odd-and-even}, but with $\mathcal{C}_{\mathrm{S}\rightarrow\mathrm{D}}=0$, i.e.
\begin{eqnarray}
g^{\mathrm{even}} =  \frac{1}{2}\int^{x}_{x^{}_0} dx'\, \left(V^{\mathrm{even}} - V^{\mathrm{Z}} \right) \,,
\label{geven}
\end{eqnarray}
which satisfies the consistency condition by construction. This makes Eq.~\eqref{darboux-ZM} a true Darboux transformation, which holds at the level of the metric perturbations by virtue of the perturbative Einstein equations.

The situation in the odd-parity sector is quite similar. The function $\Phi_{\mathrm{odd}}$ can be written as the following transformation of the Cunningham-Price-Moncrief master function:
\begin{equation}
\Phi^{}_{\mathrm{odd}} = \Psi^{}_{\mathrm{CPM},x} + {g}_{\ast}^{\mathrm{odd}} \Psi^{}_{\mathrm{CPM}} \,,
\label{darboux-like-psi-odd}
\end{equation}
where 
\begin{equation}
{g}_{\ast}^{\mathrm{odd}} = -\frac{M}{2 r^2} -\frac{r \Lambda}{3} - \frac{1}{2}\int^{x}_{x^{}_0} dx'\,V^{\mathrm{RW}} \,.
\end{equation}
Again, Eq.~\eqref{darboux-like-psi-odd} is not a Darboux transformation for the same reason as before, but we can be inserted it into Eq.~\eqref{darboux-master-functions} to find (again, for simplicity, we set $\mathcal{C} = 1$):
\begin{eqnarray}
{}^{}_{D}\Psi^{}_{\mathrm{odd}} =  \Psi^{}_{\mathrm{CPM},x} + g^{\mathrm{odd}} \Psi^{}_{\mathrm{CPM}} \,,
\label{darboux-CPM}
\end{eqnarray}
where $g^{\mathrm{odd}}$ is given by Eq.~\eqref{exp-g-odd-and-even}, again with $\mathcal{C}_{\mathrm{S}\rightarrow\mathrm{D}}=0$:
\begin{eqnarray}
g^{\mathrm{odd}} =  \frac{1}{2}\int^{x}_{x^{}_0} dx'\, \left(V^{\mathrm{odd}} - V^{\mathrm{RW}} \right) \,.
\label{godd}
\end{eqnarray}

It is important to mention that the systematic procedure of Ref.~\cite{Lenzi:2021wpc} contains the necessary consistency conditions to fix $\mathcal{C}_{\mathrm{S}\rightarrow\mathrm{D}}=0$ for both the odd- and even-parity sectors. 
This seems to happen because the consistency condition in the systematic procedure is imposed at the level of the allowed potentials, so that the generating function defined from them automatically satisfies the consistency condition.
Therefore, the master function in the Darboux branches are actual Darboux transformations of the standard branch master functions. 

Darboux covariance provides another key piece of information about the infinite landscape of master equations: It helps understanding why the Darboux branch contains only master functions and equations that result from Darboux transformations of the Cunningham-Price-Moncrief and Zerilli-Moncrief master functions, $\Psi_{\mathrm{CPM}}$ and $\Psi_{\mathrm{ZM}}$, and their associated potentials (which completely determine the master equations in the absence of matter sources). Indeed, the Darboux transformations of the other master functions in the standard branch, the Regge-Wheeler master function in the odd-parity case, $\Psi_{\mathrm{RW}}$, and the other even-parity master function [Eq.~\eqref{new-even-parity-master-function}] in the even-parity case, $\Psi_{\mathrm{NE}}$, are not present at all in the Darboux branch. 
Actually, one could in principle consider master functions that originate from $\Psi_{\mathrm{RW}}$ and $\Psi_{\mathrm{NE}}$, that is,
\begin{eqnarray}
\Psi^{\mathrm{NE}}_{\mathrm{RW},x} + g^{\mathrm{even}}_{\mathrm{odd}} \Psi^{\mathrm{NE}}_{\mathrm{RW}}\,.
\label{darboux-from-NE-and-RW}
\end{eqnarray}
In principle, these master functions, with their associated potentials, can also constitute a valid description of the perturbations. However, one can show that the master functions in Eq.~\eqref{darboux-from-NE-and-RW} contain second-order derivatives of the metric perturbations which cannot be eliminated by using Einstein perturbed equations [see Appendix~\ref{harmonic-components-perturbative-equations}]. But this is against the restrictions we imposed on the space of master functions in~\cite{Lenzi:2021wpc}, where we only considered master functions that are linear combinations of the metric perturbations and their first-order derivatives. 
This shows once more that the Cunningham-Price-Moncrief and Zerilli-Moncrief master functions seem to play a special role in BHPT.

Let us now focus on the extension to the case with sources generated by the existence of a first-order energy-momentum tensor $T_{\mu\nu}$. Here the situation is different and Eqs.~\eqref{darboux-ZM} and~\eqref{darboux-CPM} needs corrections that depend on the harmonic components of the energy-momentum tensor. The relation between the master functions in the standard and Darboux branches in the even-parity sector is now
\begin{eqnarray}
{}^{}_{D}\Psi^{}_{\mathrm{even}} = \Psi^{}_{\mathrm{ZM},x} + g^{\mathrm{even}} \Psi^{}_{\mathrm{ZM}}  - \frac{4 r^2}{\ell (\ell + 1)\lambda}t^{a} t^{b} Q^{}_{a b} \,, \qquad 
\label{darboux-ZM-source}
\end{eqnarray}
and in the case of the odd-parity sector:
\begin{eqnarray}
{}^{}_{D}\Psi^{}_{\mathrm{odd}} = \Psi^{}_{\mathrm{CPM},x} + g^{\mathrm{odd}} \Psi^{}_{\mathrm{CPM}} + \frac{4 r}{(\ell + 2)(\ell - 1)}t^{a} S^{}_{a} \,. \qquad 
\label{darboux-CPM-source}
\end{eqnarray}
The source terms appearing in these equations come from the modification of the transformations in Eqs.~\eqref{darboux-like-psi-even} and~\eqref{darboux-like-psi-odd} between the functions $\Phi^{\mathrm{even}}_{\mathrm{odd}}$  and $\Psi^{\mathrm{ZM}}_{\mathrm{CPM}}$ in the presence of matter. 
Clearly, these equations do not correspond to Darboux transformations.  
Even more, one can check that the sources $
\mathcal{F}^{\mathrm{even}}_{\mathrm{odd}} $ are not Darboux transformations of the standard branch sources $\mathcal{F}^{\mathrm{ZM}}_{\mathrm{CPM}}$ either [see Eq.~\eqref{Darboux-transformation-source-term} for the transformation of the sources of the master equations under a Darboux transformation].
At first sight, one can then think that Darboux covariance may be broken, but it turns out that we can rearrange the variables in such a way that it can be preserved. Indeed, let us define, for the even-parity sector the following function
\begin{eqnarray}
{}^{}_{D}\check{\Psi}^{}_{\mathrm{even}} & = & {}^{}_{D}\Psi^{}_{\mathrm{even}} + \frac{4 r^2}{\ell (\ell + 1) \lambda}t^{a} t^{b} Q^{}_{a b}  \nonumber \\[2mm]
& = & \Psi^{}_{\mathrm{ZM},x} + g^{\mathrm{even}} \Psi^{}_{\mathrm{ZM}}  \,,
\label{darboux-even-source}
\end{eqnarray}
where $g^{\mathrm{even}}$ is given in Eq.~\eqref{geven}.
And for the odd-parity case, let us define
\begin{eqnarray}
{}^{}_{D}\check{\Psi}^{}_{\mathrm{odd}} & = & {}^{}_{D}\Psi^{}_{\mathrm{odd}} - \frac{4 r}{(\ell + 2)(\ell - 1)}t^{a} S^{}_{a} \nonumber \\[2mm]
& = &  \Psi^{}_{\mathrm{CPM},x} + g^{\mathrm{odd}} \Psi^{}_{\mathrm{CPM}}  \,,
\label{darboux-odd-source}
\end{eqnarray}
where $g^{\mathrm{odd}}$ is given in Eq.~\eqref{godd}. 
The master functions, ${}^{}_{D}\check{\Psi}^{}_{\mathrm{even}}$ and ${}^{}_{D}\check{\Psi}^{}_{\mathrm{odd}}$, have the form of Darboux transformations of the standard branch master functions and, more importantly, they satisfy master equations with same potential as the original master functions, ${}^{}_{D}\Psi^{}_{\mathrm{even}}$ and ${}^{}_{D}\Psi^{}_{\mathrm{odd}}$,  but with a different source term. That is,
\begin{eqnarray}
\left( -\partial^2_t + \partial^2_x - V^{\mathrm{even/odd}}_{\ell} \right) {}^{}_{D} \check{\Psi}^{\mathrm{even}}_{\mathrm{odd}} = f\, \check{\mathcal{F}}^{\mathrm{even}}_{\mathrm{odd}}  \,, \qquad
\end{eqnarray}
where the expressions for the new source terms, $\check{\mathcal{F}}^{\mathrm{even}}_{\mathrm{odd}}$, are given by
\begin{eqnarray}
\check{\mathcal{F}}^{\mathrm{even}} = \mathcal{F}^{\mathrm{even}} - \left(\square^{}_2 - \frac{1}{f}V^{\mathrm{even}} \right) \frac{4 r^2 }{\ell (\ell + 1) \lambda}t^{a} t^{b} Q^{}_{a b} \,, \qquad
\end{eqnarray}
and
\begin{eqnarray}
\check{\mathcal{F}}^{\mathrm{odd}} = \mathcal{F}^{\mathrm{odd}} + \left(\square^{}_2 - \frac{1}{f}V^{\mathrm{odd}} \right) \frac{4 r }{(\ell + 2)(\ell - 1)}t^{a} S^{}_{a} \,. \qquad
\end{eqnarray}
These new source terms are now related to the standard branch ones by the Darboux condition~\eqref{Darboux-transformation-source-term}, i.e.
\begin{eqnarray}
\check{\mathcal{F}}^{\mathrm{even}} & = & \mathcal{F}^{\mathrm{ZM}}_{,x} + g^{\mathrm{even}} \mathcal{F}^{\mathrm{ZM}} \,, 
\\[2mm]
\check{\mathcal{F}}^{\mathrm{odd}} & = & \mathcal{F}^{\mathrm{CPM}}_{,x}  + g^{\mathrm{odd}} \mathcal{F}^{\mathrm{CPM}} \,.
\end{eqnarray}
Then, in the presence of matter sources in the master equations, the pairs $(\check{\Psi}, \check{\mathcal{F}})$ are the true Darboux transformations of the standard branch master functions/equations. Therefore, thanks to an adequate rearrangement of the master variables, Darboux covariance is preserved in the presence of matter sources.

\subsection{Darboux Transformations and Isospectrality}
\label{Ss:special-cases}

One of the most important consequences of Darboux covariance in vacuum is isospectrality.  This can be seen very well in the frequency domain, for instance by just considering a single frequency mode of the master function
\begin{equation}
\Psi(t,r) = e^{i k t}\,\psi(k;x) \,,
\label{time-to-frequency-domain}
\end{equation}
so that the master equation~\eqref{master-wave-equation} becomes a time-independent Schr\"odinger equation\footnote{In this subsection we assume there are no matter sources driving the perturbations, since we are precisely interested in the (quasi)normal free oscilations.}:
\begin{equation}
L^{}_V\psi = -k^2\psi \,,
\label{schrodinger}
\end{equation}
where $L_V$ is given in Eq.~\eqref{schroedinger-operator} and $k$ is the frequency of the mode.  It is simple to check that a Darboux transformation [see Eq.~\eqref{Darboux-transformation}] not only transforms this time-independent Schr\"odinger equation into another one, but it does so in such a way that the eigenvalue $-k^2$ remains unchanged.  In the asymptotically-flat case this leads to the isospectrality of quasinormal modes for all the master equations in our space.  It also implies that the transmission and reflection coefficients for scattering states are the same. However, the isospectrality between odd- and even-parity quasinormal modes is not guaranteed when $\Lambda\neq 0$ (see, e.g.~\cite{Grozdanov:2023txs,Grozdanov:2023tag}). Moreover, it may well be that isospectrality is a distinctive characteristic of General Relativity since it is broken in other theories, opening the door to perform fundamental physics tests with advanced gravitational-wave detectors~\cite{Sathyaprakash:2012jk,Barausse:2020rsu,LISA:2022kgy}.  The loss of isospectrality in other theories of gravity has been shown, for instance, in~\cite{Franciolini:2018uyq,Datta:2019npq,Ferrari:2000ep,LuisBlazquez-Salcedo:2020rqp,Chen:2021cts} for non-rotating BHs (in particular~\cite{del-Corral:2022kbk} in the context of loop quantum gravity) and~\cite{Li:2023ulk} for the case of spinning BHs, both in alternative theories of gravity.  On the other hand, the loss of isospectrality has been recently discussed in the context of the study of the instability of quasinormal modes~\cite{Jaramillo:2020tuu,Jaramillo:2021tmt} within General Relativity.

In this context, a remarkable fact, already pointed out in~\cite{Heading:1977jh,1980RSPSA.369..425C} (see also~\cite{Lenzi:2021njy}) is that the Regge-Wheeler and Zerilli potentials can be derived from a superpotential in the following way:
\begin{equation}
V^{\mathrm{RW}}_{\mathrm{Z}} =  \mp g^{}_{,x}  + g^2 + k_0^2\,, 
\label{RWZ-superpotential}
\end{equation}
where $g$ is the superpotential, which in the case of Schwarzschild looks as
\begin{eqnarray}
g(r) = i k_0 + \frac{6\,M f(r)}{\lambda(r)r^2}  = i k^{}_0 + W(r) \,,    \label{superpotential-standard}
\end{eqnarray}
and $k_0$ is the so-called algebraically special frequency, given by
\begin{equation}
i k_0 = \alpha = \frac{(\ell+2)(\ell+1)\ell(\ell-1)}{12\,M}\,,
\label{alpha-expression}
\end{equation}
where $\alpha$ was introduced before in Eq.~\eqref{definition-of-alpha}.
In Eq.~\eqref{superpotential-standard}, $W(r)$ is a potential used by Heading and Chandrasekhar to express the Regge-Wheeler and Zerilli potentials in another compact form:
\begin{equation}
V^{\mathrm{RW}}_{\mathrm{Z}} =  \mp W^{}_{,x} + \alpha W + W^2\,. 
\end{equation}
The two potentials, $W$ and $g$, differ by just a constant, but $g$ plays a special role as it is the generator of the Darboux transformation between of the two parities of the standard branch. It also corresponds to what, in the context of supersymmetric (SUSY) quantum mechanics, is called a superpotential.  The analogy with SUSY is actually exact if we absorb the $k^{2}_0$ term in Eq.~\eqref{RWZ-superpotential} into the eigenvalue in Eq.~\eqref{schrodinger}.  For more details on the role of Darboux transformations in SUSY quantum mechanics see~\cite{Witten:1981nf,Cooper:1982dm,Cooper:1994eh,GomezUllate:2004}.  It is important to notice that $g$
automatically satisfies the Riccati consistency condition~\eqref{riccati-g} with $\mathcal{C}_R = k_0^2$, which is something that highlights the difference between the Darboux tranformation as originally defined~\cite{1999physics...8003D,Darboux:89} (see also~\cite{Matveev:1991ms}) and Darboux covariance: Darboux transformations were originally introduced in the frequency domain and they are generated from a particular solution of the time-independent Schr\"odinger equation~\eqref{schrodinger}, typically an eigenfunction with eigenvalue $k_0$, to connect two equations with same eigenvalue $k$; instead, Darboux covariance is a symmetry of the full master equation that can be used both in the time domain and in the frequency domain and does not rely on having a special solution (like a bound state). This why we need the consistency condition~\eqref{riccati-g}, which in the original Darboux transformation is satisfied by construction.

The eigenfunction associated with the eigenvalue $k_0$ and $g$ in Eq.~\eqref{superpotential-standard}, is usually called a \textit{algebraically special mode} but, sometimes, in the context of scattering theory, it is usually known as an \textit{antibound} state. In Schwarzschild spacetime it can be written as
\begin{equation}
    \psi_0 = \frac{\lambda(r)}{2}e^{-i k_0 x}
    \,,
    \label{algebraically-special}
\end{equation}
so that the function Darboux generating function, $g(r)$, in Eq.~\eqref{superpotential-standard}, is obtained as the logarithmic derivative of $\psi_0$, i.e. 
\begin{equation}
g(r) = -(\ln{\psi^{}_0})^{}_{,x}\,.    
\end{equation}
However, is quite straightforward to check that it is also a special solution even with different asymptotics, i.e. in Schwarzschild-de Sitter/Schwarzschild-Anti de Sitter spacetimes (see also~\cite{Grozdanov:2023txs,Grozdanov:2023tag} for a nice discussion of the extension of this formalism to $\Lambda$-vacuum spacetimes and the special role of this solution in the \emph{pole-skipping} method to reconstruct the quasi-normal mode spectrum).

\section{Reconstruction of the Metric Perturbations: General Case} \label{metric-reconstruction-procedure}

With all the elements of BHPT presented in the previous section, we are now going to tackle the problem of reconstructing the metric perturbations in terms of the master functions and perturbative gauge functions.  Having the full knowledge of the metric perturbations is of interest for some tasks, like the computation of the gravitational self-force or in perturbative Hamiltonian formulations.

Here, we show how to carry out the metric reconstruction for any background within the family of the spherically-symmetric $\Lambda$-vacuum spacetimes, not just for the case of the Schwarzschild BH. We also consider a general situation in which the perturbations are driven by matter sources that are generated by a general energy-momentum distribution $T_{\mu\nu}$ that enters at the first perturbative order (it does not affect the background spacetime). As we have done with the previous developments, we are going to carry out the derivations in an arbitrary gauge and the expressions we obtain are valid in an arbitrary coordinate system of the two-dimensional Lorentzian spacetime $M^2$.  That is, we are going to show that the metric reconstruction can be in principle carried out in an arbitrary gauge. In order words, the metric perturbations are completely determined once the master functions are found and a perturbative gauge is fixed. This is of great relevance and gives a new perspective into BHPT computations.  It also reflects the fact that certain combinations of the metric perturbations (and their derivatives) are gauge-invariant. But this does not mean that the gauge-invariant quantities are the true degrees of freedom of the gravitational field in the perturbative regime. The true degrees of freedom are represented by the master functions which, as is well known, determine all the physically relevant quantities about the dynamics of the perturbations: Gravitational waveforms; Gravitational-waves fluxes of energy; Gravitational-wave fluxes of angular momentum; Gravitational-wave fluxes of linear momentum; etc.  Another way of looking at this fact, is to realize that while the construction of gauge-invariant quantities from the metric perturbations is a process that can be carried out independently of the Einstein equations, the reduction to the true dynamical degrees of freedom (the master functions) requires the  use of the (perturbative) Einstein field equations. One illustrative example is the reduction to the two gravitational wave polarizations in the case of the linear theory (perturbations around Minkowski spacetime; see e.g.~\cite{Misner:1973cw}).

Another important feature of our procedure for the metric reconstruction is that it is systematic, thus we expect that it can be applied to other scenarios: different backgrounds within spherical symmetry, different theories of gravity, etc. 
After the derivation of the main results we discuss their implications and scope in the light of Darboux covariance. 

To start with, let us count metric perturbations. At the level of the spacetime metric perturbations, we have ten independent metric perturbations and the generator of the perturbative gauge transformations has four components, therefore we can build six independent gauge-invariant metric perturbations. At the level of the  decomposition of the metric perturbations into spherical harmonics, we have to distinguish the two different parity sectors. In the odd-parity sector  we have three independent metric perturbations and just one gauge generator component and hence, there are 2 gauge invariant metric perturbations. In the even-parity sector we have seven independent metric perturbations, while there are three components of the gauge generator vector field, thus we can construct four gauge invariant metric perturbations. It is important to clarify, that here gauge-invariant metric perturbations means combinations of the metric perturbations and their derivatives that are invariant under a general gauge transformation.

In the case of perturbations driven by matter sources, the components of the energy-momentum tensor will also appear in the metric reconstruction.  In the vacuum case, the reconstruction can be carried out in two steps. In the first step, we identify the pure gauge metric components, those that can be canceled by a gauge transformation, and then we find how the rest of the components of the metric perturbations can be written in terms of the gauge and the gauge-invariant metric perturbations [Eqs.~\eqref{expression-hathab} and~\eqref{expression-hatK} for the even-parity case and Eq.~\eqref{expression-hta} for the odd-parity case]. The second step consists in writing the gauge-invariant metric perturbations in terms of the master functions of the standard branch.

\subsection{Metric Reconstruction in the Odd-Parity Sector}
We start with the odd-parity sector as it involves less independent metric perturbations, just three.  Then, following the plan outlined above, let us look at how the odd-parity metric perturbations transform [see Eqs.~\eqref{gauge-transformation-odd-1} and~\eqref{gauge-transformation-odd-2}].  From the transformation in Eq.~\eqref{gauge-transformation-odd-2} we see that $h_2$ is a pure gauge component of the metric perturbations, i.e. we can make it to have any value, in particular we can be put it to zero by a gauge transformation, which is what is done in the Regge-Wheeler gauge~\cite{Regge:1957td}.  Therefore, we can write the metric perturbations in the odd-parity case as follows:
\begin{eqnarray}
h^{}_a & = & \tilde{h}^{}_a + \frac{r^2}{2}\mathcal{G}^{\mathrm{odd}}_{:a} \,,
\label{reconstruction-ha-f1}\\[2mm]
h^{}_2 & = & r^2 \mathcal{G}^{\mathrm{odd}} \,,
\label{reconstruction-h2-f1}
\end{eqnarray}
where $\mathcal{G}^{\mathrm{odd}}$ is the only odd-parity gauge function.
This completes the first step. 
The second step is to write the gauge-invariant metric perturbations, $\tilde{h}_a$, in terms of the two independent master functions of the standard branch: $\Psi_{\mathrm{RW}}$ and $\Psi_{\mathrm{CPM}}$. Taking into account that the first is related to the time derivative of the second [see Eq.~\eqref{CPMdot-RW}], we can always eliminate one of them if we prefer.  Later we will comment on how to recover the metric perturbations in terms of the master functions of the Darboux branch.

There is a systematic way of carrying out the metric reconstruction of $\tilde{h}_a$. It consists in writing the gauge-invariant metric perturbations as a linear combination of the master functions and their first derivatives. In this sense, the procedure is similar to the method used in~\cite{Lenzi:2021wpc} to study the space of master functions and equations. To avoid ambiguities we have to take into account that the Regge-Wheeler master function is the time derivative of the Cunningham-Price-Moncrief master function [see Eq.~\eqref{CPMdot-RW}], which is a consequence of the Einstein perturbative field equations. Then, the most general combination is:
\begin{eqnarray}
\tilde{h}^{}_a & = & \left( A^{}_1 t^{}_a + A^{}_2 r^{}_a \right) \Psi^{}_{\mathrm{RW}} + \left( A^{}_3 t^{}_a + A^{}_4 r^{}_a \right) \Psi^{}_{\mathrm{CPM}} \nonumber \\[2mm]
& + & \left( A^{}_5 t^{}_a + A^{}_6 r^{}_a \right) t^b\Psi^{}_{{\mathrm{RW}}:b} + \left( A^{}_7 t^{}_a + A^{}_8 r^{}_a \right) r^b\Psi^{}_{{\mathrm{RW}}:b} \nonumber \\[2mm]      
& + & \left( A^{}_9 t^{}_a + A^{}_{10} r^{}_a \right) r^b\Psi^{}_{{\mathrm{CPM}}:b} \nonumber \\[2mm]
& + & \left(A^{}_{11} t^{}_a + A^{}_{12} r^{}_a\right) t^b \mathcal{S}^{}_b + \left( A^{}_{13} t^{}_a + A^{}_{14} r^{}_a\right) r^b \maketitle{S}^{}_b \nonumber \\[2mm] 
& + & \left(A^{}_{15} t^a + A^{}_{16} r^a\right) \mathcal{S} \,.
\label{ha-equation-As}
\end{eqnarray}
There are some comments in order. First, as we can see, there are no terms proportional to $t^a\Psi_{{\mathrm{CPM}}:a}$ as we mentioned that this quantity is essentially $\Psi_{\mathrm{RW}}$ by virtue of Eq.~\eqref{CPMdot-RW}.  Second, since we are going to use the perturbative Einstein equations, it is important to introduce the corresponding matter terms of the odd-parity sector, the last two lines in Eq.~\eqref{ha-equation-As}.  Then, all we have to do now is to introduce into this combination the expressions for the master functions $\Psi_{\mathrm{ZM}}$ and $\Psi_{\mathrm{CPM}}$ [Eqs.~\eqref{regge-wheeler-master-function} and~\eqref{cunningham-price-moncrief-master-function} respectively]. It is then a matter of using the perturbative Einstein field equations [see Appendix~\ref{harmonic-components-perturbative-equations}] and look for the coefficients $A_I$ ($I=1,\ldots,16$). One can see that there is a unique solution that corresponds to
\begin{eqnarray}
\tilde{h}^{}_a & = & \frac{r}{f}\Psi^{}_{\mathrm{RW}} r^{}_a - \frac{r^b}{2f}\left( r \Psi^{}_{\mathrm{CPM}} \right)^{}_{:b} t^{}_a \nonumber \\[2mm] 
& - & \frac{2r^2}{(\ell+2)(\ell-1)f}(t^b\mathcal{S}^{}_b)t^{}_a \,.
\label{reconstruction-hta}
\end{eqnarray}
As we have mentioned before, this expression can be rewritten entirely in terms of $\Psi_{\mathrm{CPM}}$ and the odd-parity source terms:
\begin{eqnarray}
\tilde{h}^{}_a = \frac{1}{2}\varepsilon_{a}{}^{b} \left(r \Psi^{}_{\mathrm{CPM}}\right)_{:b} + \frac{2r^2}{(\ell+2)(\ell-1)}\mathcal{S}^{}_a \,.
\label{reconstruction-hta-CPM}
\end{eqnarray}
Note that if we would like to express $\tilde{h}_a$ only in terms of $\Psi_{\mathrm{RW}}$, we would end up with time integrals, which may be more difficult to handle in practical situations. This shows again the preferred role of $\Psi_{\mathrm{CPM}}$ over $\Psi_{\mathrm{RW}}$. 

To sum up, combining equations~\eqref{reconstruction-ha-f1} and~\eqref{reconstruction-h2-f1} with Eq.~\eqref{reconstruction-hta} we can write the reconstruction of the odd-parity metric perturbations $(h_a,h_2)$ as follows: 
\begin{eqnarray}
\nonumber
h^{}_a & = & \frac{1}{2}\varepsilon_{a}{}^{b} \left(r \Psi^{}_{\mathrm{CPM}}\right)_{: b} +\frac{2r^2}{(\ell+2)(\ell-1)}\mathcal{S}^{}_a
\\
&+& \frac{r^2}{2}\mathcal{G}^{\mathrm{odd}}_{:a}\,,
\label{reconstruction-ha-f2}\\[2mm]
h^{}_2 & = & r^2 \mathcal{G}^{\mathrm{odd}} \,.
\label{reconstruction-h2-f2}
\end{eqnarray}
%

\subsection{Metric Reconstruction in the Even-Parity Sector}
The first step is, again, to identify the components of the metric perturbations that are pure gauge.  To that end, we have to look at the structure of the gauge transformation, which in the even-parity case is given by Eqs.~\eqref{gauge-transformation-even-1}-\eqref{gauge-transformation-even-4}.  From Eqs.~\eqref{gauge-transformation-even-2} and~\eqref{gauge-transformation-even-4}, we can immediately see that the even-parity metric perturbations $\jeven_a$ and $G$ can be canceled by a gauge transformation. Then, in the even-parity case, we can write the metric perturbations as follows:
\begin{eqnarray}
h^{}_{ab} & = & \tilde{h}^{}_{ab} + 2 \mathcal{H}^{\mathrm{even}}_{(a:b)} \,, 
\label{reconstruction-hab-f1} \\[2mm]     
\jeven^{}_a & = & \mathcal{G}^{\mathrm{even}}_a \,,
\label{reconstruction-ja-f1} \\[2mm]
K & = & \tilde{K} -\frac{\ell(\ell+1)}{2}\mathcal{G}^{\mathrm{even}} + 2 \frac{r^a}{r} \mathcal{H}^{\mathrm{even}}_a \,,
\label{reconstruction-K-f1} \\[2mm]
G & = & \mathcal{G}^{\mathrm{even}} \,.
\label{reconstruction-G-f1}
\end{eqnarray}
where
\begin{equation}
\mathcal{H}^{\mathrm{even}}_a = \mathcal{G}^{\mathrm{even}}_a - \frac{r^2}{2}\, \mathcal{G}^{\mathrm{even}}_{:a} \,,
\label{definition-Ha-gauge}
\end{equation}
and here, $(\mathcal{G}^{\mathrm{even}}_a,\mathcal{G}^{\mathrm{even}})$ are the three even-parity gauge functions that we have in this case.

The second step consists in writing the gauge-invariant metric perturbations, $\tilde{h}_{ab}$ and $\tilde{K}$, in terms of the master functions: $\Psi_{\mathrm{ZM}}$ and $\Psi_{\mathrm{NE}}$, that is, we would like to have: $\tilde{h}_{ab} = \tilde{h}_{ab}[r,\Psi_{\mathrm{ZM}},\Psi_{\mathrm{NE}}]$ and $\tilde{K} = \tilde{K}[r,\Psi_{\mathrm{ZM}},\Psi_{\mathrm{NE}}]$. We are going to use the same systematic procedure as in the odd-parity sector of the perturbations. First, we need to construct a linear combination of the master functions and their first derivatives that matches the gauge-invariant metric perturbations. To prevent ambiguities in the construction we have to take into account that the master function $\Psi_{\mathrm{NE}}$ is the time derivative of the Zerilli-Moncrief master function [see Eq.~\eqref{ZMdot-NE}], which is a consequence of the Einstein perturbative field equations. Let us start with the simplest gauge-invariant even-parity metric perturbation, i.e. $\tilde{K}$. In this case, the most general combination is:
\begin{eqnarray}
\tilde{K} & = & B^{}_1 \Psi^{}_{\mathrm{ZM}} + B^{}_2 \Psi^{}_{\mathrm{NE}} + B^{}_3 r^a \Psi^{}_{{\mathrm{ZM}}:a} + B^{}_4 t^{a} \Psi^{}_{{\mathrm{NE}}:a} \nonumber \\[2mm] 
& + & B^{}_5 r^a \Psi^{}_{{\mathrm{NE}}:a} \nonumber \\[2mm]
& + & B^{}_6 \mathcal{T} + B^{}_7 \mathcal{P} + B^{}_8 t^a\mathcal{P}_a + B^{}_9 r^a\mathcal{P}_a \nonumber \\[2mm]
& + & B^{}_{10}\mathcal{Q}^{}_{ab}t^at^b + B^{}_{11}\mathcal{Q}^{}_{ab}t^ar^b + B^{}_{12}\mathcal{Q}^{}_{ab}r^a r^b \,.  
\label{Kt-equation-Bs}
\end{eqnarray}
Notice that we have not included a term proportional to $t^{a} \Psi^{}_{{\mathrm{ZM}}:a}$ since this, by virtue of Eq.~\eqref{ZMdot-NE}, is equal to $\Psi_{\mathrm{NE}}$.  As in the treatment of the odd-parity sector, we have introduced terms proportional to the even-parity components of the energy-momentum tensor. They act as counter-terms as we have to use Einstein's perturbative equations [see Appendix~\ref{harmonic-components-perturbative-equations}] and the components of the energy-momentum tensor will appear and need to be compensated.  Then, we can determine the coefficients $B_I$ ($I=1,\ldots,12$) by introducing the expressions for the master functions $\Psi_{\mathrm{ZM}}$ and $\Psi_{\mathrm{NE}}$ in terms of the gauge-invariant metric perturbations [Eqs.~\eqref{zerilli-moncrief-master-function} and~\eqref{new-even-parity-master-function} respectively] into Eq.~\eqref{Kt-equation-Bs} and using the Einstein perturbative equations. 
As in the odd-parity sector, by forcing the vanishing of the coefficients of the remaining metric perturbations and their derivatives and of components of the energy-momentum tensor, we find that there is a unique solution for the coefficients $B_I$ so that we obtain the following expression for $\tilde{K}$:
\begin{eqnarray}
\tilde{K} & = & \left( \frac{\ell(\ell+1)}{2r} - W(r) \right) \Psi^{}_{\mathrm{ZM}} + r^a \Psi^{}_{{\mathrm{ZM}}:a}  \nonumber \\[2mm]
& - & \frac{4r^2}{\ell(\ell+1)\lambda} \mathcal{Q}^{}_{ab}t^a t^b \,,   \label{reconstruction-Kt}
\end{eqnarray}
where $W(r)$ is the superpotential shown in Eq.~\eqref{superpotential-standard}. Note that this expression for $\tilde{K}$ is consistent with the expressions we derived in the study of the Darboux covariance of the metric perturbations [See Eqs.~\eqref{darboux-like-psi-even} and~\eqref{darboux-ZM-source} in Sec.~\ref{darboux-covariance}, and Eq.~\eqref{even-new-function} for the definition of $\Phi_{\mathrm{even}}(x^a)$]. 

Let us now follow the same procedure for the reconstruction of the other gauge invariant even-parity metric perturbations, i.e. $\tilde{h}_{ab}$. The expression that we need to consider for $\tilde{h}_{ab}$ in terms of the master functions and their first-order derivatives, as well as of the even-parity components of the energy-momentum tensor, has the following structure: 
\begin{eqnarray}
\tilde{h}^{}_{ab} & = & \left( D^{}_1 t^{}_a t^{}_b + 2 D^{}_2 t^{}_{(a} r^{}_{b)} + D^{}_3 r^{}_a r^{}_b \right) \Psi^{}_{\mathrm{ZM}}  \nonumber \\[2mm]
& + & \left( D^{}_4 t^{}_a t^{}_b + 2 D^{}_5 t^{}_{(a} r^{}_{b)} + D^{}_6 r^{}_a r^{}_b \right) \Psi^{}_{\mathrm{NE}}  \nonumber \\[2mm] 
& + & \left( D^{}_7 t^{}_a t^{}_b + 2 D^{}_8 t^{}_{(a} r^{}_{b)} + D^{}_9 r^{}_a r^{}_b \right) r^c\Psi^{}_{{\mathrm{ZM}}:c}  \nonumber \\[2mm]
& + & \left( D^{}_{10} t^{}_a t^{}_b + 2 D^{}_{11} t^{}_{(a} r^{}_{b)} + D^{}_{12} r^{}_a r^{}_b \right) t^c\Psi^{}_{{\mathrm{NE}} :c}  \nonumber \\[2mm] 
& + & \left( D^{}_{13} t^{}_a t^{}_b + 2 D^{}_{14} t^{}_{(a} r^{}_{b)} + D^{}_{15} r^{}_a r^{}_b \right) r^c\Psi^{}_{{\mathrm{NE}} :c} \nonumber\\[2mm]
& + & \left(
        \begin{array}{l}
     \mbox{General combination of $T^{}_{\mu\nu}$}  \\
    \mbox{components. 21 more coefficients:}\\ 
    \mbox{$D^{}_{16}\,$,\ldots, $D^{}_{36}\,$.}    
      \end{array}
\right)^{}_{ab} \,, \quad
\label{htab-equation-Ds}
\end{eqnarray}
where again, to avoid ambiguities, we have not included terms proportional to $t^{a} \Psi^{}_{{\mathrm{ZM}}:a}$. The last line refers to a general combination of the components of the energy-momentum tensor that we have not made explicit for the sake of brevity and that will act as counter-terms when we use Einstein's perturbative equations. The only difference with the previous case, the reconstruction of $\tilde{K}$, is that now we have a tensorial equation and the number of coefficients to be determined is significantly higher: $D_I$ ($I=1,\ldots 36$). 
The final result can be written in the following form [where we have used the relationship between and $\Psi_{\mathrm{NE}}$ and the time derivative of $\Psi_{\mathrm{ZM}}$ in Eq.~\eqref{ZMdot-NE}]: 
\begin{widetext}
\begin{eqnarray}
\tilde{h}^{}_{ab} & = & \left[-\frac{r V^{}_{\mathrm{Z}}}{2f^2} \Psi^{}_{\mathrm{ZM}} + \frac{1}{\lambda f^{3/2}}r^c\left( \frac{r\lambda}{\sqrt{f}} r^d \Psi^{}_{{\mathrm{ZM}}:d} \right)^{}_{:c} \right]\left( t^{}_a t^{}_b + r^{}_a r^{}_b \right)     
- \frac{4}{\lambda f^{3/2}} r^c \left( \frac{r\lambda}{\sqrt{f}} \Psi^{}_{\mathrm{NE}} \right)^{}_{:c} \; t^{}_{(a} r^{}_{b)} \nonumber \\[2mm]
& + &\left\{ \frac{2r^2}{\ell(\ell+1)f^2\lambda^2}\left[ \left(\lambda+3f\right)^2 -2 f\left(  5\lambda + 3 - \Lambda r^2 +\frac{3f}{2}\right) \right] \mathcal{Q}^{}_{cd}t^ct^d
- \frac{4 r^3}{\ell(\ell+1)f^2\lambda} r^{c}\left(\mathcal{Q}^{}_{cd}t^c t^d \right)_{:c}\right\}\left( t^{}_a t^{}_b + r^{}_a r^{}_b \right) \nonumber \\[2mm]
& - & \frac{4r^2}{f^2\lambda}\left[ \mathcal{Q}^{}_{cd}t^c r^d + \frac{2 f}{r} t^c \mathcal{P}^{}_c \right] t^{}_{(a} r^{}_{b)} 
+ \frac{2r^2}{f} \mathcal{P} t^{}_a t^{}_b  
\,.
\label{reconstruction-htab}
\end{eqnarray}   
\end{widetext}
This expression deserves several comments. First of all, in the same way as all the previous metric reconstructions that we have carried out, it has two parts. A \textit{gravitational} part that can be expressed only in terms of the master functions and their derivatives along $r^a$, and a \textit{matter} part that can be expressed only in terms of the components of the energy-momentum tensor and its first-order derivatives. This is quite relevant and useful for different purposes. In particular, in the case of solving the master equations in the time domain, that is, evolving Eq.~\eqref{master-wave-equation} using a Cauchy evolution from initial data on an initial time slice. Then, we can always reconstruct the metric perturbations at any time sclice by taking the values of the master functions and their spatial (radial) derivatives on that slice. A particularly interesting feature of the reconstruction of $\tilde{h}_{ab}$ [Eq.~\eqref{reconstruction-htab}] is that it contains second-order radial derivatives of the master function $\Psi_{\mathrm{ZM}}$. This means we can use the master equation~\eqref{master-wave-equation} to change this expression, but this would bring second-order time derivatives of the same master function and the first term would retain the same structure but with a different coefficient.  

Another remarkable fact of the metric reconstruction of $\tilde{h}_a$, Eq.~\eqref{reconstruction-hta}, is that it uses both master functions, $\Psi_{\mathrm{ZM}}$ and $\Psi_{\mathrm{NE}}$. This change, as we already mentioned in the analysis of the odd-parity sector, if we use Eq.~\eqref{ZMdot-NE} to express the metric reconstruction in terms of $\Psi_{\mathrm{ZM}}$ only, but at the price of also having time derivatives of this master function. An expression in terms of $\Psi_{\mathrm{NE}}$ would unavoidably involve time integrals.

Finally, a quite remarkable fact about the \textit{gravitational} part of the reconstruction of $\tilde{h}_{ab}$ is that it  has only two independent components, namely one corresponding to $t^{}_a t^{}_b + r^{}_a r^{}_b$, and the other one corresponding to $2t^{}_{(a} r^{}_{b)}$. Not only that, both components are clearly traceless, and hence, it turns out that in the absence of matter sources, so it is $\tilde{h}_{ab}$:
\begin{equation}
g^{ab} \tilde{h}^{}_{ab} = 0 \,.
\end{equation}
This is obviously a consequence of the perturbative Einstein equations for the even-parity sector, but it is not completely obvious that it has to be this way.   In the general case, the trace of $\tilde{h}_{ab}$ is given by the last term in its expression [Eq.~\eqref{reconstruction-htab}], that is:
\begin{equation}
g^{ab} \tilde{h}^{}_{ab} = -2 r^2\,\mathcal{P}\,.
\label{fromEEY}
\end{equation}
This equation is one of the components of the perturbative Einstein field equations [see Eq.~\eqref{EEY} in Appendix~\ref{harmonic-components-perturbative-equations}].

To sum up, using the metric reconstructions of Eqs.~\eqref{reconstruction-Kt} and~\eqref{reconstruction-htab}, the final form of the metric reconstruction for even-parity metric perturbations is:
\begin{widetext}
\begin{eqnarray}
h^{}_{ab} & = & \left[-\frac{r V^{}_{\mathrm{Z}}}{2f^2} \Psi^{}_{\mathrm{ZM}} + \frac{1}{\lambda f^{3/2}}r^c\left( \frac{r\lambda}{\sqrt{f}} r^d \Psi^{}_{{\mathrm{ZM}}:d} \right)^{}_{:c} \right]\left( t^{}_a t^{}_b + r^{}_a r^{}_b \right)     
- \frac{4}{\lambda f^{3/2}} r^c \left( \frac{r\lambda}{\sqrt{f}} \Psi^{}_{\mathrm{NE}} \right)^{}_{:c} \; t^{}_{(a} r^{}_{b)}  + 2 \mathcal{H}^{\mathrm{even}}_{(a:b)} \nonumber \\[2mm]
& + &\left\{ \frac{2r^2}{\ell(\ell+1)f^2\lambda^2}\left[ \left(\lambda+3f\right)^2 -2 f\left(  5\lambda + 3 - \Lambda r^2 +\frac{3f}{2}\right) \right] \mathcal{Q}^{}_{cd}t^ct^d
- \frac{4 r^3}{\ell(\ell+1)f^2\lambda} r^{c}\left(\mathcal{Q}^{}_{cd}t^c t^d \right)_{:c}\right\}\left( t^{}_a t^{}_b + r^{}_a r^{}_b \right) \nonumber \\[2mm]
& - & \frac{4r^2}{f^2\lambda}\left[ \mathcal{Q}^{}_{cd}t^c r^d + \frac{2 f}{r} t^c \mathcal{P}^{}_c \right] t^{}_{(a} r^{}_{b)} 
+ \frac{2r^2}{f} \mathcal{P} t^{}_a t^{}_b    \,, 
\label{reconstruction-hab-f2} \\[2mm]     
\jeven^{}_a & = & \mathcal{G}^{\mathrm{even}}_a \,,
\label{reconstruction-ja-f2} \\[2mm]
K & = &  \left( \frac{\ell(\ell+1)}{2r} - W(r) \right) \Psi^{}_{\mathrm{ZM}} + r^a \Psi^{}_{{\mathrm{ZM}}:a} - \frac{\ell(\ell+1)}{2}\mathcal{G}^{\mathrm{even}} + 2 \frac{r^a}{r} \mathcal{H}^{\mathrm{even}}_a  - \frac{4r^2}{\ell(\ell+1)\lambda} \mathcal{Q}^{}_{ab}t^a t^b 
\,, \label{reconstruction-K-f2} \\[2mm]
G & = & \mathcal{G}^{\mathrm{even}} \,.
\label{reconstruction-G-f2}    
\end{eqnarray}    
\end{widetext}

\subsection{Metric Reconstruction from Darboux Branch Master Functions}
\label{Ss:metric-recon-darboux}

An obvious question in this construction is whether we can do the metric reconstruction shown before but, instead of in terms of the master functions of the standard branch, in terms of the master functions of the Darboux branch. The answer is positive, but given that it involves longer computations than in the case of the standard branch case that we have fully developed here, we are only going to show that is possible to do it and the steps one would need to follow. 
The main element is precisely the results we have for the standard branch. We are going to see how we can use them to get to the metric reconstruction in terms of the Darboux branch master functions.  
As we have mentioned in Sec.~\ref{darboux-covariance}, the Darboux transformation is reversible in such a way that the function generating the inverse Darboux transformation is just $-g$, being $g$ the Darboux generating function~\eqref{darboux-g-gx}. Therefore, we can just exploit this property to invert Eqs.~\eqref{darboux-even-source} and~\eqref{darboux-odd-source} to obtain:
\begin{eqnarray}
\Psi^{}_{\mathrm{CPM}}  & = & {}^{}_{D}\check{\Psi}^{}_{\mathrm{odd},x} - g^{\mathrm{odd}}\, {}^{}_{D}\check{\Psi}^{}_{\mathrm{odd}} \,, \\[2mm]
\Psi^{}_{\mathrm{ZM}}  & = &  {}^{}_{D}\check{\Psi}^{}_{\mathrm{even},x} - g^{\mathrm{even}}\, {}^{}_{D}\check{\Psi}^{}_{\mathrm{even}} \,.
\end{eqnarray}
Then, metric reconstruction in terms of ${}^{}_{D}\check{\Psi}_{\mathrm{odd}}$ and ${}^{}_{D}\check{\Psi}_{\mathrm{even}}$ can found in a straightforward way by substituting these inverted Darboux transformations into the standard branch metric reconstruction given in Eqs.~\eqref{reconstruction-ha-f2}-\eqref{reconstruction-h2-f2} and~\eqref{reconstruction-hab-f2}-\eqref{reconstruction-G-f2}. 
Notice however that, since the Darboux transformations of $\Psi_{\mathrm{RW}}$ and $\Psi_{\mathrm{NE}}$ [see Eq.~\eqref{darboux-from-NE-and-RW} and the discussion around it in Sec.~\ref{darboux-covariance}] do not belong to the space of master functions and equations that we have considered, we can do this only with the metric reconstruction expressed entirely in terms of $\Psi_{\mathrm{CPM}}$ and $\Psi_{\mathrm{ZM}}$.

\section{Discussion} \label{discussion}

In this paper, building on previous works presented in~\cite{Lenzi:2021wpc,Lenzi:2021njy,Lenzi:2022wjv,Lenzi:2023inn}, we have dealt with the problem of reconstructing the metric perturbations of $\Lambda$-vacuum spherically-spacetimes, generalizing previous results in the literature in several ways: 
(i) Our analysis includes all the spherically-symmetric background spacetimes described in Appendix~\ref{background-spacetime-metric}. 
(ii) We have included in the analysis the presence of matter sources, at the first perturbative order, by considering a completely general energy-momentum tensor. 
(iii) We have performed the metric reconstruction in a way that is completely independent of the choice of perturbative gauge. 
(iv) All the results we present have been put in the context of the space of possible master functions and equations, taking into account that all the elements of this space are connected by a hidden symmetry, Darboux covariance, which is the key ingredient to show that all the possible perturbative formulations that this space allows are physically equivalent.  Moreover, we have shown how to perform the metric reconstruction in terms of any of the possible master functions in the space we described.  
(v) All the expressions we present are covariant with respect to changes of coordinates in $M^2$ (and of course, also in $S^2$).

The emerging point of view is that, in principle (up to considerations of continuity and differentiability of the matter sources), we can always work with the master equations and functions. The master functions represent the true degrees of freedom of the gravitational field (remember the background is not dynamical in the sense that the degrees of freedom of the gravitational field are shut down).  Then, it is in the reconstruction of the metric perturbations where we can specify the particular perturbative gauge.  Once the gauge-invariant variables, which are reconstructed from the master functions, are recovered, we have to solve equations/conditions that depend on the specific choice of perturbative gauge. 

Despite all these generalizations, this was not the only aim of this work.  As it is well-known, the master functions represent the true degrees of freedom of the gravitational field described by the perturbations. Indeed, in terms of these gauge-invariant quantites we can compute any physical observable quantity that these perturbations can describe. In particular, the emission of gravitational waves and their energy-momentum content (energy, angular momentum, and linear momentum fluxes).  Then, the possibility of isolating the true degrees of freedom in BHPT, gives us some crucial advantages with respect to the full theory, where several attempts have been described in the literature as it can be the first step towards quantization.  One remarkable attempt is the introduction of the well-known Arnowitt-Deser-Misner formalism (ADM)~\cite{Arnowitt:1962hi}, which constitutes a Hamiltonian formalism for General Relativity and the mathematical basis for the development of a large part of the area of Numerical Relativity. Despite one cannot isolate the gravitational degrees of freedom, there are some analogies with BHPT that can be more appreciated in the framework we have presented here. In the ADM formulation and related ones, one evolves the intrinsic metric and extrinsic curvature of a three-dimensional initial Cauchy surfaces (encoding the physical gravitational information of the system under consideration), and we have four gauge functions (a scalar, the \emph{lapse}, and a three-dimensional vector, the \emph{shift}) that are freely specifiable. In this sense, a particularly interesting formulation of Numerical Relativity for our purpose of comparison is the Generalized Harmonic Gauge (GHG) approach introduced by Pretorius~\cite{Pretorius:2004jg,Pretorius:2005gq}, which is behind the Numerical Relativity breakthrough of solving the binary black hole problem (BBH; see also~\cite{Campanelli:2005dd,Baker:2005vv} for different Numerical Relativity formalism that also managed to solve the BBH problem shortly afterwards; these formulations are based on the formulations of Baumgarte \& Shapiro~\cite{Baumgarte:1998te} and Shibata \& Nakamura~\cite{Shibata:1995we}, the so-called BSSN formalism, an extension of the ADM formalism).  In the GHG formulation, we can obtain (non-linear) wave-type equations for a set of metric components and have four freely specifiable gauge functions.  In analogy, in this paper it becomes clear that in BHPT we can proceed to solve the master equations for the odd- and even-parity perturbations, which are valid in any gauge, and then, afterwards, we can reconstruct the full metric perturbations in any perturbative gauge of interest for the particular problem under consideration.  This provides a very interesting perspective for BHPT computations, where we can always solve master equations and, afterwards, find the gauge functions that corresponds to the particular gauge we may be interested in for our physical situation. We plan to illustrate, in a follow-up study, how this would work for the most relevant gauge choices used in the literature

Finally, the work presented in this paper can be the basis for developing applications of BHPT to different types of problems, from gravitational wave astronomy to fundamental questions of gravitation. This is particularly useful in cases where either a complete knowledge of the metric perturbations is required, or when we need to perform computations in a non-standard gauge, in particular in a gauge different from the Regge-Wheeler one, where all the gauge functions are chosen to vanish identically.  Another extension of this work is to broaden its scope.  By one hand, one can study possible extensions within General Relativity: For instance, going to second- and higher-order perturbations within spherical symmetry (see~\cite{Brizuela:2006ne,Brizuela:2007zza,Brizuela:2009qd} for systematic treatments of non-linear perturbations), or going beyond spherical symmetry, in particular to generalize the present framework to the Kerr metric, whose relevance and importance is well known. On the other hand, there is the possibility of extending these results to other theories of gravity, where the study of perturbations of spherically-symmetric systems may be of interest~\cite{Langlois:2021aji,Chen:2021cts}. A particular example in this line, where a such a perturbative scheme has been used, is in Ref.~\cite{MenaMarugan:2024qnj}, where in the context of loop quantum gravity/cosmology the authors study quantum effects in BH interiors.

\begin{acknowledgments}
ML and CFS  are supported by contracts PID2019-106515GB-I00 and PID2022-137674NB-I00 (MCIN/AEI/10.13039/501100011033) and 2017-SGR-1469 (AGAUR, Generalitat de Catalunya). 
ML is also supported by Juan de la Cierva contract FJC2021-047289-I funded by program MCIN/AEI/10.13039/501100011033 (Spanish Ministry of Science and Innovation) and by NextGenerationEU/PRTR (European Union).
This work was also partially supported by the program \textit{Unidad de Excelencia Mar\'{\i}a de Maeztu} CEX2020-001058-M (Spanish Ministry of Science and Innovation).
We have used the Mathematica software~\cite{Mathematica} to develop mathematical tools to manipulate Einstein's equations for perturbations of $\Lambda$-vacuum spherically-symmetric spacetimes (see Appendix~\ref{harmonic-components-perturbative-equations}) and also to check most of the computations that appear in this paper. 
\end{acknowledgments}

\appendix

\section{The Background Spacetime Metric}\label{background-spacetime-metric}
Vacuum spherically symmetric solutions are locally determined by Birkhoff's  uniqueness theorem~\cite{Birkhoff:1923hup} (published before by Jebsen~\cite{Jebsen:1921ori,Jebsen:2005grg}; see also~\cite{Deser:2004gi,VojeJohansen:2005nd}). The theorem can be generalized to the case of a non-vanishing cosmological constant~\cite{Eisland:1925eis} (see~\cite{Schleich:2009uj} for details), what we call a $\Lambda$-vacuum spacetime. 

The possible solutions have to be locally isometric either to one of the Schwarzschild-de Sitter (SchdS~\cite{Kottler:1918}) and Schwarzschild-anti-de Sitter (SchAdS) solutions or to the Nariai spacetime~\cite{Nariai:1950hna,Nariai:1999nar}, which can be seen as the limit of SchdS when the cosmological and event horizons coincide. We do not consider this case here as it may require a particular treatment.

This family of metrics includes, in the zero mass limit ($M\rightarrow 0$), the well-knwon maximally-symmetric solutions of Einstein equations: Minkowski flat spacetime (M; $\Lambda=0$), de Sitter (dS; $\Lambda>0$), and anti-de Sitter (AdS; $\Lambda<0$).

All these metrics are determined by the function $f(r)$ in Eq.~\eqref{f-background-function}. In the case of the Schwarzschild spacetime~\cite{Schwarzschild:1916uq} (found independently also by Droste~\cite{1917KNAB...19..197D}) we have 
\begin{equation}
f^{}_{\mathrm{Sch}} = 1-\frac{r_{s}}{r}\,,
\end{equation}
where we have introduced the Schwarzschild radius: $r_{s} = 2GM/c^{2}=2M\,$. In the case of de Sitter and anti-de Sitter spacetimes we have
\begin{equation}
f^{}_{\mathrm{dS}} = 1-\frac{r^{2}}{L^2}\,,\qquad
f^{}_{\mathrm{AdS}} = 1+\frac{r^2}{L^2}
\end{equation}
where we have introduced the length scale $L$, which is related to the absolute value of the cosmological constant by
%
\begin{equation}
L = \sqrt{\frac{3}{|\Lambda|}} \,,   
\label{cosmological-constant}
\end{equation}
%

\section{Spherical Harmonics}\label{sphericalharmonics}
Adopting the conventions of~\cite{Abramowitz:1970as,Press:1992nr}, the scalar spherical harmonics are defined by
\begin{equation}
Y^{\ell m}(\theta,\varphi) = \sqrt{\frac{2 \ell + 1}{4 \pi}
      \frac{\left(\ell - m\right)!}{\left(\ell + m\right)!}}\,
      P^m_\ell(\cos{\theta}) e^{i m \varphi}\,,
\end{equation}
where $P^{m}_{\ell}$ are the associated Legendre polynomials, which can be introduced in the following form
\begin{equation}
P^m_\ell(x) = \frac{(-1)^m}{2^\ell \ell!}(1-x^2)^{m/2}\frac{d^{\ell+m}}{dx^{\ell+m}}
(x^2-1)^\ell \,.
\end{equation}
An important property of the scalar spherical harmonics is:
\begin{equation}
\bar{Y}^{\ell m} = (-1)^{m} Y^{\ell -m} \,.
\end{equation}
From the scalar spherical harmonics, we can introduce vector and tensor spherical harmonics by the definition shown in Eqs.~(\ref{vectorharmonics})-(\ref{tensorharmonics}).

\subsection{Orthogonality properties of Spherical Harmonics}
Given the expansion of a scalar/vector/tensor in the corresponding spherical harmonics, we can extract the coefficients of each mode by using the following orthogonality properties:
\begin{eqnarray}
\int^{}_{S^2} \hspace{-3mm} d\Omega\, Y^{\ell m}\bar{Y}^{\ell' m'} & = & \delta^{\ell \ell'} \delta^{m m'}\,, 
\\
\int^{}_{S^2} \hspace{-3mm} d\Omega\,\Omega^{AB}Y^{\ell m}_A\bar{Y}^{\ell' m'}_B & = & 
\ell(\ell+1)\delta^{\ell \ell'} \delta^{m m'}\,, 
\\
\int^{}_{S^2} \hspace{-3mm} d\Omega\,\Omega^{AB}X^{\ell m}_A\bar{X}^{\ell' m'}_B & = & 
\ell(\ell+1)\delta^{\ell \ell'} \delta^{m m'}\,, 
\\
\int^{}_{S^2} \hspace{-3mm} d\Omega\,\Omega^{AB}Y^{\ell m}_A\bar{X}^{\ell' m'}_B & = & 0\,,
\\
\int^{}_{S^2} \hspace{-3mm} d\Omega\,\Omega^{AC}\Omega^{BD}T^{\ell m}_{AB}\bar{T}^{\ell' m'}_{CD} & = &  
2\delta^{\ell \ell'} \delta^{m m'}\,, 
\\
\int^{}_{S^2} \hspace{-3mm} d\Omega\,\Omega^{AC}\Omega^{BD}Y^{\ell m}_{AB}\bar{Y}^{\ell' m'}_{CD} & = &  
\frac{(\ell+2)!}{2(\ell-2)!}\delta^{\ell \ell'} \delta^{m m'}\,, 
\\
\int^{}_{S^2} \hspace{-3mm} d\Omega\,\Omega^{AC}\Omega^{BD}X^{\ell m}_{AB}\bar{X}^{\ell' m'}_{CD} & = &  
\frac{(\ell+2)!}{2(\ell-2)!}\delta^{\ell \ell'} \delta^{m m'}\,, 
\\
\int^{}_{S^2} \hspace{-3mm} d\Omega\,\Omega^{AC}\Omega^{BD}Y^{\ell m}_{AB}\bar{X}^{\ell' m'}_{CD} & = &  0\,,
\end{eqnarray}
\begin{equation}
\Omega^{AC}\Omega^{BD}Y^{\ell m}_{AB}Y^{\ell' m'}_{CD}  
=\Omega^{AC}\Omega^{BD}X^{\ell m}_{AB}T^{\ell' m'}_{CD} =0\,.
\end{equation}

\subsection{Differential properties of Spherical Harmonics}
Given that we have to introduce the expansion of the metric perturbations in spherical harmonics into the Einstein perturbative equations, it is important to consider the differential equations they satisfy. In the case of the scalar harmonics we have Eq.~\eqref{laplacian-of-Ylm} and the definitions of the other harmonics (vector and tensors) in terms of them. On the other hand, the polar vector harmonics $Y^{\ell m}_{A}$ satisfy the following differential identities:
\begin{eqnarray}
\Omega^{AB} Y^{\ell m}_{A|B} & = & -\ell(\ell+1)Y^{\ell m} \,, \\
\Omega^{BC} Y^{\ell m}_{B|CA} & = & -\ell(\ell+1)Y^{\ell m}_{A} \,,\\
\Omega^{BC} Y^{\ell m}_{A|BC} & = & \left[1 -\ell(\ell+1)\right]Y^{\ell m}_{A} \,,
\end{eqnarray}
and the axial vector harmonics $X^{\ell m}_{A}$ satisfy similar differential identities:
\begin{eqnarray}
\Omega^{AB} X^{\ell m}_{A|B} & = & 0 \,, \\
\Omega^{BC} X^{\ell m}_{A|BC} & = & \left[1 -\ell(\ell+1)\right]X^{\ell m}_{A} \,,\\
\Omega^{BC} X^{\ell m}_{B|AC} & = & X^{\ell m}_{A} \,.
\end{eqnarray}
On the other hand, the (symmetric) polar tensor harmonics $T^{\ell m}_{AB}$ and $Y^{\ell m}_{AB}$ satisfy the following differential identities:
\begin{eqnarray}
\Omega^{BC} T^{\ell m}_{BC|A} & = & 2\,Y^{\ell m}_{A} \,, \\
\Omega^{BC} T^{\ell m}_{AB|C} & = & Y^{\ell m}_{A} \,,\\
\Omega^{CD} T^{\ell m}_{AB|CD} & = & -\ell(\ell+1) T^{\ell m}_{AB} \,,\\
\Omega^{CD} T^{\ell m}_{CD|AB} & = & Y^{\ell m}_{AB} -\ell(\ell+1) T^{\ell m}_{AB} \,.
\end{eqnarray}
\begin{eqnarray}
\Omega^{BC} Y^{\ell m}_{BC|A} & = & 0 \,, \\
\Omega^{BC} Y^{\ell m}_{AB|C} & = & -\frac{(\ell+2)(\ell-1)}{2} Y^{\ell m}_{A} \,,\\
\Omega^{CD} Y^{\ell m}_{AB|CD} & = & \left[4 -\ell(\ell+1)\right] Y^{\ell m}_{AB} \,.
\end{eqnarray}
Finally, the (symmetric) axial tensor harmonics $X^{\ell m}_{AB}$ satisfy the following differential identities:
\begin{eqnarray}
\Omega^{BC} X^{\ell m}_{BC|A} & = & 0 \,, \\
\Omega^{BC} X^{\ell m}_{AB|C} & = & -\frac{(\ell+2)(\ell-1)}{2} X^{\ell m}_{A} \,,\\
\Omega^{CD} X^{\ell m}_{AB|CD} & = & \left[4 -\ell(\ell+1)\right] X^{\ell m}_{AB} \,.
\end{eqnarray}
%

\section{Equations for the Harmonic Components of the Metric Perturbations}
\label{harmonic-components-perturbative-equations}

Following~\cite{Martel:2005ir} and~\cite{Lenzi:2021wpc}, the perturbative Einstein equations [see Eq.~\eqref{efes-full}] can be decomposed in its spherical harmonic components. For a single $(\ell,m)$-harmonic, the structure of $\delta G_{\mu\nu}$ is: 
\begin{widetext}
\begin{eqnarray}
\delta G^{\ell m}_{ab}(x^{c},\Theta^{A}) & = & \mathcal{E}^{\ell m}_{ab}(x^{c})\;Y^{\ell m}(\Theta^{A})\,, 
\label{PEFEs-ab}\\[2mm]
\delta G^{\ell m}_{aA}(x^{b},\Theta^{B}) & = & \mathcal{E}^{\ell m}_{a}(x^{b})\;Y^{\ell m}_{A}(\Theta^{B}) + \mathcal{O}^{\ell m}_{a}(x^{b})\;X^{\ell m}_{A}(\Theta^{B})\,,
\label{PEFES-aA}\\[2mm]
\delta G^{\ell m}_{AB}(x^{a},\Theta^{C}) & = & \mathcal{E}^{\ell m}_{T}(x^{a})\;T^{\ell m}_{AB}(\Theta^{C}) + \mathcal{E}^{\ell m}_{Y}(x^{a})\;Y^{\ell m}_{AB}(\Theta^{C}) + \mathcal{O}^{\ell m}_{X}(x^{a})\;X^{\ell m}_{AB}(\Theta^{C})\,.
\label{PEFES-AB}
\end{eqnarray}
\end{widetext}
In this paper, we are considering the presence of matter source at first perturbative order, described by an arbitrary energy-momentum tensor $T_{\mu\nu}$ whose decomposition in spherical harmonics in given in Eqs.~\eqref{Tmunu-split}-\eqref{Tmunu-tlm}. 
Then, by looking at these decompositions, the perturbative Einstein equations are equivalent to impose the vanishing of the following harmonic components, namely: 
(a) Polar-parity sector: 
$\mathcal{EE}^{\ell m}_{ab}\equiv\mathcal{E}^{\ell m}_{ab}-\mathcal{Q}^{\ell m}_{ab}=0$, 
$\mathcal{EE}^{\ell m}_{a}\equiv\mathcal{E}^{\ell m}_{a}-\mathcal{P}^{\ell m}_{a}=0$,  
$\mathcal{EE}^{\ell m}_{T}\equiv\mathcal{E}^{\ell m}_{T}-r^2\mathcal{T}^{\ell m}=0$, 
$\mathcal{EE}^{\ell m}_{Y}\equiv\mathcal{E}^{\ell m}_{Y}-r^2\mathcal{P}^{\ell m}=0$. 
(b) Odd-parity sector: 
$\mathcal{OO}^{\ell m}_{a}\equiv\mathcal{O}^{\ell m}_{a}-\mathcal{S}^{\ell m}_{a}=0$, 
$\mathcal{OO}^{\ell m}_{X}\equiv\mathcal{O}^{\ell m}_{X}-\mathcal{S}^{\ell m}=0$.  
The result is [for simplicity we drop the harmonic numbers and these equations coincide with the components of the perturbative Einstein equations up to global factors]:

\begin{widetext}
\begin{eqnarray}
\mathcal{EE}^{}_{ab} & : &  h^{}_{c <a:b>}{}^{:c} - \frac{1}{2}h^{}_{<ab>:c}{}^{:c} - \frac{1}{2}h^{}_{:<ab>} + \frac{r^c}{r} \left( 2h^{}_{c<a:b>} - h^{}_{<ab>:c} \right) - K^{}_{:<ab>} - \frac{2}{r}r^{}_{<a}K^{}_{:b>} \nonumber \\[2mm]
& & + \frac{\ell(\ell+1)}{2 r^2} \left( h^{}_{<ab>} - 2\,\jeven^{}_{<a:b>} \right) -\Lambda h^{}_{<ab>} - \frac{1}{2}g^{}_{ab} \left[ \frac{r^c}{r}\left(2 h^{}_{cd}{}^{:d} - h^{}_{:c}\right) - K^{}_{:c}{}^{:c} - 4 \frac{r^c}{r}K^{}_{:c} + \frac{\ell(\ell+1)}{2r^2} h  \right.  \nonumber \\[2mm]
& & \left. + \frac{2}{r^2}\left(r^c r^d + r r^{:cd} \right)h^{}_{cd} - \frac{\ell(\ell+1)}{r^2}\left( \jeven^{}_c{}^{:c} + 2 \frac{r^c}{r}\jeven^{}_c \right) + \left( \frac{\ell(\ell+1)}{r^2} - \frac{2}{r^2}(r^c r^{}_c + r r^{}_{:c}{}^{:c}) - 2\Lambda  \right) K \right.   
 \nonumber \\[2mm]
& & \left. + \frac{(\ell+2)(\ell+1)\ell(\ell-1)}{2 r^2}G \right] = \mathcal{Q}^{}_{ab} \,, 
\label{EEab}\\[2mm]
\mathcal{EE}^{}_{a} & : & \frac{1}{2}h^{}_{ab}{}^{:b} - \frac{1}{2}h^{}_{:a} + \frac{r^{}_a}{2 r}h + \frac{1}{2}\jeven^c{}^{}_{:ca} - \frac{1}{2}\jeven^{}_{a:c}{}^c - \frac{r^{}_a}{r}\jeven^b{}^{}_{:b} + \frac{r^b}{r}\jeven^{}_{b:a} + \left[\left(\frac{{}^{2}R}{4} - \Lambda \right) g^{}_{ab} - \frac{r^{}_{:ab}}{r} - \frac{r^{}_a r^{}_b}{r^2}\right]\jeven^b - \frac{1}{2}K^{}_{:a}  \nonumber \\[2mm]
& & - \frac{(\ell+2)(\ell-1)}{4} G^{}_{:a} = \mathcal{P}^{}_a \,, 
\label{EEa}\\[2mm]
\mathcal{OO}^{}_{a} & : &  \frac{1}{2}h^c{}^{}_{:ca} - \frac{1}{2}h^{}_{a:c}{}^c - \frac{r_a}{r} h^b{}^{}_{:b} + \frac{r^b}{r}h^{}_{b:a} + \left[\left(\frac{{}^{2}R}{4} - \Lambda \right) g^{}_{ab} - \frac{r^{}_{:ab}}{r}-\frac{r^{}_a r^{}_b}{r^2}\right] h^b + \frac{\ell(\ell+1)}{2 r^2} h^{}_a  \nonumber \\[2mm]
& & - \frac{(\ell+2)(\ell-1)}{4 r^2}\left(h^{}_{2:a}-\frac{2 r^{}_a}{r}h^{}_2 \right) = \mathcal{S}^{}_a \,, 
\label{OOa} \\[2mm]
\mathcal{EE}^{}_{T} & : &  h^{}_{ab}{}^{:ab} - h^{}_{:a}{}^{:a} + \frac{r^b}{r}\left( 2 h^{}_{ba}{}^{:a} - h^{}_{:b}\right) + \left( \frac{\ell(\ell+1)}{2r^2} - \Lambda\right) h - \frac{\ell(\ell+1)}{r^2} \jeven^a{}^{}_{:a} - K^{}_{:a}{}^{:a} - 2\frac{r^a}{r}K^{}_{:a}  = - 2\mathcal{T} \,, 
\label{EET}  \\[2mm]
\mathcal{EE}^{}_{Y} & : & \jeven^a{}^{}_{:a} - \frac{1}{2}r^2 G^{}_{:a}{}^{:a} - r r^a G^{}_{:a} - \left( r r^{:a}{}^{}_{:a} + r^a r^{}_a - 1 + \Lambda r^2 \right) G - \frac{1}{2}h = r^2 \mathcal{P}  \,, 
\label{EEY} \\[2mm]
\mathcal{OO}^{}_{X} & :& h^a{}^{}_{:a} - \frac{1}{2}h^{}_{2:a}{}^{:a} + \frac{r^a}{r} h^{}_{2:a} + \frac{1}{r^2}\left(1-2 r^a r^{}_a- \Lambda r^2 \right)h^{}_2 = \mathcal{S} \,,
\label{OOX}
\end{eqnarray}
\end{widetext}
where in these equations we have used the trace-free notation for tensors, which in two-dimensional case of $M^2$ reads: For any symmetric tensor $A_{ab}$, we define: $A_{<ab>} = A_{ab} - (1/2)g_{ab} A\,$, with $A=g^{ab} A_{ab}$ being the trace. 

Given that the Einstein tensor (including the cosmological constant) is gauge-invariant at first-order in the perturbations (using the Stewart-Walker lemma and taking into account that it vanishes in the background), 
we can use the decomposition of the metric perturbations and rewrite them in terms of purely gauge invariant quantities. In this way the equations simplify significantly. For instance, Eq.~\eqref{EEY} becomes Eq.~\eqref{fromEEY}.

\section{Equations for the Harmonic Components of the Energy-Momentum Tensor}
\label{energy-momentum-harmonic-components-equations}

The energy-momentum conservation equations come from the second Bianchi identities [see Eq.~\eqref{perturbative-second-bianchi-identities}].  The spherical-harmonic components of these equations contain three polar-parity components and a single odd-parity component, and they are given by 
\begin{eqnarray}
\label{bianchi-tmunu-harmonic}
\mathcal{Q}^{}_{ab}{}^{:b} & = & - \frac{2r^b}{r}\mathcal{Q}^{}_{ab} +  \frac{\ell(\ell+1)}{r^2}\mathcal{P}^{}_a  +  \frac{2 r^{}_a}{r}\mathcal{T}   \,, \\[2mm]
\mathcal{P}^{}_a{}^{:a} & = & -2\frac{r^a}{r}\mathcal{P}^{}_a + \frac{(\ell+2)(\ell-1)}{2}\mathcal{P} - \mathcal{T} \,, \\[2mm]
\mathcal{S}^{}_a{}^{:a} & = & -2\frac{r^a}{r}\mathcal{S}^{}_a + \frac{(\ell+2)(\ell-1)}{2r^2}\mathcal{S} \,.
\end{eqnarray}
%

%


\begin{thebibliography}{136}%
\makeatletter
\providecommand \@ifxundefined [1]{%
 \@ifx{#1\undefined}
}%
\providecommand \@ifnum [1]{%
 \ifnum #1\expandafter \@firstoftwo
 \else \expandafter \@secondoftwo
 \fi
}%
\providecommand \@ifx [1]{%
 \ifx #1\expandafter \@firstoftwo
 \else \expandafter \@secondoftwo
 \fi
}%
\providecommand \natexlab [1]{#1}%
\providecommand \enquote  [1]{``#1''}%
\providecommand \bibnamefont  [1]{#1}%
\providecommand \bibfnamefont [1]{#1}%
\providecommand \citenamefont [1]{#1}%
\providecommand \href@noop [0]{\@secondoftwo}%
\providecommand \href [0]{\begingroup \@sanitize@url \@href}%
\providecommand \@href[1]{\@@startlink{#1}\@@href}%
\providecommand \@@href[1]{\endgroup#1\@@endlink}%
\providecommand \@sanitize@url [0]{\catcode `\\12\catcode `\$12\catcode
  `\&12\catcode `\#12\catcode `\^12\catcode `\_12\catcode `\%12\relax}%
\providecommand \@@startlink[1]{}%
\providecommand \@@endlink[0]{}%
\providecommand \url  [0]{\begingroup\@sanitize@url \@url }%
\providecommand \@url [1]{\endgroup\@href {#1}{\urlprefix }}%
\providecommand \urlprefix  [0]{URL }%
\providecommand \Eprint [0]{\href }%
\providecommand \doibase [0]{https://doi.org/}%
\providecommand \selectlanguage [0]{\@gobble}%
\providecommand \bibinfo  [0]{\@secondoftwo}%
\providecommand \bibfield  [0]{\@secondoftwo}%
\providecommand \translation [1]{[#1]}%
\providecommand \BibitemOpen [0]{}%
\providecommand \bibitemStop [0]{}%
\providecommand \bibitemNoStop [0]{.\EOS\space}%
\providecommand \EOS [0]{\spacefactor3000\relax}%
\providecommand \BibitemShut  [1]{\csname bibitem#1\endcsname}%
\let\auto@bib@innerbib\@empty
\bibitem [{\citenamefont {Abbott}\ \emph
  {et~al.}(2016{\natexlab{a}})\citenamefont {Abbott} \emph
  {et~al.}}]{LIGOScientific:2016aoc}%
  \BibitemOpen
  \bibfield  {author} {\bibinfo {author} {\bibfnamefont {B.~P.}\ \bibnamefont
  {Abbott}} \emph {et~al.} (\bibinfo {collaboration} {LIGO Scientific,
  Virgo}),\ }\bibfield  {title} {\bibinfo {title} {{Observation of
  Gravitational Waves from a Binary Black Hole Merger}},\ }\href
  {https://doi.org/10.1103/PhysRevLett.116.061102} {\bibfield  {journal}
  {\bibinfo  {journal} {Phys. Rev. Lett.}\ }\textbf {\bibinfo {volume} {116}},\
  \bibinfo {pages} {061102} (\bibinfo {year} {2016}{\natexlab{a}})},\ \Eprint
  {https://arxiv.org/abs/1602.03837} {arXiv:1602.03837 [gr-qc]} \BibitemShut
  {NoStop}%
\bibitem [{\citenamefont {Abbott}\ \emph {et~al.}(2019)\citenamefont {Abbott}
  \emph {et~al.}}]{LIGOScientific:2018mvr}%
  \BibitemOpen
  \bibfield  {author} {\bibinfo {author} {\bibfnamefont {B.~P.}\ \bibnamefont
  {Abbott}} \emph {et~al.} (\bibinfo {collaboration} {LIGO Scientific,
  Virgo}),\ }\bibfield  {title} {\bibinfo {title} {{GWTC-1: A
  Gravitational-Wave Transient Catalog of Compact Binary Mergers Observed by
  LIGO and Virgo during the First and Second Observing Runs}},\ }\href
  {https://doi.org/10.1103/PhysRevX.9.031040} {\bibfield  {journal} {\bibinfo
  {journal} {Phys. Rev. X}\ }\textbf {\bibinfo {volume} {9}},\ \bibinfo {pages}
  {031040} (\bibinfo {year} {2019})},\ \Eprint
  {https://arxiv.org/abs/1811.12907} {arXiv:1811.12907 [astro-ph.HE]}
  \BibitemShut {NoStop}%
\bibitem [{\citenamefont {Abbott}\ \emph {et~al.}(2024)\citenamefont {Abbott}
  \emph {et~al.}}]{LIGOScientific:2021usb}%
  \BibitemOpen
  \bibfield  {author} {\bibinfo {author} {\bibfnamefont {R.}~\bibnamefont
  {Abbott}} \emph {et~al.} (\bibinfo {collaboration} {LIGO Scientific,
  VIRGO}),\ }\bibfield  {title} {\bibinfo {title} {{GWTC-2.1: Deep extended
  catalog of compact binary coalescences observed by LIGO and Virgo during the
  first half of the third observing run}},\ }\href
  {https://doi.org/10.1103/PhysRevD.109.022001} {\bibfield  {journal} {\bibinfo
   {journal} {Phys. Rev. D}\ }\textbf {\bibinfo {volume} {109}},\ \bibinfo
  {pages} {022001} (\bibinfo {year} {2024})},\ \Eprint
  {https://arxiv.org/abs/2108.01045} {arXiv:2108.01045 [gr-qc]} \BibitemShut
  {NoStop}%
\bibitem [{\citenamefont {Abbott}\ \emph {et~al.}(2023)\citenamefont {Abbott}
  \emph {et~al.}}]{KAGRA:2021vkt}%
  \BibitemOpen
  \bibfield  {author} {\bibinfo {author} {\bibfnamefont {R.}~\bibnamefont
  {Abbott}} \emph {et~al.} (\bibinfo {collaboration} {KAGRA, VIRGO, LIGO
  Scientific}),\ }\bibfield  {title} {\bibinfo {title} {{GWTC-3: Compact Binary
  Coalescences Observed by LIGO and Virgo during the Second Part of the Third
  Observing Run}},\ }\href {https://doi.org/10.1103/PhysRevX.13.041039}
  {\bibfield  {journal} {\bibinfo  {journal} {Phys. Rev. X}\ }\textbf {\bibinfo
  {volume} {13}},\ \bibinfo {pages} {041039} (\bibinfo {year} {2023})},\
  \Eprint {https://arxiv.org/abs/2111.03606} {arXiv:2111.03606 [gr-qc]}
  \BibitemShut {NoStop}%
\bibitem [{\citenamefont {Abbott}\ \emph
  {et~al.}(2016{\natexlab{b}})\citenamefont {Abbott} \emph
  {et~al.}}]{LIGOScientific:2016lio}%
  \BibitemOpen
  \bibfield  {author} {\bibinfo {author} {\bibfnamefont {B.~P.}\ \bibnamefont
  {Abbott}} \emph {et~al.} (\bibinfo {collaboration} {LIGO Scientific,
  Virgo}),\ }\bibfield  {title} {\bibinfo {title} {{Tests of general relativity
  with GW150914}},\ }\href {https://doi.org/10.1103/PhysRevLett.116.221101}
  {\bibfield  {journal} {\bibinfo  {journal} {Phys. Rev. Lett.}\ }\textbf
  {\bibinfo {volume} {116}},\ \bibinfo {pages} {221101} (\bibinfo {year}
  {2016}{\natexlab{b}})},\ \bibinfo {note} {[Erratum: Phys.Rev.Lett. 121,
  129902 (2018)]},\ \Eprint {https://arxiv.org/abs/1602.03841}
  {arXiv:1602.03841 [gr-qc]} \BibitemShut {NoStop}%
\bibitem [{\citenamefont {Abbott}\ \emph
  {et~al.}(2016{\natexlab{c}})\citenamefont {Abbott} \emph
  {et~al.}}]{LIGOScientific:2016vlm}%
  \BibitemOpen
  \bibfield  {author} {\bibinfo {author} {\bibfnamefont {B.~P.}\ \bibnamefont
  {Abbott}} \emph {et~al.} (\bibinfo {collaboration} {LIGO Scientific,
  Virgo}),\ }\bibfield  {title} {\bibinfo {title} {{Properties of the Binary
  Black Hole Merger GW150914}},\ }\href
  {https://doi.org/10.1103/PhysRevLett.116.241102} {\bibfield  {journal}
  {\bibinfo  {journal} {Phys. Rev. Lett.}\ }\textbf {\bibinfo {volume} {116}},\
  \bibinfo {pages} {241102} (\bibinfo {year} {2016}{\natexlab{c}})},\ \Eprint
  {https://arxiv.org/abs/1602.03840} {arXiv:1602.03840 [gr-qc]} \BibitemShut
  {NoStop}%
\bibitem [{\citenamefont {Sathyaprakash}\ \emph {et~al.}(2012)\citenamefont
  {Sathyaprakash} \emph {et~al.}}]{Sathyaprakash:2012jk}%
  \BibitemOpen
  \bibfield  {author} {\bibinfo {author} {\bibfnamefont {B.}~\bibnamefont
  {Sathyaprakash}} \emph {et~al.},\ }\bibfield  {title} {\bibinfo {title}
  {{Scientific Objectives of Einstein Telescope}},\ }\href
  {https://doi.org/10.1088/0264-9381/29/12/124013} {\bibfield  {journal}
  {\bibinfo  {journal} {Class. Quant. Grav.}\ }\textbf {\bibinfo {volume}
  {29}},\ \bibinfo {pages} {124013} (\bibinfo {year} {2012})},\ \bibinfo {note}
  {[Erratum: Class.Quant.Grav. 30, 079501 (2013)]},\ \Eprint
  {https://arxiv.org/abs/1206.0331} {arXiv:1206.0331 [gr-qc]} \BibitemShut
  {NoStop}%
\bibitem [{\citenamefont {Evans}\ \emph {et~al.}(2021)\citenamefont {Evans}
  \emph {et~al.}}]{Evans:2021gyd}%
  \BibitemOpen
  \bibfield  {author} {\bibinfo {author} {\bibfnamefont {M.}~\bibnamefont
  {Evans}} \emph {et~al.},\ }\bibfield  {title} {\bibinfo {title} {{A Horizon
  Study for Cosmic Explorer: Science, Observatories, and Community}},\
  }\href@noop {} {\  (\bibinfo {year} {2021})},\ \Eprint
  {https://arxiv.org/abs/2109.09882} {arXiv:2109.09882 [astro-ph.IM]}
  \BibitemShut {NoStop}%
\bibitem [{\citenamefont {Amaro-Seoane}\ \emph {et~al.}(2017)\citenamefont
  {Amaro-Seoane} \emph {et~al.}}]{LISA:2017pwj}%
  \BibitemOpen
  \bibfield  {author} {\bibinfo {author} {\bibfnamefont {P.}~\bibnamefont
  {Amaro-Seoane}} \emph {et~al.} (\bibinfo {collaboration} {LISA}),\ }\bibfield
   {title} {\bibinfo {title} {{Laser Interferometer Space Antenna}},\
  }\href@noop {} {\  (\bibinfo {year} {2017})},\ \Eprint
  {https://arxiv.org/abs/1702.00786} {arXiv:1702.00786 [astro-ph.IM]}
  \BibitemShut {NoStop}%
\bibitem [{\citenamefont {Barausse}\ \emph {et~al.}(2020)\citenamefont
  {Barausse} \emph {et~al.}}]{Barausse:2020rsu}%
  \BibitemOpen
  \bibfield  {author} {\bibinfo {author} {\bibfnamefont {E.}~\bibnamefont
  {Barausse}} \emph {et~al.},\ }\bibfield  {title} {\bibinfo {title}
  {{Prospects for Fundamental Physics with LISA}},\ }\href
  {https://doi.org/10.1007/s10714-020-02691-1} {\bibfield  {journal} {\bibinfo
  {journal} {Gen. Rel. Grav.}\ }\textbf {\bibinfo {volume} {52}},\ \bibinfo
  {pages} {81} (\bibinfo {year} {2020})},\ \Eprint
  {https://arxiv.org/abs/2001.09793} {arXiv:2001.09793 [gr-qc]} \BibitemShut
  {NoStop}%
\bibitem [{\citenamefont {Arun}\ \emph {et~al.}(2022)\citenamefont {Arun} \emph
  {et~al.}}]{LISA:2022kgy}%
  \BibitemOpen
  \bibfield  {author} {\bibinfo {author} {\bibfnamefont {K.~G.}\ \bibnamefont
  {Arun}} \emph {et~al.} (\bibinfo {collaboration} {LISA}),\ }\bibfield
  {title} {\bibinfo {title} {{New horizons for fundamental physics with
  LISA}},\ }\href {https://doi.org/10.1007/s41114-022-00036-9} {\bibfield
  {journal} {\bibinfo  {journal} {Living Rev. Rel.}\ }\textbf {\bibinfo
  {volume} {25}},\ \bibinfo {pages} {4} (\bibinfo {year} {2022})},\ \Eprint
  {https://arxiv.org/abs/2205.01597} {arXiv:2205.01597 [gr-qc]} \BibitemShut
  {NoStop}%
\bibitem [{\citenamefont {Colpi}\ \emph {et~al.}(2024)\citenamefont {Colpi}
  \emph {et~al.}}]{Colpi:2024xhw}%
  \BibitemOpen
  \bibfield  {author} {\bibinfo {author} {\bibfnamefont {M.}~\bibnamefont
  {Colpi}} \emph {et~al.},\ }\bibfield  {title} {\bibinfo {title} {{LISA
  Definition Study Report}},\ }\href@noop {} {\  (\bibinfo {year} {2024})},\
  \Eprint {https://arxiv.org/abs/2402.07571} {arXiv:2402.07571 [astro-ph.CO]}
  \BibitemShut {NoStop}%
\bibitem [{\citenamefont {Afshordi}\ \emph {et~al.}(2023)\citenamefont
  {Afshordi} \emph {et~al.}}]{LISAConsortiumWaveformWorkingGroup:2023arg}%
  \BibitemOpen
  \bibfield  {author} {\bibinfo {author} {\bibfnamefont {N.}~\bibnamefont
  {Afshordi}} \emph {et~al.} (\bibinfo {collaboration} {LISA Consortium
  Waveform Working Group}),\ }\bibfield  {title} {\bibinfo {title} {{Waveform
  Modelling for the Laser Interferometer Space Antenna}},\ }\href@noop {} {\
  (\bibinfo {year} {2023})},\ \Eprint {https://arxiv.org/abs/2311.01300}
  {arXiv:2311.01300 [gr-qc]} \BibitemShut {NoStop}%
\bibitem [{\citenamefont {Misner}\ \emph {et~al.}(1973)\citenamefont {Misner},
  \citenamefont {Thorne},\ and\ \citenamefont {Wheeler}}]{Misner:1973cw}%
  \BibitemOpen
  \bibfield  {author} {\bibinfo {author} {\bibfnamefont {C.~W.}\ \bibnamefont
  {Misner}}, \bibinfo {author} {\bibfnamefont {K.}~\bibnamefont {Thorne}},\
  and\ \bibinfo {author} {\bibfnamefont {J.~A.}\ \bibnamefont {Wheeler}},\
  }\href@noop {} {\emph {\bibinfo {title} {{Gravitation}}}}\ (\bibinfo
  {publisher} {W. H. Freeman \& Co.},\ \bibinfo {address} {San Francisco},\
  \bibinfo {year} {1973})\BibitemShut {NoStop}%
\bibitem [{\citenamefont {Gerlach}\ and\ \citenamefont
  {Sengupta}(1979)}]{Gerlach:1979rw}%
  \BibitemOpen
  \bibfield  {author} {\bibinfo {author} {\bibfnamefont {U.~H.}\ \bibnamefont
  {Gerlach}}\ and\ \bibinfo {author} {\bibfnamefont {U.~K.}\ \bibnamefont
  {Sengupta}},\ }\bibfield  {title} {\bibinfo {title} {{Gauge invariant
  perturbations on most general spherically symmetric space-times}},\ }\href
  {https://doi.org/10.1103/PhysRevD.19.2268} {\bibfield  {journal} {\bibinfo
  {journal} {Phys. Rev. D}\ }\textbf {\bibinfo {volume} {19}},\ \bibinfo
  {pages} {2268} (\bibinfo {year} {1979})}\BibitemShut {NoStop}%
\bibitem [{\citenamefont {Gerlach}\ and\ \citenamefont
  {Sengupta}(1980)}]{Gerlach:1980tx}%
  \BibitemOpen
  \bibfield  {author} {\bibinfo {author} {\bibfnamefont {U.~H.}\ \bibnamefont
  {Gerlach}}\ and\ \bibinfo {author} {\bibfnamefont {U.~K.}\ \bibnamefont
  {Sengupta}},\ }\bibfield  {title} {\bibinfo {title} {{Gauge invariant coupled
  gravitational, acoustical, and electromagnetic modes on most general
  spherical space-times}},\ }\href {https://doi.org/10.1103/PhysRevD.22.1300}
  {\bibfield  {journal} {\bibinfo  {journal} {Phys. Rev. D}\ }\textbf {\bibinfo
  {volume} {22}},\ \bibinfo {pages} {1300} (\bibinfo {year}
  {1980})}\BibitemShut {NoStop}%
\bibitem [{\citenamefont {Sarbach}\ and\ \citenamefont
  {Tiglio}(2001)}]{Sarbach:2001qq}%
  \BibitemOpen
  \bibfield  {author} {\bibinfo {author} {\bibfnamefont {O.}~\bibnamefont
  {Sarbach}}\ and\ \bibinfo {author} {\bibfnamefont {M.}~\bibnamefont
  {Tiglio}},\ }\bibfield  {title} {\bibinfo {title} {{Gauge invariant
  perturbations of Schwarzschild black holes in horizon-penetrating
  coordinates}},\ }\href@noop {} {\bibfield  {journal} {\bibinfo  {journal}
  {Phys. Rev. D}\ }\textbf {\bibinfo {volume} {64}},\ \bibinfo {pages} {084016}
  (\bibinfo {year} {2001})},\ \Eprint {https://arxiv.org/abs/gr-qc/0104061}
  {gr-qc/0104061} \BibitemShut {NoStop}%
\bibitem [{\citenamefont {Clarkson}\ and\ \citenamefont
  {Barrett}(2003)}]{Clarkson:2002jz}%
  \BibitemOpen
  \bibfield  {author} {\bibinfo {author} {\bibfnamefont {C.~A.}\ \bibnamefont
  {Clarkson}}\ and\ \bibinfo {author} {\bibfnamefont {R.~K.}\ \bibnamefont
  {Barrett}},\ }\bibfield  {title} {\bibinfo {title} {{Covariant perturbations
  of Schwarzschild black holes}},\ }\href
  {https://doi.org/10.1088/0264-9381/20/18/301} {\bibfield  {journal} {\bibinfo
   {journal} {Class. Quant. Grav.}\ }\textbf {\bibinfo {volume} {20}},\
  \bibinfo {pages} {3855} (\bibinfo {year} {2003})},\ \Eprint
  {https://arxiv.org/abs/gr-qc/0209051} {arXiv:gr-qc/0209051} \BibitemShut
  {NoStop}%
\bibitem [{\citenamefont {Mukohyama}(2000)}]{Mukohyama:2000ui}%
  \BibitemOpen
  \bibfield  {author} {\bibinfo {author} {\bibfnamefont {S.}~\bibnamefont
  {Mukohyama}},\ }\bibfield  {title} {\bibinfo {title} {{Gauge invariant
  gravitational perturbations of maximally symmetric space-times}},\ }\href
  {https://doi.org/10.1103/PhysRevD.62.084015} {\bibfield  {journal} {\bibinfo
  {journal} {Phys. Rev. D}\ }\textbf {\bibinfo {volume} {62}},\ \bibinfo
  {pages} {084015} (\bibinfo {year} {2000})},\ \Eprint
  {https://arxiv.org/abs/hep-th/0004067} {arXiv:hep-th/0004067} \BibitemShut
  {NoStop}%
\bibitem [{\citenamefont {Kodama}\ and\ \citenamefont
  {Ishibashi}(2003)}]{Kodama:2003jz}%
  \BibitemOpen
  \bibfield  {author} {\bibinfo {author} {\bibfnamefont {H.}~\bibnamefont
  {Kodama}}\ and\ \bibinfo {author} {\bibfnamefont {A.}~\bibnamefont
  {Ishibashi}},\ }\bibfield  {title} {\bibinfo {title} {{A Master equation for
  gravitational perturbations of maximally symmetric black holes in higher
  dimensions}},\ }\href {https://doi.org/10.1143/PTP.110.701} {\bibfield
  {journal} {\bibinfo  {journal} {Prog. Theor. Phys.}\ }\textbf {\bibinfo
  {volume} {110}},\ \bibinfo {pages} {701} (\bibinfo {year} {2003})},\ \Eprint
  {https://arxiv.org/abs/hep-th/0305147} {arXiv:hep-th/0305147 [hep-th]}
  \BibitemShut {NoStop}%
\bibitem [{\citenamefont {{Nollert}}(1999)}]{Nollert:1999re}%
  \BibitemOpen
  \bibfield  {author} {\bibinfo {author} {\bibfnamefont {H.-P.}\ \bibnamefont
  {{Nollert}}},\ }\bibfield  {title} {\bibinfo {title} {{Quasinormal modes: the
  characteristic `sound' of black holes and neutron stars}},\ }\href@noop {}
  {\bibfield  {journal} {\bibinfo  {journal} {Class. Quantum Grav.}\ }\textbf
  {\bibinfo {volume} {16}},\ \bibinfo {pages} {159} (\bibinfo {year}
  {1999})}\BibitemShut {NoStop}%
\bibitem [{\citenamefont {Kokkotas}\ and\ \citenamefont
  {Schmidt}(1999)}]{Kokkotas:1999bd}%
  \BibitemOpen
  \bibfield  {author} {\bibinfo {author} {\bibfnamefont {K.~D.}\ \bibnamefont
  {Kokkotas}}\ and\ \bibinfo {author} {\bibfnamefont {B.~G.}\ \bibnamefont
  {Schmidt}},\ }\bibfield  {title} {\bibinfo {title} {{Quasinormal modes of
  stars and black holes}},\ }\href {https://doi.org/10.12942/lrr-1999-2}
  {\bibfield  {journal} {\bibinfo  {journal} {Living Rev. Rel.}\ }\textbf
  {\bibinfo {volume} {2}},\ \bibinfo {pages} {2} (\bibinfo {year} {1999})},\
  \Eprint {https://arxiv.org/abs/gr-qc/9909058} {arXiv:gr-qc/9909058}
  \BibitemShut {NoStop}%
\bibitem [{\citenamefont {Andersson}\ and\ \citenamefont
  {Kokkotas}(2001)}]{Andersson:2000mf}%
  \BibitemOpen
  \bibfield  {author} {\bibinfo {author} {\bibfnamefont {N.}~\bibnamefont
  {Andersson}}\ and\ \bibinfo {author} {\bibfnamefont {K.~D.}\ \bibnamefont
  {Kokkotas}},\ }\bibfield  {title} {\bibinfo {title} {{The r-mode instability
  in rotating neutron stars}},\ }\href
  {https://doi.org/10.1142/S0218271801001062} {\bibfield  {journal} {\bibinfo
  {journal} {Int. J. Mod. Phys. D}\ }\textbf {\bibinfo {volume} {10}},\
  \bibinfo {pages} {381} (\bibinfo {year} {2001})},\ \Eprint
  {https://arxiv.org/abs/gr-qc/0010102} {arXiv:gr-qc/0010102} \BibitemShut
  {NoStop}%
\bibitem [{\citenamefont {Kokkotas}\ and\ \citenamefont
  {Schutz}(1992)}]{Kokkotas:2003mh}%
  \BibitemOpen
  \bibfield  {author} {\bibinfo {author} {\bibfnamefont {K.~D.}\ \bibnamefont
  {Kokkotas}}\ and\ \bibinfo {author} {\bibfnamefont {B.~F.}\ \bibnamefont
  {Schutz}},\ }\bibfield  {title} {\bibinfo {title} {{W-modes: A New family of
  normal modes of pulsating relativistic stars}},\ }\href@noop {} {\bibfield
  {journal} {\bibinfo  {journal} {Mon. Not. Roy. Astron. Soc.}\ }\textbf
  {\bibinfo {volume} {255}},\ \bibinfo {pages} {119} (\bibinfo {year}
  {1992})}\BibitemShut {NoStop}%
\bibitem [{\citenamefont {Sasaki}\ and\ \citenamefont
  {Tagoshi}(2003)}]{Sasaki:2003xr}%
  \BibitemOpen
  \bibfield  {author} {\bibinfo {author} {\bibfnamefont {M.}~\bibnamefont
  {Sasaki}}\ and\ \bibinfo {author} {\bibfnamefont {H.}~\bibnamefont
  {Tagoshi}},\ }\bibfield  {title} {\bibinfo {title} {{Analytic black hole
  perturbation approach to gravitational radiation}},\ }\href
  {https://doi.org/10.12942/lrr-2003-6} {\bibfield  {journal} {\bibinfo
  {journal} {Living Rev. Rel.}\ }\textbf {\bibinfo {volume} {6}},\ \bibinfo
  {pages} {6} (\bibinfo {year} {2003})},\ \Eprint
  {https://arxiv.org/abs/gr-qc/0306120} {arXiv:gr-qc/0306120} \BibitemShut
  {NoStop}%
\bibitem [{\citenamefont {Ferrari}\ and\ \citenamefont
  {Gualtieri}(2008)}]{Ferrari:2007dd}%
  \BibitemOpen
  \bibfield  {author} {\bibinfo {author} {\bibfnamefont {V.}~\bibnamefont
  {Ferrari}}\ and\ \bibinfo {author} {\bibfnamefont {L.}~\bibnamefont
  {Gualtieri}},\ }\bibfield  {title} {\bibinfo {title} {{Quasi-Normal Modes and
  Gravitational Wave Astronomy}},\ }\href
  {https://doi.org/10.1007/s10714-007-0585-1} {\bibfield  {journal} {\bibinfo
  {journal} {Gen. Rel. Grav.}\ }\textbf {\bibinfo {volume} {40}},\ \bibinfo
  {pages} {945} (\bibinfo {year} {2008})},\ \Eprint
  {https://arxiv.org/abs/0709.0657} {arXiv:0709.0657 [gr-qc]} \BibitemShut
  {NoStop}%
\bibitem [{\citenamefont {Berti}\ \emph {et~al.}(2018)\citenamefont {Berti},
  \citenamefont {Yagi}, \citenamefont {Yang},\ and\ \citenamefont
  {Yunes}}]{Berti:2018vdi}%
  \BibitemOpen
  \bibfield  {author} {\bibinfo {author} {\bibfnamefont {E.}~\bibnamefont
  {Berti}}, \bibinfo {author} {\bibfnamefont {K.}~\bibnamefont {Yagi}},
  \bibinfo {author} {\bibfnamefont {H.}~\bibnamefont {Yang}},\ and\ \bibinfo
  {author} {\bibfnamefont {N.}~\bibnamefont {Yunes}},\ }\bibfield  {title}
  {\bibinfo {title} {{Extreme Gravity Tests with Gravitational Waves from
  Compact Binary Coalescences: (II) Ringdown}},\ }\href
  {https://doi.org/10.1007/s10714-018-2372-6} {\bibfield  {journal} {\bibinfo
  {journal} {Gen. Rel. Grav.}\ }\textbf {\bibinfo {volume} {50}},\ \bibinfo
  {pages} {49} (\bibinfo {year} {2018})},\ \Eprint
  {https://arxiv.org/abs/1801.03587} {arXiv:1801.03587 [gr-qc]} \BibitemShut
  {NoStop}%
\bibitem [{\citenamefont {Cardoso}\ \emph {et~al.}(2012)\citenamefont
  {Cardoso}, \citenamefont {Gualtieri}, \citenamefont {Herdeiro}, \citenamefont
  {Sperhake}, \citenamefont {Chesler} \emph {et~al.}}]{Cardoso:2012qm}%
  \BibitemOpen
  \bibfield  {author} {\bibinfo {author} {\bibfnamefont {V.}~\bibnamefont
  {Cardoso}}, \bibinfo {author} {\bibfnamefont {L.}~\bibnamefont {Gualtieri}},
  \bibinfo {author} {\bibfnamefont {C.}~\bibnamefont {Herdeiro}}, \bibinfo
  {author} {\bibfnamefont {U.}~\bibnamefont {Sperhake}}, \bibinfo {author}
  {\bibfnamefont {P.~M.}\ \bibnamefont {Chesler}}, \emph {et~al.},\ }\bibfield
  {title} {\bibinfo {title} {{NR/HEP: roadmap for the future}},\ }\href
  {https://doi.org/10.1088/0264-9381/29/24/244001} {\bibfield  {journal}
  {\bibinfo  {journal} {Class. Quant. Grav.}\ }\textbf {\bibinfo {volume}
  {29}},\ \bibinfo {pages} {244001} (\bibinfo {year} {2012})},\ \Eprint
  {https://arxiv.org/abs/1201.5118} {arXiv:1201.5118 [hep-th]} \BibitemShut
  {NoStop}%
\bibitem [{\citenamefont {Brito}\ \emph {et~al.}(2015)\citenamefont {Brito},
  \citenamefont {Cardoso},\ and\ \citenamefont {Pani}}]{Brito:2015oca}%
  \BibitemOpen
  \bibfield  {author} {\bibinfo {author} {\bibfnamefont {R.}~\bibnamefont
  {Brito}}, \bibinfo {author} {\bibfnamefont {V.}~\bibnamefont {Cardoso}},\
  and\ \bibinfo {author} {\bibfnamefont {P.}~\bibnamefont {Pani}},\ }\href
  {https://doi.org/10.1007/978-3-319-19000-6} {\emph {\bibinfo {title}
  {{Superradiance}: {New Frontiers in Black Hole Physics}}}},\ Vol.\ \bibinfo
  {volume} {906}\ (\bibinfo  {publisher} {Springer},\ \bibinfo {year} {2015})\
  \Eprint {https://arxiv.org/abs/1501.06570} {arXiv:1501.06570 [gr-qc]}
  \BibitemShut {NoStop}%
\bibitem [{\citenamefont {Barack}\ \emph {et~al.}(2019)\citenamefont {Barack}
  \emph {et~al.}}]{Barack:2018yly}%
  \BibitemOpen
  \bibfield  {author} {\bibinfo {author} {\bibfnamefont {L.}~\bibnamefont
  {Barack}} \emph {et~al.},\ }\bibfield  {title} {\bibinfo {title} {{Black
  holes, gravitational waves and fundamental physics: a roadmap}},\ }\href
  {https://doi.org/10.1088/1361-6382/ab0587} {\bibfield  {journal} {\bibinfo
  {journal} {Class. Quant. Grav.}\ }\textbf {\bibinfo {volume} {36}},\ \bibinfo
  {pages} {143001} (\bibinfo {year} {2019})},\ \Eprint
  {https://arxiv.org/abs/1806.05195} {arXiv:1806.05195 [gr-qc]} \BibitemShut
  {NoStop}%
\bibitem [{\citenamefont {Amaro-Seoane}(2018)}]{Amaro-Seoane:2012lgq}%
  \BibitemOpen
  \bibfield  {author} {\bibinfo {author} {\bibfnamefont {P.}~\bibnamefont
  {Amaro-Seoane}},\ }\bibfield  {title} {\bibinfo {title} {{Relativistic
  dynamics and extreme mass ratio inspirals}},\ }\href
  {https://doi.org/10.1007/s41114-018-0013-8} {\bibfield  {journal} {\bibinfo
  {journal} {Living Rev. Rel.}\ }\textbf {\bibinfo {volume} {21}},\ \bibinfo
  {pages} {4} (\bibinfo {year} {2018})},\ \Eprint
  {https://arxiv.org/abs/1205.5240} {arXiv:1205.5240 [astro-ph.CO]}
  \BibitemShut {NoStop}%
\bibitem [{\citenamefont {Babak}\ \emph {et~al.}(2017)\citenamefont {Babak},
  \citenamefont {Gair}, \citenamefont {Sesana}, \citenamefont {Barausse},
  \citenamefont {Sopuerta}, \citenamefont {Berry}, \citenamefont {Berti},
  \citenamefont {Amaro-Seoane}, \citenamefont {Petiteau},\ and\ \citenamefont
  {Klein}}]{Babak:2017tow}%
  \BibitemOpen
  \bibfield  {author} {\bibinfo {author} {\bibfnamefont {S.}~\bibnamefont
  {Babak}}, \bibinfo {author} {\bibfnamefont {J.}~\bibnamefont {Gair}},
  \bibinfo {author} {\bibfnamefont {A.}~\bibnamefont {Sesana}}, \bibinfo
  {author} {\bibfnamefont {E.}~\bibnamefont {Barausse}}, \bibinfo {author}
  {\bibfnamefont {C.~F.}\ \bibnamefont {Sopuerta}}, \bibinfo {author}
  {\bibfnamefont {C.~P.~L.}\ \bibnamefont {Berry}}, \bibinfo {author}
  {\bibfnamefont {E.}~\bibnamefont {Berti}}, \bibinfo {author} {\bibfnamefont
  {P.}~\bibnamefont {Amaro-Seoane}}, \bibinfo {author} {\bibfnamefont
  {A.}~\bibnamefont {Petiteau}},\ and\ \bibinfo {author} {\bibfnamefont
  {A.}~\bibnamefont {Klein}},\ }\bibfield  {title} {\bibinfo {title} {{Science
  with the space-based interferometer LISA. V: Extreme mass-ratio inspirals}},\
  }\href {https://doi.org/10.1103/PhysRevD.95.103012} {\bibfield  {journal}
  {\bibinfo  {journal} {Phys. Rev. D}\ }\textbf {\bibinfo {volume} {95}},\
  \bibinfo {pages} {103012} (\bibinfo {year} {2017})},\ \Eprint
  {https://arxiv.org/abs/1703.09722} {arXiv:1703.09722 [gr-qc]} \BibitemShut
  {NoStop}%
\bibitem [{\citenamefont {Berry}\ \emph {et~al.}(2019)\citenamefont {Berry},
  \citenamefont {Hughes}, \citenamefont {Sopuerta}, \citenamefont {Chua},
  \citenamefont {Heffernan}, \citenamefont {Holley-Bockelmann}, \citenamefont
  {Mihaylov}, \citenamefont {Miller},\ and\ \citenamefont
  {Sesana}}]{Berry:2019wgg}%
  \BibitemOpen
  \bibfield  {author} {\bibinfo {author} {\bibfnamefont {C.~P.~L.}\
  \bibnamefont {Berry}}, \bibinfo {author} {\bibfnamefont {S.~A.}\ \bibnamefont
  {Hughes}}, \bibinfo {author} {\bibfnamefont {C.~F.}\ \bibnamefont
  {Sopuerta}}, \bibinfo {author} {\bibfnamefont {A.~J.~K.}\ \bibnamefont
  {Chua}}, \bibinfo {author} {\bibfnamefont {A.}~\bibnamefont {Heffernan}},
  \bibinfo {author} {\bibfnamefont {K.}~\bibnamefont {Holley-Bockelmann}},
  \bibinfo {author} {\bibfnamefont {D.~P.}\ \bibnamefont {Mihaylov}}, \bibinfo
  {author} {\bibfnamefont {M.~C.}\ \bibnamefont {Miller}},\ and\ \bibinfo
  {author} {\bibfnamefont {A.}~\bibnamefont {Sesana}},\ }\bibfield  {title}
  {\bibinfo {title} {{The unique potential of extreme mass-ratio inspirals for
  gravitational-wave astronomy}},\ }\href@noop {} {\  (\bibinfo {year}
  {2019})},\ \Eprint {https://arxiv.org/abs/1903.03686} {arXiv:1903.03686
  [astro-ph.HE]} \BibitemShut {NoStop}%
\bibitem [{\citenamefont {Amaro-Seoane}\ \emph {et~al.}(2022)\citenamefont
  {Amaro-Seoane} \emph {et~al.}}]{Amaro-Seoane:2022rxf}%
  \BibitemOpen
  \bibfield  {author} {\bibinfo {author} {\bibfnamefont {P.}~\bibnamefont
  {Amaro-Seoane}} \emph {et~al.},\ }\bibfield  {title} {\bibinfo {title}
  {{Astrophysics with the Laser Interferometer Space Antenna}},\ }\href@noop {}
  {\  (\bibinfo {year} {2022})},\ \Eprint {https://arxiv.org/abs/2203.06016}
  {arXiv:2203.06016 [gr-qc]} \BibitemShut {NoStop}%
\bibitem [{\citenamefont {C\'ardenas-Avenda\~no}\ and\ \citenamefont
  {Sopuerta}(2024)}]{Cardenas-Avendano:2024mqp}%
  \BibitemOpen
  \bibfield  {author} {\bibinfo {author} {\bibfnamefont {A.}~\bibnamefont
  {C\'ardenas-Avenda\~no}}\ and\ \bibinfo {author} {\bibfnamefont {C.~F.}\
  \bibnamefont {Sopuerta}},\ }\bibfield  {title} {\bibinfo {title} {{Testing
  gravity with Extreme-Mass-Ratio Inspirals}},\ }\href@noop {} {\  (\bibinfo
  {year} {2024})},\ \Eprint {https://arxiv.org/abs/2401.08085}
  {arXiv:2401.08085 [gr-qc]} \BibitemShut {NoStop}%
\bibitem [{\citenamefont {Mino}\ \emph {et~al.}(1997)\citenamefont {Mino},
  \citenamefont {Sasaki}, \citenamefont {Shibata}, \citenamefont {Tagoshi},\
  and\ \citenamefont {Tanaka}}]{Mino:1997bx}%
  \BibitemOpen
  \bibfield  {author} {\bibinfo {author} {\bibfnamefont {Y.}~\bibnamefont
  {Mino}}, \bibinfo {author} {\bibfnamefont {M.}~\bibnamefont {Sasaki}},
  \bibinfo {author} {\bibfnamefont {M.}~\bibnamefont {Shibata}}, \bibinfo
  {author} {\bibfnamefont {H.}~\bibnamefont {Tagoshi}},\ and\ \bibinfo {author}
  {\bibfnamefont {T.}~\bibnamefont {Tanaka}},\ }\bibfield  {title} {\bibinfo
  {title} {{Black hole perturbation: Chapter 1}},\ }\href
  {https://doi.org/10.1143/PTPS.128.1} {\bibfield  {journal} {\bibinfo
  {journal} {Prog. Theor. Phys. Suppl.}\ }\textbf {\bibinfo {volume} {128}},\
  \bibinfo {pages} {1} (\bibinfo {year} {1997})},\ \Eprint
  {https://arxiv.org/abs/gr-qc/9712057} {arXiv:gr-qc/9712057} \BibitemShut
  {NoStop}%
\bibitem [{\citenamefont {Poisson}\ \emph {et~al.}(2011)\citenamefont
  {Poisson}, \citenamefont {Pound},\ and\ \citenamefont
  {Vega}}]{Poisson:2011nh}%
  \BibitemOpen
  \bibfield  {author} {\bibinfo {author} {\bibfnamefont {E.}~\bibnamefont
  {Poisson}}, \bibinfo {author} {\bibfnamefont {A.}~\bibnamefont {Pound}},\
  and\ \bibinfo {author} {\bibfnamefont {I.}~\bibnamefont {Vega}},\ }\bibfield
  {title} {\bibinfo {title} {{The Motion of point particles in curved
  spacetime}},\ }\href {https://doi.org/10.12942/lrr-2011-7} {\bibfield
  {journal} {\bibinfo  {journal} {Living Rev. Rel.}\ }\textbf {\bibinfo
  {volume} {14}},\ \bibinfo {pages} {7} (\bibinfo {year} {2011})},\ \Eprint
  {https://arxiv.org/abs/1102.0529} {arXiv:1102.0529 [gr-qc]} \BibitemShut
  {NoStop}%
\bibitem [{\citenamefont {Barack}(2009)}]{Barack:2009ux}%
  \BibitemOpen
  \bibfield  {author} {\bibinfo {author} {\bibfnamefont {L.}~\bibnamefont
  {Barack}},\ }\bibfield  {title} {\bibinfo {title} {{Gravitational self force
  in extreme mass-ratio inspirals}},\ }\href
  {https://doi.org/10.1088/0264-9381/26/21/213001} {\bibfield  {journal}
  {\bibinfo  {journal} {Class. Quant. Grav.}\ }\textbf {\bibinfo {volume}
  {26}},\ \bibinfo {pages} {213001} (\bibinfo {year} {2009})},\ \Eprint
  {https://arxiv.org/abs/0908.1664} {arXiv:0908.1664 [gr-qc]} \BibitemShut
  {NoStop}%
\bibitem [{\citenamefont {Barack}\ and\ \citenamefont
  {Pound}(2019)}]{Barack:2018yvs}%
  \BibitemOpen
  \bibfield  {author} {\bibinfo {author} {\bibfnamefont {L.}~\bibnamefont
  {Barack}}\ and\ \bibinfo {author} {\bibfnamefont {A.}~\bibnamefont {Pound}},\
  }\bibfield  {title} {\bibinfo {title} {{Self-force and radiation reaction in
  general relativity}},\ }\href {https://doi.org/10.1088/1361-6633/aae552}
  {\bibfield  {journal} {\bibinfo  {journal} {Rept. Prog. Phys.}\ }\textbf
  {\bibinfo {volume} {82}},\ \bibinfo {pages} {016904} (\bibinfo {year}
  {2019})},\ \Eprint {https://arxiv.org/abs/1805.10385} {arXiv:1805.10385
  [gr-qc]} \BibitemShut {NoStop}%
\bibitem [{\citenamefont {Pound}\ and\ \citenamefont
  {Wardell}(2021)}]{Pound:2021qin}%
  \BibitemOpen
  \bibfield  {author} {\bibinfo {author} {\bibfnamefont {A.}~\bibnamefont
  {Pound}}\ and\ \bibinfo {author} {\bibfnamefont {B.}~\bibnamefont
  {Wardell}},\ }\bibfield  {title} {\bibinfo {title} {{Black hole perturbation
  theory and gravitational self-force}}\ }\href
  {https://doi.org/10.1007/978-981-15-4702-7\_38-1}
  {10.1007/978-981-15-4702-7\_38-1} (\bibinfo {year} {2021}),\ \Eprint
  {https://arxiv.org/abs/2101.04592} {arXiv:2101.04592 [gr-qc]} \BibitemShut
  {NoStop}%
\bibitem [{\citenamefont {Regge}\ and\ \citenamefont
  {Wheeler}(1957)}]{Regge:1957td}%
  \BibitemOpen
  \bibfield  {author} {\bibinfo {author} {\bibfnamefont {T.}~\bibnamefont
  {Regge}}\ and\ \bibinfo {author} {\bibfnamefont {J.~A.}\ \bibnamefont
  {Wheeler}},\ }\bibfield  {title} {\bibinfo {title} {{Stability of a
  Schwarzschild singularity}},\ }\href
  {https://doi.org/10.1103/PhysRev.108.1063} {\bibfield  {journal} {\bibinfo
  {journal} {Phys. Rev.}\ }\textbf {\bibinfo {volume} {108}},\ \bibinfo {pages}
  {1063} (\bibinfo {year} {1957})}\BibitemShut {NoStop}%
\bibitem [{\citenamefont {Schwarzschild}(1916)}]{Schwarzschild:1916uq}%
  \BibitemOpen
  \bibfield  {author} {\bibinfo {author} {\bibfnamefont {K.}~\bibnamefont
  {Schwarzschild}},\ }\bibfield  {title} {\bibinfo {title} {{On the
  gravitational field of a mass point according to Einstein's theory}},\
  }\href@noop {} {\bibfield  {journal} {\bibinfo  {journal} {Sitzungsber.
  Preuss. Akad. Wiss. Berlin (Math. Phys.)}\ }\textbf {\bibinfo {volume}
  {1916}},\ \bibinfo {pages} {189} (\bibinfo {year} {1916})},\ \Eprint
  {https://arxiv.org/abs/physics/9905030} {arXiv:physics/9905030 [physics]}
  \BibitemShut {NoStop}%
\bibitem [{\citenamefont {Zerilli}(1970{\natexlab{a}})}]{Zerilli:1970se}%
  \BibitemOpen
  \bibfield  {author} {\bibinfo {author} {\bibfnamefont {F.~J.}\ \bibnamefont
  {Zerilli}},\ }\bibfield  {title} {\bibinfo {title} {{Effective potential for
  even parity Regge-Wheeler gravitational perturbation equations}},\ }\href
  {https://doi.org/10.1103/PhysRevLett.24.737} {\bibfield  {journal} {\bibinfo
  {journal} {Phys. Rev. Lett.}\ }\textbf {\bibinfo {volume} {24}},\ \bibinfo
  {pages} {737} (\bibinfo {year} {1970}{\natexlab{a}})}\BibitemShut {NoStop}%
\bibitem [{\citenamefont {Zerilli}(1970{\natexlab{b}})}]{Zerilli:1970la}%
  \BibitemOpen
  \bibfield  {author} {\bibinfo {author} {\bibfnamefont {F.~J.}\ \bibnamefont
  {Zerilli}},\ }\bibfield  {title} {\bibinfo {title} {{Gravitational Field of a
  Particle Falling in a Schwarzschild Geometry Analyzed in Tensor Harmonics}},\
  }\href@noop {} {\bibfield  {journal} {\bibinfo  {journal} {Phys. Rev. D}\
  }\textbf {\bibinfo {volume} {2}},\ \bibinfo {pages} {2141} (\bibinfo {year}
  {1970}{\natexlab{b}})}\BibitemShut {NoStop}%
\bibitem [{\citenamefont {Moncrief}(1974)}]{Moncrief:1974vm}%
  \BibitemOpen
  \bibfield  {author} {\bibinfo {author} {\bibfnamefont {V.}~\bibnamefont
  {Moncrief}},\ }\bibfield  {title} {\bibinfo {title} {{Gravitational
  perturbations of spherically symmetric systems. I. The exterior problem}},\
  }\href@noop {} {\bibfield  {journal} {\bibinfo  {journal} {Ann. Phys.
  (N.Y.)}\ }\textbf {\bibinfo {volume} {88}},\ \bibinfo {pages} {323} (\bibinfo
  {year} {1974})}\BibitemShut {NoStop}%
\bibitem [{\citenamefont
  {Vishveshwara}(1970{\natexlab{a}})}]{Vishveshwara:1970cc}%
  \BibitemOpen
  \bibfield  {author} {\bibinfo {author} {\bibfnamefont {C.}~\bibnamefont
  {Vishveshwara}},\ }\bibfield  {title} {\bibinfo {title} {{Stability of the
  Schwarzschild metric}},\ }\href {https://doi.org/10.1103/PhysRevD.1.2870}
  {\bibfield  {journal} {\bibinfo  {journal} {Phys. Rev. D}\ }\textbf {\bibinfo
  {volume} {1}},\ \bibinfo {pages} {2870} (\bibinfo {year}
  {1970}{\natexlab{a}})}\BibitemShut {NoStop}%
\bibitem [{\citenamefont
  {Vishveshwara}(1970{\natexlab{b}})}]{Vishveshwara:1970zz}%
  \BibitemOpen
  \bibfield  {author} {\bibinfo {author} {\bibfnamefont {C.}~\bibnamefont
  {Vishveshwara}},\ }\bibfield  {title} {\bibinfo {title} {{Scattering of
  Gravitational Radiation by a Schwarzschild Black-hole}},\ }\href
  {https://doi.org/10.1038/227936a0} {\bibfield  {journal} {\bibinfo  {journal}
  {Nature}\ }\textbf {\bibinfo {volume} {227}},\ \bibinfo {pages} {936}
  (\bibinfo {year} {1970}{\natexlab{b}})}\BibitemShut {NoStop}%
\bibitem [{\citenamefont {Kerr}(1963)}]{Kerr:1963ud}%
  \BibitemOpen
  \bibfield  {author} {\bibinfo {author} {\bibfnamefont {R.~P.}\ \bibnamefont
  {Kerr}},\ }\bibfield  {title} {\bibinfo {title} {{Gravitational field of a
  spinning mass as an example of algebraically special metrics}},\ }\href
  {https://doi.org/10.1103/PhysRevLett.11.237} {\bibfield  {journal} {\bibinfo
  {journal} {Phys. Rev. Lett.}\ }\textbf {\bibinfo {volume} {11}},\ \bibinfo
  {pages} {237} (\bibinfo {year} {1963})}\BibitemShut {NoStop}%
\bibitem [{\citenamefont {{Teukolsky}}(1972)}]{Teukolsky:1972le}%
  \BibitemOpen
  \bibfield  {author} {\bibinfo {author} {\bibfnamefont {S.~A.}\ \bibnamefont
  {{Teukolsky}}},\ }\bibfield  {title} {\bibinfo {title} {{Rotating Black
  Holes: Separable Wave Equations for Gravitational and Electromagnetic
  Perturbations}},\ }\href {https://doi.org/10.1103/PhysRevLett.29.1114}
  {\bibfield  {journal} {\bibinfo  {journal} {Phys. Rev. Lett.}\ }\textbf
  {\bibinfo {volume} {29}},\ \bibinfo {pages} {1114} (\bibinfo {year}
  {1972})}\BibitemShut {NoStop}%
\bibitem [{\citenamefont {Teukolsky}(1973)}]{Teukolsky:1973ha}%
  \BibitemOpen
  \bibfield  {author} {\bibinfo {author} {\bibfnamefont {S.~A.}\ \bibnamefont
  {Teukolsky}},\ }\bibfield  {title} {\bibinfo {title} {{Perturbations of a
  rotating black hole. 1. Fundamental equations for gravitational
  electromagnetic and neutrino field perturbations}},\ }\href
  {https://doi.org/10.1086/152444} {\bibfield  {journal} {\bibinfo  {journal}
  {Astrophys. J.}\ }\textbf {\bibinfo {volume} {185}},\ \bibinfo {pages} {635}
  (\bibinfo {year} {1973})}\BibitemShut {NoStop}%
\bibitem [{\citenamefont {Newman}\ and\ \citenamefont
  {Penrose}(1962)}]{Newman:1961qr}%
  \BibitemOpen
  \bibfield  {author} {\bibinfo {author} {\bibfnamefont {E.}~\bibnamefont
  {Newman}}\ and\ \bibinfo {author} {\bibfnamefont {R.}~\bibnamefont
  {Penrose}},\ }\bibfield  {title} {\bibinfo {title} {{An Approach to
  gravitational radiation by a method of spin coefficients}},\ }\href
  {https://doi.org/10.1063/1.1724257} {\bibfield  {journal} {\bibinfo
  {journal} {J. Math. Phys.}\ }\textbf {\bibinfo {volume} {3}},\ \bibinfo
  {pages} {566} (\bibinfo {year} {1962})}\BibitemShut {NoStop}%
\bibitem [{\citenamefont {Stephani}\ \emph {et~al.}(2003)\citenamefont
  {Stephani}, \citenamefont {Kramer}, \citenamefont {MacCallum}, \citenamefont
  {Hoenselaers},\ and\ \citenamefont {Herlt}}]{Stephani:2003tm}%
  \BibitemOpen
  \bibfield  {author} {\bibinfo {author} {\bibfnamefont {H.}~\bibnamefont
  {Stephani}}, \bibinfo {author} {\bibfnamefont {D.}~\bibnamefont {Kramer}},
  \bibinfo {author} {\bibfnamefont {M.}~\bibnamefont {MacCallum}}, \bibinfo
  {author} {\bibfnamefont {C.}~\bibnamefont {Hoenselaers}},\ and\ \bibinfo
  {author} {\bibfnamefont {E.}~\bibnamefont {Herlt}},\ }\href@noop {} {\emph
  {\bibinfo {title} {Exact solutions of Einstein's field equations}}}\
  (\bibinfo  {publisher} {Cambridge University Press},\ \bibinfo {address}
  {Cambridge},\ \bibinfo {year} {2003})\BibitemShut {NoStop}%
\bibitem [{\citenamefont {Spiers}\ \emph {et~al.}(2023)\citenamefont {Spiers},
  \citenamefont {Pound},\ and\ \citenamefont {Wardell}}]{Spiers:2023mor}%
  \BibitemOpen
  \bibfield  {author} {\bibinfo {author} {\bibfnamefont {A.}~\bibnamefont
  {Spiers}}, \bibinfo {author} {\bibfnamefont {A.}~\bibnamefont {Pound}},\ and\
  \bibinfo {author} {\bibfnamefont {B.}~\bibnamefont {Wardell}},\ }\bibfield
  {title} {\bibinfo {title} {{Second-order perturbations of the Schwarzschild
  spacetime: practical, covariant and gauge-invariant formalisms}},\
  }\href@noop {} {\  (\bibinfo {year} {2023})},\ \Eprint
  {https://arxiv.org/abs/2306.17847} {arXiv:2306.17847 [gr-qc]} \BibitemShut
  {NoStop}%
\bibitem [{\citenamefont {Thompson}\ \emph {et~al.}(2019)\citenamefont
  {Thompson}, \citenamefont {Wardell},\ and\ \citenamefont
  {Whiting}}]{Thompson:2018lgb}%
  \BibitemOpen
  \bibfield  {author} {\bibinfo {author} {\bibfnamefont {J.~E.}\ \bibnamefont
  {Thompson}}, \bibinfo {author} {\bibfnamefont {B.}~\bibnamefont {Wardell}},\
  and\ \bibinfo {author} {\bibfnamefont {B.~F.}\ \bibnamefont {Whiting}},\
  }\bibfield  {title} {\bibinfo {title} {{Gravitational Self-Force
  Regularization in the Regge-Wheeler and Easy Gauges}},\ }\href
  {https://doi.org/10.1103/PhysRevD.99.124046} {\bibfield  {journal} {\bibinfo
  {journal} {Phys. Rev. D}\ }\textbf {\bibinfo {volume} {99}},\ \bibinfo
  {pages} {124046} (\bibinfo {year} {2019})},\ \Eprint
  {https://arxiv.org/abs/1811.04432} {arXiv:1811.04432 [gr-qc]} \BibitemShut
  {NoStop}%
\bibitem [{\citenamefont {Hopper}\ and\ \citenamefont
  {Evans}(2010)}]{Hopper:2010uv}%
  \BibitemOpen
  \bibfield  {author} {\bibinfo {author} {\bibfnamefont {S.}~\bibnamefont
  {Hopper}}\ and\ \bibinfo {author} {\bibfnamefont {C.~R.}\ \bibnamefont
  {Evans}},\ }\bibfield  {title} {\bibinfo {title} {{Gravitational
  perturbations and metric reconstruction: Method of extended homogeneous
  solutions applied to eccentric orbits on a Schwarzschild black hole}},\
  }\href {https://doi.org/10.1103/PhysRevD.82.084010} {\bibfield  {journal}
  {\bibinfo  {journal} {Phys. Rev. D}\ }\textbf {\bibinfo {volume} {82}},\
  \bibinfo {pages} {084010} (\bibinfo {year} {2010})},\ \Eprint
  {https://arxiv.org/abs/1006.4907} {arXiv:1006.4907 [gr-qc]} \BibitemShut
  {NoStop}%
\bibitem [{\citenamefont {Brizuela}\ \emph {et~al.}(2009)\citenamefont
  {Brizuela}, \citenamefont {Martin-Garcia},\ and\ \citenamefont
  {Tiglio}}]{Brizuela:2009qd}%
  \BibitemOpen
  \bibfield  {author} {\bibinfo {author} {\bibfnamefont {D.}~\bibnamefont
  {Brizuela}}, \bibinfo {author} {\bibfnamefont {J.~M.}\ \bibnamefont
  {Martin-Garcia}},\ and\ \bibinfo {author} {\bibfnamefont {M.}~\bibnamefont
  {Tiglio}},\ }\bibfield  {title} {\bibinfo {title} {{A Complete
  gauge-invariant formalism for arbitrary second-order perturbations of a
  Schwarzschild black hole}},\ }\href
  {https://doi.org/10.1103/PhysRevD.80.024021} {\bibfield  {journal} {\bibinfo
  {journal} {Phys. Rev. D}\ }\textbf {\bibinfo {volume} {80}},\ \bibinfo
  {pages} {024021} (\bibinfo {year} {2009})},\ \Eprint
  {https://arxiv.org/abs/0903.1134} {arXiv:0903.1134 [gr-qc]} \BibitemShut
  {NoStop}%
\bibitem [{\citenamefont {Martel}(2004)}]{Martel:2003jj}%
  \BibitemOpen
  \bibfield  {author} {\bibinfo {author} {\bibfnamefont {K.}~\bibnamefont
  {Martel}},\ }\bibfield  {title} {\bibinfo {title} {{Gravitational waveforms
  from a point particle orbiting a Schwarzschild black hole}},\ }\href@noop {}
  {\bibfield  {journal} {\bibinfo  {journal} {Phys. Rev. D}\ }\textbf {\bibinfo
  {volume} {69}},\ \bibinfo {pages} {044025} (\bibinfo {year} {2004})},\
  \Eprint {https://arxiv.org/abs/gr-qc/0311017} {gr-qc/0311017} \BibitemShut
  {NoStop}%
\bibitem [{\citenamefont {Lenzi}\ and\ \citenamefont
  {Sopuerta}(2021{\natexlab{a}})}]{Lenzi:2021wpc}%
  \BibitemOpen
  \bibfield  {author} {\bibinfo {author} {\bibfnamefont {M.}~\bibnamefont
  {Lenzi}}\ and\ \bibinfo {author} {\bibfnamefont {C.~F.}\ \bibnamefont
  {Sopuerta}},\ }\bibfield  {title} {\bibinfo {title} {{Master functions and
  equations for perturbations of vacuum spherically symmetric spacetimes}},\
  }\href {https://doi.org/10.1103/PhysRevD.104.084053} {\bibfield  {journal}
  {\bibinfo  {journal} {Phys. Rev. D}\ }\textbf {\bibinfo {volume} {104}},\
  \bibinfo {pages} {084053} (\bibinfo {year} {2021}{\natexlab{a}})},\ \Eprint
  {https://arxiv.org/abs/2108.08668} {arXiv:2108.08668 [gr-qc]} \BibitemShut
  {NoStop}%
\bibitem [{\citenamefont {Lenzi}\ and\ \citenamefont
  {Sopuerta}(2021{\natexlab{b}})}]{Lenzi:2021njy}%
  \BibitemOpen
  \bibfield  {author} {\bibinfo {author} {\bibfnamefont {M.}~\bibnamefont
  {Lenzi}}\ and\ \bibinfo {author} {\bibfnamefont {C.~F.}\ \bibnamefont
  {Sopuerta}},\ }\bibfield  {title} {\bibinfo {title} {{Darboux covariance: A
  hidden symmetry of perturbed Schwarzschild black holes}},\ }\href
  {https://doi.org/10.1103/PhysRevD.104.124068} {\bibfield  {journal} {\bibinfo
   {journal} {Phys. Rev. D}\ }\textbf {\bibinfo {volume} {104}},\ \bibinfo
  {pages} {124068} (\bibinfo {year} {2021}{\natexlab{b}})},\ \Eprint
  {https://arxiv.org/abs/2109.00503} {arXiv:2109.00503 [gr-qc]} \BibitemShut
  {NoStop}%
\bibitem [{\citenamefont {Glampedakis}\ \emph {et~al.}(2017)\citenamefont
  {Glampedakis}, \citenamefont {Johnson},\ and\ \citenamefont
  {Kennefick}}]{Glampedakis:2017rar}%
  \BibitemOpen
  \bibfield  {author} {\bibinfo {author} {\bibfnamefont {K.}~\bibnamefont
  {Glampedakis}}, \bibinfo {author} {\bibfnamefont {A.~D.}\ \bibnamefont
  {Johnson}},\ and\ \bibinfo {author} {\bibfnamefont {D.}~\bibnamefont
  {Kennefick}},\ }\bibfield  {title} {\bibinfo {title} {{Darboux transformation
  in black hole perturbation theory}},\ }\href
  {https://doi.org/10.1103/PhysRevD.96.024036} {\bibfield  {journal} {\bibinfo
  {journal} {Phys. Rev. D}\ }\textbf {\bibinfo {volume} {96}},\ \bibinfo
  {pages} {024036} (\bibinfo {year} {2017})},\ \Eprint
  {https://arxiv.org/abs/1702.06459} {arXiv:1702.06459 [gr-qc]} \BibitemShut
  {NoStop}%
\bibitem [{\citenamefont {{Chandrasekhar}}(1980)}]{1980RSPSA.369..425C}%
  \BibitemOpen
  \bibfield  {author} {\bibinfo {author} {\bibfnamefont {S.}~\bibnamefont
  {{Chandrasekhar}}},\ }\bibfield  {title} {\bibinfo {title} {{On
  One-Dimensional Potential Barriers Having Equal Reflexion and Transmission
  Coefficients}},\ }\href {https://doi.org/10.1098/rspa.1980.0008} {\bibfield
  {journal} {\bibinfo  {journal} {Proc. Roy. Soc. Lond. A}\ }\textbf {\bibinfo
  {volume} {369}},\ \bibinfo {pages} {425} (\bibinfo {year}
  {1980})}\BibitemShut {NoStop}%
\bibitem [{\citenamefont {{Chandrasekhar}}(1992)}]{Chandrasekhar:1992bo}%
  \BibitemOpen
  \bibfield  {author} {\bibinfo {author} {\bibfnamefont {S.}~\bibnamefont
  {{Chandrasekhar}}},\ }\href@noop {} {\emph {\bibinfo {title} {{The
  Mathematical Theory of Black Holes}}}}\ (\bibinfo  {publisher} {Oxford
  University Press},\ \bibinfo {address} {New York},\ \bibinfo {year}
  {1992})\BibitemShut {NoStop}%
\bibitem [{\citenamefont {Miura}(1976)}]{doi:10.1137/1018076}%
  \BibitemOpen
  \bibfield  {author} {\bibinfo {author} {\bibfnamefont {R.~M.}\ \bibnamefont
  {Miura}},\ }\bibfield  {title} {\bibinfo {title} {{The Korteweg–de Vries
  Equation: A Survey of Results}},\ }\href {https://doi.org/10.1137/1018076}
  {\bibfield  {journal} {\bibinfo  {journal} {SIAM Review}\ }\textbf {\bibinfo
  {volume} {18}},\ \bibinfo {pages} {412} (\bibinfo {year} {1976})},\ \Eprint
  {https://arxiv.org/abs/https://doi.org/10.1137/1018076}
  {https://doi.org/10.1137/1018076} \BibitemShut {NoStop}%
\bibitem [{\citenamefont {{Miura}}\ \emph {et~al.}(1968)\citenamefont
  {{Miura}}, \citenamefont {{Gardner}},\ and\ \citenamefont
  {{Kruskal}}}]{Miura:1968JMP.....9.1204M}%
  \BibitemOpen
  \bibfield  {author} {\bibinfo {author} {\bibfnamefont {R.~M.}\ \bibnamefont
  {{Miura}}}, \bibinfo {author} {\bibfnamefont {C.~S.}\ \bibnamefont
  {{Gardner}}},\ and\ \bibinfo {author} {\bibfnamefont {M.~D.}\ \bibnamefont
  {{Kruskal}}},\ }\bibfield  {title} {\bibinfo {title} {{Korteweg-de Vries
  Equation and Generalizations. II. Existence of Conservation Laws and
  Constants of Motion}},\ }\href {https://doi.org/10.1063/1.1664701} {\bibfield
   {journal} {\bibinfo  {journal} {J. Math. Phys.}\ }\textbf {\bibinfo {volume}
  {9}},\ \bibinfo {pages} {1204} (\bibinfo {year} {1968})}\BibitemShut
  {NoStop}%
\bibitem [{\citenamefont {Zakharov}\ and\ \citenamefont
  {Faddeev}(1971)}]{Zakharov:1971faa}%
  \BibitemOpen
  \bibfield  {author} {\bibinfo {author} {\bibfnamefont {V.~E.}\ \bibnamefont
  {Zakharov}}\ and\ \bibinfo {author} {\bibfnamefont {L.~D.}\ \bibnamefont
  {Faddeev}},\ }\bibfield  {title} {\bibinfo {title} {{Korteweg-de Vries
  equation: A completely integrable Hamiltonian system}},\ }\href
  {https://doi.org/10.1007/BF01086739} {\bibfield  {journal} {\bibinfo
  {journal} {Functional Analysis and Its Applications}\ }\textbf {\bibinfo
  {volume} {5}},\ \bibinfo {pages} {280} (\bibinfo {year} {1971})}\BibitemShut
  {NoStop}%
\bibitem [{\citenamefont {Lax}(1968)}]{Lax:1968fm}%
  \BibitemOpen
  \bibfield  {author} {\bibinfo {author} {\bibfnamefont {P.~D.}\ \bibnamefont
  {Lax}},\ }\bibfield  {title} {\bibinfo {title} {{Integrals of Nonlinear
  Equations of Evolution and Solitary Waves}},\ }\href@noop {} {\bibfield
  {journal} {\bibinfo  {journal} {Commun. Pure Appl. Math.}\ }\textbf {\bibinfo
  {volume} {21}},\ \bibinfo {pages} {467} (\bibinfo {year} {1968})}\BibitemShut
  {NoStop}%
\bibitem [{\citenamefont {Lenzi}\ and\ \citenamefont
  {Sopuerta}(2022)}]{Lenzi:2022wjv}%
  \BibitemOpen
  \bibfield  {author} {\bibinfo {author} {\bibfnamefont {M.}~\bibnamefont
  {Lenzi}}\ and\ \bibinfo {author} {\bibfnamefont {C.~F.}\ \bibnamefont
  {Sopuerta}},\ }\bibfield  {title} {\bibinfo {title} {{Black Hole Greybody
  Factors from Korteweg-de Vries Integrals: Theory}},\ }\href@noop {} {\
  (\bibinfo {year} {2022})},\ \Eprint {https://arxiv.org/abs/2212.03732}
  {arXiv:2212.03732 [gr-qc]} \BibitemShut {NoStop}%
\bibitem [{\citenamefont {Lenzi}\ and\ \citenamefont
  {Sopuerta}(2023)}]{Lenzi:2023inn}%
  \BibitemOpen
  \bibfield  {author} {\bibinfo {author} {\bibfnamefont {M.}~\bibnamefont
  {Lenzi}}\ and\ \bibinfo {author} {\bibfnamefont {C.~F.}\ \bibnamefont
  {Sopuerta}},\ }\bibfield  {title} {\bibinfo {title} {{Black Hole Greybody
  Factors from Korteweg-de Vries Integrals: Computation}},\ }\href@noop {} {\
  (\bibinfo {year} {2023})},\ \Eprint {https://arxiv.org/abs/2301.01096}
  {arXiv:2301.01096 [gr-qc]} \BibitemShut {NoStop}%
\bibitem [{\citenamefont {Brizuela}\ and\ \citenamefont
  {Martin-Garcia}(2009)}]{Brizuela:2008sk}%
  \BibitemOpen
  \bibfield  {author} {\bibinfo {author} {\bibfnamefont {D.}~\bibnamefont
  {Brizuela}}\ and\ \bibinfo {author} {\bibfnamefont {J.~M.}\ \bibnamefont
  {Martin-Garcia}},\ }\bibfield  {title} {\bibinfo {title} {{Hamiltonian theory
  for the axial perturbations of a dynamical spherical background}},\ }\href
  {https://doi.org/10.1088/0264-9381/26/1/015003} {\bibfield  {journal}
  {\bibinfo  {journal} {Class. Quant. Grav.}\ }\textbf {\bibinfo {volume}
  {26}},\ \bibinfo {pages} {015003} (\bibinfo {year} {2009})},\ \Eprint
  {https://arxiv.org/abs/0810.4786} {arXiv:0810.4786 [gr-qc]} \BibitemShut
  {NoStop}%
\bibitem [{\citenamefont {Bernar}\ \emph {et~al.}(2016)\citenamefont {Bernar},
  \citenamefont {Crispino},\ and\ \citenamefont {Higuchi}}]{Bernar:2016zgq}%
  \BibitemOpen
  \bibfield  {author} {\bibinfo {author} {\bibfnamefont {R.~P.}\ \bibnamefont
  {Bernar}}, \bibinfo {author} {\bibfnamefont {L.~C.~B.}\ \bibnamefont
  {Crispino}},\ and\ \bibinfo {author} {\bibfnamefont {A.}~\bibnamefont
  {Higuchi}},\ }\bibfield  {title} {\bibinfo {title} {{Graviton two-point
  function in 3 + 1 static de Sitter spacetime}},\ }\bibfield  {booktitle}
  {\emph {\bibinfo {booktitle} {{Proceedings, 3rd Amazonian Symposium on
  Physics: Belem, Brazil, September 28-October 2, 2015}}},\ }\href
  {https://doi.org/10.1142/S0218271816410169} {\bibfield  {journal} {\bibinfo
  {journal} {Int. J. Mod. Phys. D}\ }\textbf {\bibinfo {volume} {25}},\
  \bibinfo {pages} {1641016} (\bibinfo {year} {2016})}\BibitemShut {NoStop}%
\bibitem [{\citenamefont {Bernar}\ \emph {et~al.}(2017)\citenamefont {Bernar},
  \citenamefont {Crispino},\ and\ \citenamefont {Higuchi}}]{Bernar:2017kug}%
  \BibitemOpen
  \bibfield  {author} {\bibinfo {author} {\bibfnamefont {R.~P.}\ \bibnamefont
  {Bernar}}, \bibinfo {author} {\bibfnamefont {L.~C.~B.}\ \bibnamefont
  {Crispino}},\ and\ \bibinfo {author} {\bibfnamefont {A.}~\bibnamefont
  {Higuchi}},\ }\bibfield  {title} {\bibinfo {title} {{Gravitational waves
  emitted by a particle rotating around a Schwarzschild black hole: A
  semiclassical approach}},\ }\href
  {https://doi.org/10.1103/PhysRevD.95.064042} {\bibfield  {journal} {\bibinfo
  {journal} {Phys. Rev. D}\ }\textbf {\bibinfo {volume} {95}},\ \bibinfo
  {pages} {064042} (\bibinfo {year} {2017})},\ \Eprint
  {https://arxiv.org/abs/1703.10648} {arXiv:1703.10648 [gr-qc]} \BibitemShut
  {NoStop}%
\bibitem [{\citenamefont {Bernar}\ \emph {et~al.}(2018)\citenamefont {Bernar},
  \citenamefont {Crispino},\ and\ \citenamefont {Higuchi}}]{Bernar:2018nww}%
  \BibitemOpen
  \bibfield  {author} {\bibinfo {author} {\bibfnamefont {R.~P.}\ \bibnamefont
  {Bernar}}, \bibinfo {author} {\bibfnamefont {L.~C.~B.}\ \bibnamefont
  {Crispino}},\ and\ \bibinfo {author} {\bibfnamefont {A.}~\bibnamefont
  {Higuchi}},\ }\bibfield  {title} {\bibinfo {title} {{Gibbons-Hawking
  radiation of gravitons in the Poincar\'e and static patches of de Sitter
  spacetime}},\ }\href {https://doi.org/10.1103/PhysRevD.97.085005} {\bibfield
  {journal} {\bibinfo  {journal} {Phys. Rev. D}\ }\textbf {\bibinfo {volume}
  {97}},\ \bibinfo {pages} {085005} (\bibinfo {year} {2018})},\ \Eprint
  {https://arxiv.org/abs/1803.01204} {arXiv:1803.01204 [gr-qc]} \BibitemShut
  {NoStop}%
\bibitem [{\citenamefont {{Stewart}}\ and\ \citenamefont
  {{Walker}}(1974)}]{Stewart:1974uz}%
  \BibitemOpen
  \bibfield  {author} {\bibinfo {author} {\bibfnamefont {J.~M.}\ \bibnamefont
  {{Stewart}}}\ and\ \bibinfo {author} {\bibfnamefont {M.}~\bibnamefont
  {{Walker}}},\ }\bibfield  {title} {\bibinfo {title} {{Perturbations of
  space-times in general relativity}},\ }\href@noop {} {\bibfield  {journal}
  {\bibinfo  {journal} {Royal Society of London Proceedings Series A}\ }\textbf
  {\bibinfo {volume} {341}},\ \bibinfo {pages} {49} (\bibinfo {year}
  {1974})}\BibitemShut {NoStop}%
\bibitem [{\citenamefont {Wald}(1984)}]{Wald:1984cw}%
  \BibitemOpen
  \bibfield  {author} {\bibinfo {author} {\bibfnamefont {R.}~\bibnamefont
  {Wald}},\ }\href@noop {} {\emph {\bibinfo {title} {{General Relativity}}}}\
  (\bibinfo  {publisher} {The University of Chicago Press},\ \bibinfo {address}
  {Chicago},\ \bibinfo {year} {1984})\BibitemShut {NoStop}%
\bibitem [{\citenamefont {Bruni}\ \emph {et~al.}(1997)\citenamefont {Bruni},
  \citenamefont {Matarrese}, \citenamefont {Mollerach},\ and\ \citenamefont
  {Sonego}}]{Bruni:1996im}%
  \BibitemOpen
  \bibfield  {author} {\bibinfo {author} {\bibfnamefont {M.}~\bibnamefont
  {Bruni}}, \bibinfo {author} {\bibfnamefont {S.}~\bibnamefont {Matarrese}},
  \bibinfo {author} {\bibfnamefont {S.}~\bibnamefont {Mollerach}},\ and\
  \bibinfo {author} {\bibfnamefont {S.}~\bibnamefont {Sonego}},\ }\bibfield
  {title} {\bibinfo {title} {{Perturbations of space-time: Gauge
  transformations and gauge invariance at second order and beyond}},\ }\href
  {https://doi.org/10.1088/0264-9381/14/9/014} {\bibfield  {journal} {\bibinfo
  {journal} {Class. Quant. Grav.}\ }\textbf {\bibinfo {volume} {14}},\ \bibinfo
  {pages} {2585} (\bibinfo {year} {1997})},\ \Eprint
  {https://arxiv.org/abs/gr-qc/9609040} {arXiv:gr-qc/9609040} \BibitemShut
  {NoStop}%
\bibitem [{\citenamefont {Bruni}\ \emph {et~al.}(2003)\citenamefont {Bruni},
  \citenamefont {Gualtieri},\ and\ \citenamefont {Sopuerta}}]{Bruni:2002sma}%
  \BibitemOpen
  \bibfield  {author} {\bibinfo {author} {\bibfnamefont {M.}~\bibnamefont
  {Bruni}}, \bibinfo {author} {\bibfnamefont {L.}~\bibnamefont {Gualtieri}},\
  and\ \bibinfo {author} {\bibfnamefont {C.~F.}\ \bibnamefont {Sopuerta}},\
  }\bibfield  {title} {\bibinfo {title} {{Two parameter nonlinear space-time
  perturbations: Gauge transformations and gauge invariance}},\ }\href
  {https://doi.org/10.1088/0264-9381/20/3/310} {\bibfield  {journal} {\bibinfo
  {journal} {Class. Quant. Grav.}\ }\textbf {\bibinfo {volume} {20}},\ \bibinfo
  {pages} {535} (\bibinfo {year} {2003})},\ \Eprint
  {https://arxiv.org/abs/gr-qc/0207105} {arXiv:gr-qc/0207105} \BibitemShut
  {NoStop}%
\bibitem [{\citenamefont {Sopuerta}\ \emph {et~al.}(2004)\citenamefont
  {Sopuerta}, \citenamefont {Bruni},\ and\ \citenamefont
  {Gualtieri}}]{Sopuerta:2003rg}%
  \BibitemOpen
  \bibfield  {author} {\bibinfo {author} {\bibfnamefont {C.~F.}\ \bibnamefont
  {Sopuerta}}, \bibinfo {author} {\bibfnamefont {M.}~\bibnamefont {Bruni}},\
  and\ \bibinfo {author} {\bibfnamefont {L.}~\bibnamefont {Gualtieri}},\
  }\bibfield  {title} {\bibinfo {title} {{Nonlinear N-parameter space-time
  perturbations: Gauge transformations}},\ }\href
  {https://doi.org/10.1103/PhysRevD.70.064002} {\bibfield  {journal} {\bibinfo
  {journal} {Phys. Rev. D}\ }\textbf {\bibinfo {volume} {70}},\ \bibinfo
  {pages} {064002} (\bibinfo {year} {2004})},\ \Eprint
  {https://arxiv.org/abs/gr-qc/0306027} {arXiv:gr-qc/0306027} \BibitemShut
  {NoStop}%
\bibitem [{\citenamefont {Wald}(1974)}]{Wald:1974np}%
  \BibitemOpen
  \bibfield  {author} {\bibinfo {author} {\bibfnamefont {R.~M.}\ \bibnamefont
  {Wald}},\ }\bibfield  {title} {\bibinfo {title} {{Black hole in a uniform
  magnetic field}},\ }\href {https://doi.org/10.1103/PhysRevD.10.1680}
  {\bibfield  {journal} {\bibinfo  {journal} {Phys. Rev. D}\ }\textbf {\bibinfo
  {volume} {10}},\ \bibinfo {pages} {1680} (\bibinfo {year}
  {1974})}\BibitemShut {NoStop}%
\bibitem [{\citenamefont {{Cunningham}}\ \emph {et~al.}(1978)\citenamefont
  {{Cunningham}}, \citenamefont {{Price}},\ and\ \citenamefont
  {{Moncrief}}}]{Cunningham:1978cp}%
  \BibitemOpen
  \bibfield  {author} {\bibinfo {author} {\bibfnamefont {C.~T.}\ \bibnamefont
  {{Cunningham}}}, \bibinfo {author} {\bibfnamefont {R.~H.}\ \bibnamefont
  {{Price}}},\ and\ \bibinfo {author} {\bibfnamefont {V.}~\bibnamefont
  {{Moncrief}}},\ }\bibfield  {title} {\bibinfo {title} {{Radiation from
  collapsing relativistic stars. I - Linearized odd-parity radiation}},\
  }\href@noop {} {\bibfield  {journal} {\bibinfo  {journal} {Astrophys. J.}\
  }\textbf {\bibinfo {volume} {224}},\ \bibinfo {pages} {643} (\bibinfo {year}
  {1978})}\BibitemShut {NoStop}%
\bibitem [{\citenamefont {Cunningham}\ \emph {et~al.}(1979)\citenamefont
  {Cunningham}, \citenamefont {Price},\ and\ \citenamefont
  {Moncrief}}]{Cunningham:1979px}%
  \BibitemOpen
  \bibfield  {author} {\bibinfo {author} {\bibfnamefont {C.~T.}\ \bibnamefont
  {Cunningham}}, \bibinfo {author} {\bibfnamefont {R.~H.}\ \bibnamefont
  {Price}},\ and\ \bibinfo {author} {\bibfnamefont {V.}~\bibnamefont
  {Moncrief}},\ }\bibfield  {title} {\bibinfo {title} {{Radiation from
  collapsing relativistic stars. II. Linearized even parity radiation}},\
  }\href@noop {} {\bibfield  {journal} {\bibinfo  {journal} {Astrophys. J.}\
  }\textbf {\bibinfo {volume} {230}},\ \bibinfo {pages} {870} (\bibinfo {year}
  {1979})}\BibitemShut {NoStop}%
\bibitem [{\citenamefont {{Cunningham}}\ \emph {et~al.}(1980)\citenamefont
  {{Cunningham}}, \citenamefont {{Price}},\ and\ \citenamefont
  {{Moncrief}}}]{Cunningham:1980cp}%
  \BibitemOpen
  \bibfield  {author} {\bibinfo {author} {\bibfnamefont {C.~T.}\ \bibnamefont
  {{Cunningham}}}, \bibinfo {author} {\bibfnamefont {R.~H.}\ \bibnamefont
  {{Price}}},\ and\ \bibinfo {author} {\bibfnamefont {V.}~\bibnamefont
  {{Moncrief}}},\ }\bibfield  {title} {\bibinfo {title} {{Radiation from
  collapsing relativistic stars. III - Second order perturbations of collapse
  with rotation}},\ }\href@noop {} {\bibfield  {journal} {\bibinfo  {journal}
  {Astrophys. J.}\ }\textbf {\bibinfo {volume} {236}},\ \bibinfo {pages} {674}
  (\bibinfo {year} {1980})}\BibitemShut {NoStop}%
\bibitem [{\citenamefont {Jhingan}\ and\ \citenamefont
  {Tanaka}(2003)}]{Jhingan:2002kb}%
  \BibitemOpen
  \bibfield  {author} {\bibinfo {author} {\bibfnamefont {S.}~\bibnamefont
  {Jhingan}}\ and\ \bibinfo {author} {\bibfnamefont {T.}~\bibnamefont
  {Tanaka}},\ }\bibfield  {title} {\bibinfo {title} {{Improvement on the metric
  reconstruction scheme in Regge-Wheeler-Zerilli formalism}},\ }\href@noop {}
  {\bibfield  {journal} {\bibinfo  {journal} {Phys. Rev. D}\ }\textbf {\bibinfo
  {volume} {67}},\ \bibinfo {pages} {104018} (\bibinfo {year} {2003})},\
  \Eprint {https://arxiv.org/abs/gr-qc/0211060} {gr-qc/0211060} \BibitemShut
  {NoStop}%
\bibitem [{\citenamefont {Martel}\ and\ \citenamefont
  {Poisson}(2005)}]{Martel:2005ir}%
  \BibitemOpen
  \bibfield  {author} {\bibinfo {author} {\bibfnamefont {K.}~\bibnamefont
  {Martel}}\ and\ \bibinfo {author} {\bibfnamefont {E.}~\bibnamefont
  {Poisson}},\ }\bibfield  {title} {\bibinfo {title} {{Gravitational
  perturbations of the Schwarzschild spacetime: A practical covariant and
  gauge-invariant formalism}},\ }\href@noop {} {\bibfield  {journal} {\bibinfo
  {journal} {Phys. Rev. D}\ }\textbf {\bibinfo {volume} {71}},\ \bibinfo
  {pages} {104003} (\bibinfo {year} {2005})},\ \Eprint
  {https://arxiv.org/abs/gr-qc/0502028} {gr-qc/0502028} \BibitemShut {NoStop}%
\bibitem [{\citenamefont {Lousto}\ and\ \citenamefont
  {Price}(1997)}]{Lousto:1996sx}%
  \BibitemOpen
  \bibfield  {author} {\bibinfo {author} {\bibfnamefont {C.~O.}\ \bibnamefont
  {Lousto}}\ and\ \bibinfo {author} {\bibfnamefont {R.~H.}\ \bibnamefont
  {Price}},\ }\bibfield  {title} {\bibinfo {title} {{Head-on collisions of
  black holes: The Particle limit}},\ }\href
  {https://doi.org/10.1103/PhysRevD.55.2124} {\bibfield  {journal} {\bibinfo
  {journal} {Phys. Rev. D}\ }\textbf {\bibinfo {volume} {55}},\ \bibinfo
  {pages} {2124} (\bibinfo {year} {1997})},\ \Eprint
  {https://arxiv.org/abs/gr-qc/9609012} {arXiv:gr-qc/9609012} \BibitemShut
  {NoStop}%
\bibitem [{\citenamefont {Gleiser}\ \emph {et~al.}(2000)\citenamefont
  {Gleiser}, \citenamefont {Nicasio}, \citenamefont {Price},\ and\
  \citenamefont {Pullin}}]{Gleiser:1998rw}%
  \BibitemOpen
  \bibfield  {author} {\bibinfo {author} {\bibfnamefont {R.~J.}\ \bibnamefont
  {Gleiser}}, \bibinfo {author} {\bibfnamefont {C.~O.}\ \bibnamefont
  {Nicasio}}, \bibinfo {author} {\bibfnamefont {R.~H.}\ \bibnamefont {Price}},\
  and\ \bibinfo {author} {\bibfnamefont {J.}~\bibnamefont {Pullin}},\
  }\bibfield  {title} {\bibinfo {title} {{Gravitational radiation from
  Schwarzschild black holes: The Second order perturbation formalism}},\ }\href
  {https://doi.org/10.1016/S0370-1573(99)00048-4} {\bibfield  {journal}
  {\bibinfo  {journal} {Phys. Rept.}\ }\textbf {\bibinfo {volume} {325}},\
  \bibinfo {pages} {41} (\bibinfo {year} {2000})},\ \Eprint
  {https://arxiv.org/abs/gr-qc/9807077} {arXiv:gr-qc/9807077} \BibitemShut
  {NoStop}%
\bibitem [{\citenamefont {Price}\ and\ \citenamefont
  {Pullin}(1994)}]{Price:1994pm}%
  \BibitemOpen
  \bibfield  {author} {\bibinfo {author} {\bibfnamefont {R.~H.}\ \bibnamefont
  {Price}}\ and\ \bibinfo {author} {\bibfnamefont {J.}~\bibnamefont {Pullin}},\
  }\bibfield  {title} {\bibinfo {title} {{Colliding black holes: The Close
  limit}},\ }\href {https://doi.org/10.1103/PhysRevLett.72.3297} {\bibfield
  {journal} {\bibinfo  {journal} {Phys. Rev. Lett.}\ }\textbf {\bibinfo
  {volume} {72}},\ \bibinfo {pages} {3297} (\bibinfo {year} {1994})},\ \Eprint
  {https://arxiv.org/abs/gr-qc/9402039} {arXiv:gr-qc/9402039} \BibitemShut
  {NoStop}%
\bibitem [{\citenamefont {Gleiser}\ \emph {et~al.}(1996)\citenamefont
  {Gleiser}, \citenamefont {Nicasio}, \citenamefont {Price},\ and\
  \citenamefont {Pullin}}]{Gleiser:1996yc}%
  \BibitemOpen
  \bibfield  {author} {\bibinfo {author} {\bibfnamefont {R.~J.}\ \bibnamefont
  {Gleiser}}, \bibinfo {author} {\bibfnamefont {C.~O.}\ \bibnamefont
  {Nicasio}}, \bibinfo {author} {\bibfnamefont {R.~H.}\ \bibnamefont {Price}},\
  and\ \bibinfo {author} {\bibfnamefont {J.}~\bibnamefont {Pullin}},\
  }\bibfield  {title} {\bibinfo {title} {{Colliding black holes: How far can
  the close approximation go?}},\ }\href
  {https://doi.org/10.1103/PhysRevLett.77.4483} {\bibfield  {journal} {\bibinfo
   {journal} {Phys. Rev. Lett.}\ }\textbf {\bibinfo {volume} {77}},\ \bibinfo
  {pages} {4483} (\bibinfo {year} {1996})},\ \Eprint
  {https://arxiv.org/abs/gr-qc/9609022} {arXiv:gr-qc/9609022} \BibitemShut
  {NoStop}%
\bibitem [{\citenamefont {Khanna}\ \emph {et~al.}(1999)\citenamefont {Khanna},
  \citenamefont {Baker}, \citenamefont {Gleiser}, \citenamefont {Laguna},
  \citenamefont {Nicasio}, \citenamefont {Nollert}, \citenamefont {Price},\
  and\ \citenamefont {Pullin}}]{Khanna:1999mh}%
  \BibitemOpen
  \bibfield  {author} {\bibinfo {author} {\bibfnamefont {G.}~\bibnamefont
  {Khanna}}, \bibinfo {author} {\bibfnamefont {J.~G.}\ \bibnamefont {Baker}},
  \bibinfo {author} {\bibfnamefont {R.~J.}\ \bibnamefont {Gleiser}}, \bibinfo
  {author} {\bibfnamefont {P.}~\bibnamefont {Laguna}}, \bibinfo {author}
  {\bibfnamefont {C.~O.}\ \bibnamefont {Nicasio}}, \bibinfo {author}
  {\bibfnamefont {H.-P.}\ \bibnamefont {Nollert}}, \bibinfo {author}
  {\bibfnamefont {R.}~\bibnamefont {Price}},\ and\ \bibinfo {author}
  {\bibfnamefont {J.}~\bibnamefont {Pullin}},\ }\bibfield  {title} {\bibinfo
  {title} {{Inspiralling black holes: The Close limit}},\ }\href
  {https://doi.org/10.1103/PhysRevLett.83.3581} {\bibfield  {journal} {\bibinfo
   {journal} {Phys. Rev. Lett.}\ }\textbf {\bibinfo {volume} {83}},\ \bibinfo
  {pages} {3581} (\bibinfo {year} {1999})},\ \Eprint
  {https://arxiv.org/abs/gr-qc/9905081} {arXiv:gr-qc/9905081} \BibitemShut
  {NoStop}%
\bibitem [{\citenamefont {Pullin}(1999)}]{Pullin:1999rg}%
  \BibitemOpen
  \bibfield  {author} {\bibinfo {author} {\bibfnamefont {J.}~\bibnamefont
  {Pullin}},\ }\bibfield  {title} {\bibinfo {title} {{The Close limit of
  colliding black holes: An Update}},\ }\href
  {https://doi.org/10.1143/PTPS.136.107} {\bibfield  {journal} {\bibinfo
  {journal} {Prog. Theor. Phys. Suppl.}\ }\textbf {\bibinfo {volume} {136}},\
  \bibinfo {pages} {107} (\bibinfo {year} {1999})},\ \Eprint
  {https://arxiv.org/abs/gr-qc/9909021} {arXiv:gr-qc/9909021} \BibitemShut
  {NoStop}%
\bibitem [{\citenamefont {Hopper}(2018)}]{Hopper:2017iyq}%
  \BibitemOpen
  \bibfield  {author} {\bibinfo {author} {\bibfnamefont {S.}~\bibnamefont
  {Hopper}},\ }\bibfield  {title} {\bibinfo {title} {{Unbound motion on a
  Schwarzschild background: Practical approaches to frequency domain
  computations}},\ }\href {https://doi.org/10.1103/PhysRevD.97.064007}
  {\bibfield  {journal} {\bibinfo  {journal} {Phys. Rev. D}\ }\textbf {\bibinfo
  {volume} {97}},\ \bibinfo {pages} {064007} (\bibinfo {year} {2018})},\
  \Eprint {https://arxiv.org/abs/1706.05455} {arXiv:1706.05455 [gr-qc]}
  \BibitemShut {NoStop}%
\bibitem [{\citenamefont {Darboux}(1889)}]{Darboux:89}%
  \BibitemOpen
  \bibfield  {author} {\bibinfo {author} {\bibfnamefont {G.}~\bibnamefont
  {Darboux}},\ }\href@noop {} {\emph {\bibinfo {title} {{Le\c{c}ons sur la
  th\'eorie g\'en\'erale des surfaces et les application g\'eom\'etriques du
  calcul infinit\'esimal. Deuxi\`eme partie}}}}\ (\bibinfo  {publisher}
  {Gauthier Villars et fils},\ \bibinfo {address} {Paris},\ \bibinfo {year}
  {1889})\BibitemShut {NoStop}%
\bibitem [{\citenamefont {{Darboux}}(1882)}]{1999physics...8003D}%
  \BibitemOpen
  \bibfield  {author} {\bibinfo {author} {\bibfnamefont {G.}~\bibnamefont
  {{Darboux}}},\ }\bibfield  {title} {\bibinfo {title} {{On a proposition
  relative to linear equations}},\ }\href@noop {} {\bibfield  {journal}
  {\bibinfo  {journal} {C.R. Acad. Sci. Paris}\ }\textbf {\bibinfo {volume}
  {94}},\ \bibinfo {pages} {1456} (\bibinfo {year} {1882})},\ \Eprint
  {https://arxiv.org/abs/physics/9908003} {arXiv:physics/9908003
  [physics.hist-ph]} \BibitemShut {NoStop}%
\bibitem [{\citenamefont {Matveev}\ and\ \citenamefont
  {Salle}(1991)}]{Matveev:1991ms}%
  \BibitemOpen
  \bibfield  {author} {\bibinfo {author} {\bibfnamefont {V.~B.}\ \bibnamefont
  {Matveev}}\ and\ \bibinfo {author} {\bibfnamefont {M.~A.}\ \bibnamefont
  {Salle}},\ }\href@noop {} {\emph {\bibinfo {title} {{Darboux Transformations
  and Solitons}}}}\ (\bibinfo  {publisher} {{Springer-Verlag}},\ \bibinfo
  {address} {New York, Berlin, Heidelberg},\ \bibinfo {year}
  {1991})\BibitemShut {NoStop}%
\bibitem [{\citenamefont {{Gendenshte{\"\i}n}}(1983)}]{Gendenshtein:1983jt}%
  \BibitemOpen
  \bibfield  {author} {\bibinfo {author} {\bibfnamefont {L.~{\'E}.}\
  \bibnamefont {{Gendenshte{\"\i}n}}},\ }\bibfield  {title} {\bibinfo {title}
  {{Derivation of exact spectra of the Schr{\"o}dinger equation by means of
  supersymmetry}},\ }\href@noop {} {\bibfield  {journal} {\bibinfo  {journal}
  {Soviet Journal of Experimental and Theoretical Physics Letters}\ }\textbf
  {\bibinfo {volume} {38}},\ \bibinfo {pages} {356} (\bibinfo {year}
  {1983})}\BibitemShut {NoStop}%
\bibitem [{\citenamefont {Organista}\ \emph {et~al.}(2006)\citenamefont
  {Organista}, \citenamefont {Nowakowski},\ and\ \citenamefont
  {Rosu}}]{Organista:2006jo}%
  \BibitemOpen
  \bibfield  {author} {\bibinfo {author} {\bibfnamefont {J.~O.}\ \bibnamefont
  {Organista}}, \bibinfo {author} {\bibfnamefont {M.}~\bibnamefont
  {Nowakowski}},\ and\ \bibinfo {author} {\bibfnamefont {H.~C.}\ \bibnamefont
  {Rosu}},\ }\bibfield  {title} {\bibinfo {title} {{Shape invariance through
  Crum transformation}},\ }\href {https://doi.org/10.1063/1.2397556} {\bibfield
   {journal} {\bibinfo  {journal} {Journal of Mathematical Physics}\ }\textbf
  {\bibinfo {volume} {47}},\ \bibinfo {pages} {122104} (\bibinfo {year}
  {2006})}\BibitemShut {NoStop}%
\bibitem [{\citenamefont {Grozdanov}\ and\ \citenamefont
  {Vrbica}(2023)}]{Grozdanov:2023txs}%
  \BibitemOpen
  \bibfield  {author} {\bibinfo {author} {\bibfnamefont {S.}~\bibnamefont
  {Grozdanov}}\ and\ \bibinfo {author} {\bibfnamefont {M.}~\bibnamefont
  {Vrbica}},\ }\bibfield  {title} {\bibinfo {title} {{Pole-skipping of
  gravitational waves in the backgrounds of four-dimensional massive black
  holes}},\ }\href@noop {} {\  (\bibinfo {year} {2023})},\ \Eprint
  {https://arxiv.org/abs/2303.15921} {arXiv:2303.15921 [hep-th]} \BibitemShut
  {NoStop}%
\bibitem [{\citenamefont {Grozdanov}\ \emph {et~al.}(2023)\citenamefont
  {Grozdanov}, \citenamefont {Lemut},\ and\ \citenamefont
  {Pedraza}}]{Grozdanov:2023tag}%
  \BibitemOpen
  \bibfield  {author} {\bibinfo {author} {\bibfnamefont {S.}~\bibnamefont
  {Grozdanov}}, \bibinfo {author} {\bibfnamefont {T.}~\bibnamefont {Lemut}},\
  and\ \bibinfo {author} {\bibfnamefont {J.~F.}\ \bibnamefont {Pedraza}},\
  }\bibfield  {title} {\bibinfo {title} {{Reconstruction of the quasinormal
  spectrum from pole skipping}},\ }\href
  {https://doi.org/10.1103/PhysRevD.108.L101901} {\bibfield  {journal}
  {\bibinfo  {journal} {Phys. Rev. D}\ }\textbf {\bibinfo {volume} {108}},\
  \bibinfo {pages} {L101901} (\bibinfo {year} {2023})},\ \Eprint
  {https://arxiv.org/abs/2308.01371} {arXiv:2308.01371 [hep-th]} \BibitemShut
  {NoStop}%
\bibitem [{\citenamefont {Franciolini}\ \emph {et~al.}(2019)\citenamefont
  {Franciolini}, \citenamefont {Hui}, \citenamefont {Penco}, \citenamefont
  {Santoni},\ and\ \citenamefont {Trincherini}}]{Franciolini:2018uyq}%
  \BibitemOpen
  \bibfield  {author} {\bibinfo {author} {\bibfnamefont {G.}~\bibnamefont
  {Franciolini}}, \bibinfo {author} {\bibfnamefont {L.}~\bibnamefont {Hui}},
  \bibinfo {author} {\bibfnamefont {R.}~\bibnamefont {Penco}}, \bibinfo
  {author} {\bibfnamefont {L.}~\bibnamefont {Santoni}},\ and\ \bibinfo {author}
  {\bibfnamefont {E.}~\bibnamefont {Trincherini}},\ }\bibfield  {title}
  {\bibinfo {title} {{Effective Field Theory of Black Hole Quasinormal Modes in
  Scalar-Tensor Theories}},\ }\href {https://doi.org/10.1007/JHEP02(2019)127}
  {\bibfield  {journal} {\bibinfo  {journal} {JHEP}\ }\textbf {\bibinfo
  {volume} {02}},\ \bibinfo {pages} {127}},\ \Eprint
  {https://arxiv.org/abs/1810.07706} {arXiv:1810.07706 [hep-th]} \BibitemShut
  {NoStop}%
\bibitem [{\citenamefont {Datta}\ and\ \citenamefont
  {Bose}(2020)}]{Datta:2019npq}%
  \BibitemOpen
  \bibfield  {author} {\bibinfo {author} {\bibfnamefont {S.}~\bibnamefont
  {Datta}}\ and\ \bibinfo {author} {\bibfnamefont {S.}~\bibnamefont {Bose}},\
  }\bibfield  {title} {\bibinfo {title} {{Quasi-normal Modes of Static
  Spherically Symmetric Black Holes in $f(R)$ Theory}},\ }\href
  {https://doi.org/10.1140/epjc/s10052-019-7546-1} {\bibfield  {journal}
  {\bibinfo  {journal} {Eur. Phys. J. C}\ }\textbf {\bibinfo {volume} {80}},\
  \bibinfo {pages} {14} (\bibinfo {year} {2020})},\ \Eprint
  {https://arxiv.org/abs/1904.01519} {arXiv:1904.01519 [gr-qc]} \BibitemShut
  {NoStop}%
\bibitem [{\citenamefont {Ferrari}\ \emph {et~al.}(2001)\citenamefont
  {Ferrari}, \citenamefont {Pauri},\ and\ \citenamefont
  {Piazza}}]{Ferrari:2000ep}%
  \BibitemOpen
  \bibfield  {author} {\bibinfo {author} {\bibfnamefont {V.}~\bibnamefont
  {Ferrari}}, \bibinfo {author} {\bibfnamefont {M.}~\bibnamefont {Pauri}},\
  and\ \bibinfo {author} {\bibfnamefont {F.}~\bibnamefont {Piazza}},\
  }\bibfield  {title} {\bibinfo {title} {{Quasinormal modes of charged, dilaton
  black holes}},\ }\href {https://doi.org/10.1103/PhysRevD.63.064009}
  {\bibfield  {journal} {\bibinfo  {journal} {Phys. Rev. D}\ }\textbf {\bibinfo
  {volume} {63}},\ \bibinfo {pages} {064009} (\bibinfo {year} {2001})},\
  \Eprint {https://arxiv.org/abs/gr-qc/0005125} {arXiv:gr-qc/0005125}
  \BibitemShut {NoStop}%
\bibitem [{\citenamefont {Luis Bl\'azquez-Salcedo}\ \emph
  {et~al.}(2021)\citenamefont {Luis Bl\'azquez-Salcedo}, \citenamefont
  {Herdeiro}, \citenamefont {Kahlen}, \citenamefont {Kunz}, \citenamefont
  {Pombo},\ and\ \citenamefont {Radu}}]{LuisBlazquez-Salcedo:2020rqp}%
  \BibitemOpen
  \bibfield  {author} {\bibinfo {author} {\bibfnamefont {J.}~\bibnamefont {Luis
  Bl\'azquez-Salcedo}}, \bibinfo {author} {\bibfnamefont {C.~A.~R.}\
  \bibnamefont {Herdeiro}}, \bibinfo {author} {\bibfnamefont {S.}~\bibnamefont
  {Kahlen}}, \bibinfo {author} {\bibfnamefont {J.}~\bibnamefont {Kunz}},
  \bibinfo {author} {\bibfnamefont {A.~M.}\ \bibnamefont {Pombo}},\ and\
  \bibinfo {author} {\bibfnamefont {E.}~\bibnamefont {Radu}},\ }\bibfield
  {title} {\bibinfo {title} {{Quasinormal modes of hot, cold and bald
  Einstein\textendash{}Maxwell-scalar black holes}},\ }\href
  {https://doi.org/10.1140/epjc/s10052-021-08952-w} {\bibfield  {journal}
  {\bibinfo  {journal} {Eur. Phys. J. C}\ }\textbf {\bibinfo {volume} {81}},\
  \bibinfo {pages} {155} (\bibinfo {year} {2021})},\ \Eprint
  {https://arxiv.org/abs/2008.11744} {arXiv:2008.11744 [gr-qc]} \BibitemShut
  {NoStop}%
\bibitem [{\citenamefont {Chen}\ \emph {et~al.}(2021)\citenamefont {Chen},
  \citenamefont {Bouhmadi-L\'opez},\ and\ \citenamefont {Chen}}]{Chen:2021cts}%
  \BibitemOpen
  \bibfield  {author} {\bibinfo {author} {\bibfnamefont {C.-Y.}\ \bibnamefont
  {Chen}}, \bibinfo {author} {\bibfnamefont {M.}~\bibnamefont
  {Bouhmadi-L\'opez}},\ and\ \bibinfo {author} {\bibfnamefont {P.}~\bibnamefont
  {Chen}},\ }\bibfield  {title} {\bibinfo {title} {{Lessons from black hole
  quasinormal modes in modified gravity}},\ }\href
  {https://doi.org/10.1140/epjp/s13360-021-01227-z} {\bibfield  {journal}
  {\bibinfo  {journal} {Eur. Phys. J. Plus}\ }\textbf {\bibinfo {volume}
  {136}},\ \bibinfo {pages} {253} (\bibinfo {year} {2021})},\ \Eprint
  {https://arxiv.org/abs/2103.01249} {arXiv:2103.01249 [gr-qc]} \BibitemShut
  {NoStop}%
\bibitem [{\citenamefont {del Corral}\ and\ \citenamefont
  {Olmedo}(2022)}]{del-Corral:2022kbk}%
  \BibitemOpen
  \bibfield  {author} {\bibinfo {author} {\bibfnamefont {D.}~\bibnamefont {del
  Corral}}\ and\ \bibinfo {author} {\bibfnamefont {J.}~\bibnamefont {Olmedo}},\
  }\bibfield  {title} {\bibinfo {title} {{Breaking of isospectrality of
  quasinormal modes in nonrotating loop quantum gravity black holes}},\ }\href
  {https://doi.org/10.1103/PhysRevD.105.064053} {\bibfield  {journal} {\bibinfo
   {journal} {Phys. Rev. D}\ }\textbf {\bibinfo {volume} {105}},\ \bibinfo
  {pages} {064053} (\bibinfo {year} {2022})},\ \Eprint
  {https://arxiv.org/abs/2201.09584} {arXiv:2201.09584 [gr-qc]} \BibitemShut
  {NoStop}%
\bibitem [{\citenamefont {Li}\ \emph {et~al.}(2023)\citenamefont {Li},
  \citenamefont {Hussain}, \citenamefont {Wagle}, \citenamefont {Chen},
  \citenamefont {Yunes},\ and\ \citenamefont {Zimmerman}}]{Li:2023ulk}%
  \BibitemOpen
  \bibfield  {author} {\bibinfo {author} {\bibfnamefont {D.}~\bibnamefont
  {Li}}, \bibinfo {author} {\bibfnamefont {A.}~\bibnamefont {Hussain}},
  \bibinfo {author} {\bibfnamefont {P.}~\bibnamefont {Wagle}}, \bibinfo
  {author} {\bibfnamefont {Y.}~\bibnamefont {Chen}}, \bibinfo {author}
  {\bibfnamefont {N.}~\bibnamefont {Yunes}},\ and\ \bibinfo {author}
  {\bibfnamefont {A.}~\bibnamefont {Zimmerman}},\ }\bibfield  {title} {\bibinfo
  {title} {{Isospectrality breaking in the Teukolsky formalism}},\ }\href@noop
  {} {\  (\bibinfo {year} {2023})},\ \Eprint {https://arxiv.org/abs/2310.06033}
  {arXiv:2310.06033 [gr-qc]} \BibitemShut {NoStop}%
\bibitem [{\citenamefont {Jaramillo}\ \emph {et~al.}(2021)\citenamefont
  {Jaramillo}, \citenamefont {Panosso~Macedo},\ and\ \citenamefont
  {Al~Sheikh}}]{Jaramillo:2020tuu}%
  \BibitemOpen
  \bibfield  {author} {\bibinfo {author} {\bibfnamefont {J.~L.}\ \bibnamefont
  {Jaramillo}}, \bibinfo {author} {\bibfnamefont {R.}~\bibnamefont
  {Panosso~Macedo}},\ and\ \bibinfo {author} {\bibfnamefont {L.}~\bibnamefont
  {Al~Sheikh}},\ }\bibfield  {title} {\bibinfo {title} {{Pseudospectrum and
  Black Hole Quasinormal Mode Instability}},\ }\href
  {https://doi.org/10.1103/PhysRevX.11.031003} {\bibfield  {journal} {\bibinfo
  {journal} {Phys. Rev. X}\ }\textbf {\bibinfo {volume} {11}},\ \bibinfo
  {pages} {031003} (\bibinfo {year} {2021})},\ \Eprint
  {https://arxiv.org/abs/2004.06434} {arXiv:2004.06434 [gr-qc]} \BibitemShut
  {NoStop}%
\bibitem [{\citenamefont {Jaramillo}\ \emph {et~al.}(2022)\citenamefont
  {Jaramillo}, \citenamefont {Panosso~Macedo},\ and\ \citenamefont
  {Sheikh}}]{Jaramillo:2021tmt}%
  \BibitemOpen
  \bibfield  {author} {\bibinfo {author} {\bibfnamefont {J.~L.}\ \bibnamefont
  {Jaramillo}}, \bibinfo {author} {\bibfnamefont {R.}~\bibnamefont
  {Panosso~Macedo}},\ and\ \bibinfo {author} {\bibfnamefont {L.~A.}\
  \bibnamefont {Sheikh}},\ }\bibfield  {title} {\bibinfo {title}
  {{Gravitational Wave Signatures of Black Hole Quasinormal Mode
  Instability}},\ }\href {https://doi.org/10.1103/PhysRevLett.128.211102}
  {\bibfield  {journal} {\bibinfo  {journal} {Phys. Rev. Lett.}\ }\textbf
  {\bibinfo {volume} {128}},\ \bibinfo {pages} {211102} (\bibinfo {year}
  {2022})},\ \Eprint {https://arxiv.org/abs/2105.03451} {arXiv:2105.03451
  [gr-qc]} \BibitemShut {NoStop}%
\bibitem [{\citenamefont {Heading}(1977)}]{Heading:1977jh}%
  \BibitemOpen
  \bibfield  {author} {\bibinfo {author} {\bibfnamefont {J.}~\bibnamefont
  {Heading}},\ }\bibfield  {title} {\bibinfo {title} {{Resolution of the
  mystery behind Chandrasekhar{\textquotesingle}s black hole
  transformations}},\ }\href {https://doi.org/10.1088/0305-4470/10/6/011}
  {\bibfield  {journal} {\bibinfo  {journal} {J. Phys. A: Math. Gen.}\ }\textbf
  {\bibinfo {volume} {10}},\ \bibinfo {pages} {885} (\bibinfo {year}
  {1977})}\BibitemShut {NoStop}%
\bibitem [{\citenamefont {Witten}(1981)}]{Witten:1981nf}%
  \BibitemOpen
  \bibfield  {author} {\bibinfo {author} {\bibfnamefont {E.}~\bibnamefont
  {Witten}},\ }\bibfield  {title} {\bibinfo {title} {{Dynamical Breaking of
  Supersymmetry}},\ }\href {https://doi.org/10.1016/0550-3213(81)90006-7}
  {\bibfield  {journal} {\bibinfo  {journal} {Nucl. Phys. B}\ }\textbf
  {\bibinfo {volume} {188}},\ \bibinfo {pages} {513} (\bibinfo {year}
  {1981})}\BibitemShut {NoStop}%
\bibitem [{\citenamefont {Cooper}\ and\ \citenamefont
  {Freedman}(1983)}]{Cooper:1982dm}%
  \BibitemOpen
  \bibfield  {author} {\bibinfo {author} {\bibfnamefont {F.}~\bibnamefont
  {Cooper}}\ and\ \bibinfo {author} {\bibfnamefont {B.}~\bibnamefont
  {Freedman}},\ }\bibfield  {title} {\bibinfo {title} {{Aspects of
  Supersymmetric Quantum Mechanics}},\ }\href
  {https://doi.org/10.1016/0003-4916(83)90034-9} {\bibfield  {journal}
  {\bibinfo  {journal} {Annals Phys.}\ }\textbf {\bibinfo {volume} {146}},\
  \bibinfo {pages} {262} (\bibinfo {year} {1983})}\BibitemShut {NoStop}%
\bibitem [{\citenamefont {Cooper}\ \emph {et~al.}(1995)\citenamefont {Cooper},
  \citenamefont {Khare},\ and\ \citenamefont {Sukhatme}}]{Cooper:1994eh}%
  \BibitemOpen
  \bibfield  {author} {\bibinfo {author} {\bibfnamefont {F.}~\bibnamefont
  {Cooper}}, \bibinfo {author} {\bibfnamefont {A.}~\bibnamefont {Khare}},\ and\
  \bibinfo {author} {\bibfnamefont {U.}~\bibnamefont {Sukhatme}},\ }\bibfield
  {title} {\bibinfo {title} {{Supersymmetry and quantum mechanics}},\ }\href
  {https://doi.org/10.1016/0370-1573(94)00080-M} {\bibfield  {journal}
  {\bibinfo  {journal} {Phys. Rept.}\ }\textbf {\bibinfo {volume} {251}},\
  \bibinfo {pages} {267} (\bibinfo {year} {1995})},\ \Eprint
  {https://arxiv.org/abs/hep-th/9405029} {arXiv:hep-th/9405029} \BibitemShut
  {NoStop}%
\bibitem [{\citenamefont {{G{\'o}mez-Ullate}}\ \emph
  {et~al.}(2004)\citenamefont {{G{\'o}mez-Ullate}}, \citenamefont {{Kamran}},\
  and\ \citenamefont {{Milson}}}]{GomezUllate:2004}%
  \BibitemOpen
  \bibfield  {author} {\bibinfo {author} {\bibfnamefont {D.}~\bibnamefont
  {{G{\'o}mez-Ullate}}}, \bibinfo {author} {\bibfnamefont {N.}~\bibnamefont
  {{Kamran}}},\ and\ \bibinfo {author} {\bibfnamefont {R.}~\bibnamefont
  {{Milson}}},\ }\bibfield  {title} {\bibinfo {title} {{Supersymmetry and
  algebraic Darboux transformations}},\ }\href
  {https://doi.org/10.1088/0305-4470/37/43/004} {\bibfield  {journal} {\bibinfo
   {journal} {Journal of Physics A Mathematical General}\ }\textbf {\bibinfo
  {volume} {37}},\ \bibinfo {pages} {10065} (\bibinfo {year} {2004})},\ \Eprint
  {https://arxiv.org/abs/nlin/0402052} {arXiv:nlin/0402052 [nlin.SI]}
  \BibitemShut {NoStop}%
\bibitem [{\citenamefont {Arnowitt}\ \emph {et~al.}(2008)\citenamefont
  {Arnowitt}, \citenamefont {Deser},\ and\ \citenamefont
  {Misner}}]{Arnowitt:1962hi}%
  \BibitemOpen
  \bibfield  {author} {\bibinfo {author} {\bibfnamefont {R.~L.}\ \bibnamefont
  {Arnowitt}}, \bibinfo {author} {\bibfnamefont {S.}~\bibnamefont {Deser}},\
  and\ \bibinfo {author} {\bibfnamefont {C.~W.}\ \bibnamefont {Misner}},\
  }\bibfield  {title} {\bibinfo {title} {{The Dynamics of general
  relativity}},\ }\href {https://doi.org/10.1007/s10714-008-0661-1} {\bibfield
  {journal} {\bibinfo  {journal} {Gen. Rel. Grav.}\ }\textbf {\bibinfo {volume}
  {40}},\ \bibinfo {pages} {1997} (\bibinfo {year} {2008})},\ \Eprint
  {https://arxiv.org/abs/gr-qc/0405109} {arXiv:gr-qc/0405109 [gr-qc]}
  \BibitemShut {NoStop}%
\bibitem [{\citenamefont {Pretorius}(2005{\natexlab{a}})}]{Pretorius:2004jg}%
  \BibitemOpen
  \bibfield  {author} {\bibinfo {author} {\bibfnamefont {F.}~\bibnamefont
  {Pretorius}},\ }\bibfield  {title} {\bibinfo {title} {{Numerical relativity
  using a generalized harmonic decomposition}},\ }\href
  {https://doi.org/10.1088/0264-9381/22/2/014} {\bibfield  {journal} {\bibinfo
  {journal} {Class. Quant. Grav.}\ }\textbf {\bibinfo {volume} {22}},\ \bibinfo
  {pages} {425} (\bibinfo {year} {2005}{\natexlab{a}})},\ \Eprint
  {https://arxiv.org/abs/gr-qc/0407110} {arXiv:gr-qc/0407110} \BibitemShut
  {NoStop}%
\bibitem [{\citenamefont {Pretorius}(2005{\natexlab{b}})}]{Pretorius:2005gq}%
  \BibitemOpen
  \bibfield  {author} {\bibinfo {author} {\bibfnamefont {F.}~\bibnamefont
  {Pretorius}},\ }\bibfield  {title} {\bibinfo {title} {{Evolution of binary
  black hole spacetimes}},\ }\href
  {https://doi.org/10.1103/PhysRevLett.95.121101} {\bibfield  {journal}
  {\bibinfo  {journal} {Phys. Rev. Lett.}\ }\textbf {\bibinfo {volume} {95}},\
  \bibinfo {pages} {121101} (\bibinfo {year} {2005}{\natexlab{b}})},\ \Eprint
  {https://arxiv.org/abs/gr-qc/0507014} {arXiv:gr-qc/0507014} \BibitemShut
  {NoStop}%
\bibitem [{\citenamefont {Campanelli}\ \emph {et~al.}(2006)\citenamefont
  {Campanelli}, \citenamefont {Lousto}, \citenamefont {Marronetti},\ and\
  \citenamefont {Zlochower}}]{Campanelli:2005dd}%
  \BibitemOpen
  \bibfield  {author} {\bibinfo {author} {\bibfnamefont {M.}~\bibnamefont
  {Campanelli}}, \bibinfo {author} {\bibfnamefont {C.~O.}\ \bibnamefont
  {Lousto}}, \bibinfo {author} {\bibfnamefont {P.}~\bibnamefont {Marronetti}},\
  and\ \bibinfo {author} {\bibfnamefont {Y.}~\bibnamefont {Zlochower}},\
  }\bibfield  {title} {\bibinfo {title} {{Accurate evolutions of orbiting
  black-hole binaries without excision}},\ }\href
  {https://doi.org/10.1103/PhysRevLett.96.111101} {\bibfield  {journal}
  {\bibinfo  {journal} {Phys. Rev. Lett.}\ }\textbf {\bibinfo {volume} {96}},\
  \bibinfo {pages} {111101} (\bibinfo {year} {2006})},\ \Eprint
  {https://arxiv.org/abs/gr-qc/0511048} {arXiv:gr-qc/0511048} \BibitemShut
  {NoStop}%
\bibitem [{\citenamefont {Baker}\ \emph {et~al.}(2006)\citenamefont {Baker},
  \citenamefont {Centrella}, \citenamefont {Choi}, \citenamefont {Koppitz},\
  and\ \citenamefont {van Meter}}]{Baker:2005vv}%
  \BibitemOpen
  \bibfield  {author} {\bibinfo {author} {\bibfnamefont {J.~G.}\ \bibnamefont
  {Baker}}, \bibinfo {author} {\bibfnamefont {J.}~\bibnamefont {Centrella}},
  \bibinfo {author} {\bibfnamefont {D.-I.}\ \bibnamefont {Choi}}, \bibinfo
  {author} {\bibfnamefont {M.}~\bibnamefont {Koppitz}},\ and\ \bibinfo {author}
  {\bibfnamefont {J.}~\bibnamefont {van Meter}},\ }\bibfield  {title} {\bibinfo
  {title} {{Gravitational wave extraction from an inspiraling configuration of
  merging black holes}},\ }\href
  {https://doi.org/10.1103/PhysRevLett.96.111102} {\bibfield  {journal}
  {\bibinfo  {journal} {Phys. Rev. Lett.}\ }\textbf {\bibinfo {volume} {96}},\
  \bibinfo {pages} {111102} (\bibinfo {year} {2006})},\ \Eprint
  {https://arxiv.org/abs/gr-qc/0511103} {arXiv:gr-qc/0511103} \BibitemShut
  {NoStop}%
\bibitem [{\citenamefont {Baumgarte}\ and\ \citenamefont
  {Shapiro}(1998)}]{Baumgarte:1998te}%
  \BibitemOpen
  \bibfield  {author} {\bibinfo {author} {\bibfnamefont {T.~W.}\ \bibnamefont
  {Baumgarte}}\ and\ \bibinfo {author} {\bibfnamefont {S.~L.}\ \bibnamefont
  {Shapiro}},\ }\bibfield  {title} {\bibinfo {title} {{On the numerical
  integration of Einstein's field equations}},\ }\href
  {https://doi.org/10.1103/PhysRevD.59.024007} {\bibfield  {journal} {\bibinfo
  {journal} {Phys. Rev. D}\ }\textbf {\bibinfo {volume} {59}},\ \bibinfo
  {pages} {024007} (\bibinfo {year} {1998})},\ \Eprint
  {https://arxiv.org/abs/gr-qc/9810065} {arXiv:gr-qc/9810065} \BibitemShut
  {NoStop}%
\bibitem [{\citenamefont {Shibata}\ and\ \citenamefont
  {Nakamura}(1995)}]{Shibata:1995we}%
  \BibitemOpen
  \bibfield  {author} {\bibinfo {author} {\bibfnamefont {M.}~\bibnamefont
  {Shibata}}\ and\ \bibinfo {author} {\bibfnamefont {T.}~\bibnamefont
  {Nakamura}},\ }\bibfield  {title} {\bibinfo {title} {{Evolution of
  three-dimensional gravitational waves: Harmonic slicing case}},\ }\href
  {https://doi.org/10.1103/PhysRevD.52.5428} {\bibfield  {journal} {\bibinfo
  {journal} {Phys. Rev. D}\ }\textbf {\bibinfo {volume} {52}},\ \bibinfo
  {pages} {5428} (\bibinfo {year} {1995})}\BibitemShut {NoStop}%
\bibitem [{\citenamefont {Brizuela}\ \emph {et~al.}(2006)\citenamefont
  {Brizuela}, \citenamefont {Martin-Garcia},\ and\ \citenamefont
  {Mena~Marugan}}]{Brizuela:2006ne}%
  \BibitemOpen
  \bibfield  {author} {\bibinfo {author} {\bibfnamefont {D.}~\bibnamefont
  {Brizuela}}, \bibinfo {author} {\bibfnamefont {J.~M.}\ \bibnamefont
  {Martin-Garcia}},\ and\ \bibinfo {author} {\bibfnamefont {G.~A.}\
  \bibnamefont {Mena~Marugan}},\ }\bibfield  {title} {\bibinfo {title} {{Second
  and higher-order perturbations of a spherical spacetime}},\ }\href
  {https://doi.org/10.1103/PhysRevD.74.044039} {\bibfield  {journal} {\bibinfo
  {journal} {Phys. Rev. D}\ }\textbf {\bibinfo {volume} {74}},\ \bibinfo
  {pages} {044039} (\bibinfo {year} {2006})},\ \Eprint
  {https://arxiv.org/abs/gr-qc/0607025} {arXiv:gr-qc/0607025} \BibitemShut
  {NoStop}%
\bibitem [{\citenamefont {Brizuela}\ \emph {et~al.}(2007)\citenamefont
  {Brizuela}, \citenamefont {Martin-Garcia},\ and\ \citenamefont
  {Marugan}}]{Brizuela:2007zza}%
  \BibitemOpen
  \bibfield  {author} {\bibinfo {author} {\bibfnamefont {D.}~\bibnamefont
  {Brizuela}}, \bibinfo {author} {\bibfnamefont {J.~M.}\ \bibnamefont
  {Martin-Garcia}},\ and\ \bibinfo {author} {\bibfnamefont {G.~A.~M.}\
  \bibnamefont {Marugan}},\ }\bibfield  {title} {\bibinfo {title} {{High-order
  gauge-invariant perturbations of a spherical spacetime}},\ }\href
  {https://doi.org/10.1103/PhysRevD.76.024004} {\bibfield  {journal} {\bibinfo
  {journal} {Phys. Rev. D}\ }\textbf {\bibinfo {volume} {76}},\ \bibinfo
  {pages} {024004} (\bibinfo {year} {2007})},\ \Eprint
  {https://arxiv.org/abs/gr-qc/0703069} {arXiv:gr-qc/0703069} \BibitemShut
  {NoStop}%
\bibitem [{\citenamefont {Langlois}\ \emph {et~al.}(2021)\citenamefont
  {Langlois}, \citenamefont {Noui},\ and\ \citenamefont
  {Roussille}}]{Langlois:2021aji}%
  \BibitemOpen
  \bibfield  {author} {\bibinfo {author} {\bibfnamefont {D.}~\bibnamefont
  {Langlois}}, \bibinfo {author} {\bibfnamefont {K.}~\bibnamefont {Noui}},\
  and\ \bibinfo {author} {\bibfnamefont {H.}~\bibnamefont {Roussille}},\
  }\bibfield  {title} {\bibinfo {title} {{Black hole perturbations in modified
  gravity}},\ }\href {https://doi.org/10.1103/PhysRevD.104.124044} {\bibfield
  {journal} {\bibinfo  {journal} {Phys. Rev. D}\ }\textbf {\bibinfo {volume}
  {104}},\ \bibinfo {pages} {124044} (\bibinfo {year} {2021})},\ \Eprint
  {https://arxiv.org/abs/2103.14750} {arXiv:2103.14750 [gr-qc]} \BibitemShut
  {NoStop}%
\bibitem [{\citenamefont {Mena~Marug\'an}\ and\ \citenamefont
  {M\'\i{}nguez-S\'anchez}(2024)}]{MenaMarugan:2024qnj}%
  \BibitemOpen
  \bibfield  {author} {\bibinfo {author} {\bibfnamefont {G.~A.}\ \bibnamefont
  {Mena~Marug\'an}}\ and\ \bibinfo {author} {\bibfnamefont {A.}~\bibnamefont
  {M\'\i{}nguez-S\'anchez}},\ }\bibfield  {title} {\bibinfo {title} {{Axial
  perturbations in Kantowski-Sachs spacetimes and hybrid quantum cosmology}},\
  }\href@noop {} {\  (\bibinfo {year} {2024})},\ \Eprint
  {https://arxiv.org/abs/2402.08307} {arXiv:2402.08307 [gr-qc]} \BibitemShut
  {NoStop}%
\bibitem [{\citenamefont {{Wolfram Research}}(2020)}]{Mathematica}%
  \BibitemOpen
  \bibfield  {author} {\bibinfo {author} {\bibnamefont {{Wolfram Research}}},\
  }\href@noop {} {\bibinfo {title} {{Mathematica 12}}} (\bibinfo {year}
  {2020})\BibitemShut {NoStop}%
\bibitem [{\citenamefont {Birkhoff}(1923)}]{Birkhoff:1923hup}%
  \BibitemOpen
  \bibfield  {author} {\bibinfo {author} {\bibfnamefont {G.~D.}\ \bibnamefont
  {Birkhoff}},\ }\href {https://doi.org/10.4159/harvard.9780674734487} {\emph
  {\bibinfo {title} {{Relativity and Modern Physics}}}}\ (\bibinfo  {publisher}
  {Harvard University Press},\ \bibinfo {year} {1923})\BibitemShut {NoStop}%
\bibitem [{\citenamefont {Jebsen}(1921)}]{Jebsen:1921ori}%
  \BibitemOpen
  \bibfield  {author} {\bibinfo {author} {\bibfnamefont {J.}~\bibnamefont
  {Jebsen}},\ }\bibfield  {title} {\bibinfo {title} {{\"Uber die allgemeinen
  kugelsymmetrischen L\"osungen der Einsteinschen Gravitationsgleichungen im
  Vakuum}},\ }\href@noop {} {\bibfield  {journal} {\bibinfo  {journal} {Ark.
  Mat. Ast. Fys.(Stockholm)}\ }\textbf {\bibinfo {volume} {15}},\ \bibinfo
  {pages} {1} (\bibinfo {year} {1921})}\BibitemShut {NoStop}%
\bibitem [{\citenamefont {Jebsen}(2005)}]{Jebsen:2005grg}%
  \BibitemOpen
  \bibfield  {author} {\bibinfo {author} {\bibfnamefont {J.}~\bibnamefont
  {Jebsen}},\ }\bibfield  {title} {\bibinfo {title} {{On the general
  spherically symmetric solutions of Einstein's gravitational equations in
  vacuo}},\ }\href {https://doi.org/10.1007/s10714-005-0168-y} {\bibfield
  {journal} {\bibinfo  {journal} {Gen. Rel. Grav.}\ }\textbf {\bibinfo {volume}
  {37}},\ \bibinfo {pages} {2253–2259} (\bibinfo {year} {2005})}\BibitemShut
  {NoStop}%
\bibitem [{\citenamefont {Deser}\ and\ \citenamefont
  {Franklin}(2005)}]{Deser:2004gi}%
  \BibitemOpen
  \bibfield  {author} {\bibinfo {author} {\bibfnamefont {S.}~\bibnamefont
  {Deser}}\ and\ \bibinfo {author} {\bibfnamefont {J.}~\bibnamefont
  {Franklin}},\ }\bibfield  {title} {\bibinfo {title} {{Schwarzschild and
  Birkhoff a la Weyl}},\ }\href {https://doi.org/10.1119/1.1830505} {\bibfield
  {journal} {\bibinfo  {journal} {Am. J. Phys.}\ }\textbf {\bibinfo {volume}
  {73}},\ \bibinfo {pages} {261} (\bibinfo {year} {2005})},\ \Eprint
  {https://arxiv.org/abs/gr-qc/0408067} {arXiv:gr-qc/0408067} \BibitemShut
  {NoStop}%
\bibitem [{\citenamefont {Voje~Johansen}\ and\ \citenamefont
  {Ravndal}(2006)}]{VojeJohansen:2005nd}%
  \BibitemOpen
  \bibfield  {author} {\bibinfo {author} {\bibfnamefont {N.}~\bibnamefont
  {Voje~Johansen}}\ and\ \bibinfo {author} {\bibfnamefont {F.}~\bibnamefont
  {Ravndal}},\ }\bibfield  {title} {\bibinfo {title} {{On the discovery of
  Birkhoff's theorem}},\ }\href {https://doi.org/10.1007/s10714-006-0242-0}
  {\bibfield  {journal} {\bibinfo  {journal} {Gen. Rel. Grav.}\ }\textbf
  {\bibinfo {volume} {38}},\ \bibinfo {pages} {537} (\bibinfo {year} {2006})},\
  \Eprint {https://arxiv.org/abs/physics/0508163} {arXiv:physics/0508163}
  \BibitemShut {NoStop}%
\bibitem [{\citenamefont {Eiesland}(1925)}]{Eisland:1925eis}%
  \BibitemOpen
  \bibfield  {author} {\bibinfo {author} {\bibfnamefont {J.}~\bibnamefont
  {Eiesland}},\ }\bibfield  {title} {\bibinfo {title} {{The group of motions of
  an Einstein space}},\ }\href
  {https://doi.org/10.1090/S0002-9947-1925-1501308-7} {\bibfield  {journal}
  {\bibinfo  {journal} {Trans. Amer. Math. Soc.}\ }\textbf {\bibinfo {volume}
  {27}},\ \bibinfo {pages} {213} (\bibinfo {year} {1925})}\BibitemShut
  {NoStop}%
\bibitem [{\citenamefont {Schleich}\ and\ \citenamefont
  {Witt}(2010)}]{Schleich:2009uj}%
  \BibitemOpen
  \bibfield  {author} {\bibinfo {author} {\bibfnamefont {K.}~\bibnamefont
  {Schleich}}\ and\ \bibinfo {author} {\bibfnamefont {D.~M.}\ \bibnamefont
  {Witt}},\ }\bibfield  {title} {\bibinfo {title} {{A simple proof of
  Birkhoff's theorem for cosmological constant}},\ }\href
  {https://doi.org/10.1063/1.3503447} {\bibfield  {journal} {\bibinfo
  {journal} {J. Math. Phys.}\ }\textbf {\bibinfo {volume} {51}},\ \bibinfo
  {pages} {112502} (\bibinfo {year} {2010})},\ \Eprint
  {https://arxiv.org/abs/0908.4110} {arXiv:0908.4110 [gr-qc]} \BibitemShut
  {NoStop}%
\bibitem [{\citenamefont {Kottler}(1918)}]{Kottler:1918}%
  \BibitemOpen
  \bibfield  {author} {\bibinfo {author} {\bibfnamefont {F.}~\bibnamefont
  {Kottler}},\ }\bibfield  {title} {\bibinfo {title} {{{\"U}ber die
  physikalischen Grundlagen der Einsteinschen Gravitationstheorie}},\
  }\href@noop {} {\bibfield  {journal} {\bibinfo  {journal} {{Ann. Phys.
  (Germany)}}\ }\textbf {\bibinfo {volume} {56}},\ \bibinfo {pages} {401–462}
  (\bibinfo {year} {1918})}\BibitemShut {NoStop}%
\bibitem [{\citenamefont {Nariai}(1950)}]{Nariai:1950hna}%
  \BibitemOpen
  \bibfield  {author} {\bibinfo {author} {\bibfnamefont {H.}~\bibnamefont
  {Nariai}},\ }\bibfield  {title} {\bibinfo {title} {{On some static solutions
  of Einstein's gravitational field equations in a spherically symmetric
  case}},\ }\href@noop {} {\bibfield  {journal} {\bibinfo  {journal} {Sci. Rep.
  Tohoku Univ. Series I}\ }\textbf {\bibinfo {volume} {34}},\ \bibinfo {pages}
  {160} (\bibinfo {year} {1950})}\BibitemShut {NoStop}%
\bibitem [{\citenamefont {Nariai}(1999)}]{Nariai:1999nar}%
  \BibitemOpen
  \bibfield  {author} {\bibinfo {author} {\bibfnamefont {H.}~\bibnamefont
  {Nariai}},\ }\bibfield  {title} {\bibinfo {title} {{On a New Cosmological
  Solution of Einstein's Field Equations of Gravitation}},\ }\href
  {https://doi.org/10.1023/A:1026602724948} {\bibfield  {journal} {\bibinfo
  {journal} {Gen. Rel. Grav.}\ }\textbf {\bibinfo {volume} {31}},\ \bibinfo
  {pages} {963} (\bibinfo {year} {1999})}\BibitemShut {NoStop}%
\bibitem [{\citenamefont {{Droste}}(1917)}]{1917KNAB...19..197D}%
  \BibitemOpen
  \bibfield  {author} {\bibinfo {author} {\bibfnamefont {J.}~\bibnamefont
  {{Droste}}},\ }\bibfield  {title} {\bibinfo {title} {{The field of a single
  centre in Einstein's theory of gravitation, and the motion of a particle in
  that field}},\ }\href@noop {} {\bibfield  {journal} {\bibinfo  {journal}
  {Koninklijke Nederlandse Akademie van Wetenschappen Proceedings Series B
  Physical Sciences}\ }\textbf {\bibinfo {volume} {19}},\ \bibinfo {pages}
  {197} (\bibinfo {year} {1917})}\BibitemShut {NoStop}%
\bibitem [{\citenamefont {Abramowitz}\ and\ \citenamefont
  {Stegun}(1972)}]{Abramowitz:1970as}%
  \BibitemOpen
  \bibfield  {author} {\bibinfo {author} {\bibfnamefont {M.}~\bibnamefont
  {Abramowitz}}\ and\ \bibinfo {author} {\bibfnamefont {I.~A.}\ \bibnamefont
  {Stegun}},\ }\href@noop {} {\emph {\bibinfo {title} {{Handbook of
  Mathematical Functions with Formulas, Graphs, and Mathematical Tables}}}}\
  (\bibinfo  {publisher} {Dover},\ \bibinfo {address} {New York},\ \bibinfo
  {year} {1972})\BibitemShut {NoStop}%
\bibitem [{\citenamefont {Press}\ \emph {et~al.}(1992)\citenamefont {Press},
  \citenamefont {Flannery}, \citenamefont {Teukolsky},\ and\ \citenamefont
  {Vetterling}}]{Press:1992nr}%
  \BibitemOpen
  \bibfield  {author} {\bibinfo {author} {\bibfnamefont {W.~H.}\ \bibnamefont
  {Press}}, \bibinfo {author} {\bibfnamefont {B.~P.}\ \bibnamefont {Flannery}},
  \bibinfo {author} {\bibfnamefont {S.~A.}\ \bibnamefont {Teukolsky}},\ and\
  \bibinfo {author} {\bibfnamefont {W.~T.}\ \bibnamefont {Vetterling}},\
  }\href@noop {} {\emph {\bibinfo {title} {{Numerical Recipes: The Art of
  Scientific Computing}}}}\ (\bibinfo  {publisher} {Cambridge University
  Press},\ \bibinfo {address} {Cambridge (UK) and New York},\ \bibinfo {year}
  {1992})\BibitemShut {NoStop}%
\end{thebibliography}
\end{document}